\documentclass[aps,twocolumn,superscriptaddress,floatfix,nofootinbib,showpacs,amsmath,amssymb,altaffilletter]{revtex4-2}
\pdfoutput=1
\usepackage{graphicx}
\usepackage{bm}
\usepackage{url}
\usepackage[hyperindex]{hyperref}
\usepackage{color}
\usepackage[ddmmyy,24hr]{datetime}
\usepackage{bigdelim}
\usepackage{booktabs}
\usepackage{dcolumn}
\usepackage{multirow}
\usepackage[large]{subfigure}
\usepackage{cancel}
\usepackage{stackrel}
\usepackage{paralist}
\usepackage{xspace}
\usepackage{slashed}
\usepackage{amsmath}
\usepackage{cleveref}
\usepackage[normalem]{ulem}

\newcommand{\nua}[1]{\ensuremath{\rlap{\kern-2.5pt\ensuremath{\overset{\scriptscriptstyle(-)}{\phantom{\nu}}}}{\ensuremath{{\nu}_{#1}}}}}
\newcommand{\vet}[1]{\ensuremath{\hskip-1pt\vec{\hskip1pt#1}}}
\usepackage{enumitem}

\newcommand{\QW}{$Q_{W}$}

\newcommand{\cenns}{CE$\nu$NS\xspace}
\newcommand{\hourglass}{\mathbin{\;{\rotatebox{90}{$\small\bowtie$}}}}

\begin{document}

\title{New insights into nuclear physics and weak mixing angle using electroweak probes}

\author{M.~Cadeddu}
\email{matteo.cadeddu@ca.infn.it}
\affiliation{Istituto Nazionale di Fisica Nucleare (INFN), Sezione di Cagliari,
Complesso Universitario di Monserrato - S.P. per Sestu Km 0.700,
09042 Monserrato (Cagliari), Italy}

\author{N.~Cargioli}
\email{nicola.cargioli@ca.infn.it}
\affiliation{Dipartimento di Fisica, Universit\`{a} degli Studi di Cagliari,
and
INFN, Sezione di Cagliari,
Complesso Universitario di Monserrato - S.P. per Sestu Km 0.700,
09042 Monserrato (Cagliari), Italy}
\affiliation{Istituto Nazionale di Fisica Nucleare (INFN), Sezione di Cagliari,
Complesso Universitario di Monserrato - S.P. per Sestu Km 0.700,
09042 Monserrato (Cagliari), Italy}

\author{F.~Dordei}
\email{francesca.dordei@cern.ch}
\affiliation{Istituto Nazionale di Fisica Nucleare (INFN), Sezione di Cagliari,
Complesso Universitario di Monserrato - S.P. per Sestu Km 0.700,
09042 Monserrato (Cagliari), Italy}

\author{C.~Giunti}
\email{carlo.giunti@to.infn.it}
\affiliation{Istituto Nazionale di Fisica Nucleare (INFN), Sezione di Torino, Via P. Giuria 1, I--10125 Torino, Italy}

\author{Y.F.~Li}
\email{liyufeng@ihep.ac.cn}
\affiliation{Institute of High Energy Physics,
Chinese Academy of Sciences, Beijing 100049, China}
\affiliation{School of Physical Sciences, University of Chinese Academy of Sciences, Beijing 100049, China}

\author{E.~Picciau}
\email{emmanuele.picciau@ca.infn.it}
\affiliation{Dipartimento di Fisica, Universit\`{a} degli Studi di Cagliari,
and
INFN, Sezione di Cagliari,
Complesso Universitario di Monserrato - S.P. per Sestu Km 0.700,
09042 Monserrato (Cagliari), Italy}
\affiliation{Istituto Nazionale di Fisica Nucleare (INFN), Sezione di Cagliari,
Complesso Universitario di Monserrato - S.P. per Sestu Km 0.700,
09042 Monserrato (Cagliari), Italy}

\author{C.A.~Ternes}
\email{ternes@to.infn.it}
\affiliation{Istituto Nazionale di Fisica Nucleare (INFN), Sezione di Torino, Via P. Giuria 1, I--10125 Torino, Italy}

\author{Y.Y.~Zhang}
\email{zhangyiyu@ihep.ac.cn}
\affiliation{Institute of High Energy Physics,
Chinese Academy of Sciences, Beijing 100049, China}
\affiliation{School of Physical Sciences, University of Chinese Academy of Sciences, Beijing 100049, China}

\date{11 September 2021}

\begin{abstract}
Using the new results on coherent elastic neutrino-nucleus scattering data in cesium-iodide provided by the COHERENT experiment, we determine a new measurement of the average neutron rms radius of $^{133}\text{Cs}$ and $^{127}\text{I}$. In combination with the atomic parity violation (APV)  experimental  result,  we  derive the most precise measurement of the neutron  rms  radii  of $^{133}\text{Cs}$ and $^{127}\text{I}$, disentangling for the first time the contributions of the two nuclei. By exploiting these measurements we determine the corresponding neutron skin values for $^{133}\text{Cs}$ and $^{127}\text{I}$. These results suggest a preference for models which predict large neutron skin values, as corroborated by the only other electroweak measurements of the neutron skin of $^{208}\text{Pb}$ performed by PREX experiments.
Moreover, for the first time, we obtain a data-driven APV+COHERENT  measurement of the low-energy weak mixing angle with a percent uncertainty, independent of the value of the average neutron rms radius of $^{133}\text{Cs}$ and $^{127}\text{I}$,
that is allowed to vary freely in the fit. The value of the low-energy weak mixing angle that we found is slightly larger than the standard model prediction. 
\end{abstract}

\maketitle

\section{Introduction}
\label{sec:intro}

The detection of coherent elastic neutrino-nucleus scattering (\cenns) in 2017  in cesium-iodide (CsI) by the COHERENT experiment~\cite{Akimov:2017ade,Akimov:2018vzs} 
motivated a burst of studies of diverse physical phenomena, with important implications for high-energy physics, astrophysics, nuclear physics, and beyond~\cite{Cadeddu:2017etk,Papoulias:2019lfi,Coloma:2017ncl,Liao:2017uzy,Kosmas:2017tsq,Denton:2018xmq,AristizabalSierra:2018eqm,Cadeddu:2018dux,Dutta:2019eml,Cadeddu:2018izq,Dutta:2019nbn, Cadeddu:2019eta,Papoulias_2020,Khan_2019, Cadeddu:2020nbr}.
After a fruitful discovery period, recently enriched by the observation of \cenns in argon~\cite{Akimov:2020pdx, Cadeddu:2020lky, Miranda_2020}, a new era of precision measurements has now begun, thanks to the new data recorded by the COHERENT experiment using a CsI target~\cite{Pershey:MCENNS20}. Indeed, the larger \cenns statistics collected together with a refined quenching factor determination allow us to perform stringent tests of the Standard Model (SM).

In previous works~\cite{Cadeddu:2017etk,Cadeddu:2019eta,Khan_2019, Huang:2019ene,Papoulias_2020,Cadeddu:2020lky, Miranda_2020}, it has been shown that the \cenns process gives model-independent information on the neutron nuclear form factor, which is more difficult to obtain than the  proton one.
Form factors represent the Fourier transform of the corresponding nucleon distribution, necessary for obtaining in turn measurements of the neutron rms radius, $R_n$, which is a crucial ingredient of the nuclear matter equation of state (EOS). The latter plays an essential role in understanding nuclei in laboratory experiments and several processes, like heavy ion collisions, and the
structure and evolution of compact  astrophysical  objects  as  neutron  stars~\cite{Lattimer:2004pg, Steiner:2004fi,PhysRevLett.85.5296, PhysRevC.64.027302}. However, while the proton form factor is well known since it can be measured through electromagnetic processes~\cite{Fricke,Angeli}, the same cannot be said for the neutron one. 
Indeed, despite its importance, $R_n$ is still unknown for many nuclei, especially in a model independent way, since the interpretation of hadron scattering experiments depends on the model used to describe nonperturbative strong interactions~\cite{Thiel:2019tkm}.

The \cenns process can also give information on the weak mixing angle, usually referred to as $\sin^2{\vartheta_W}$, a fundamental parameter of the electroweak theory of the SM. However, in the low-energy sector, the most precise $\sin^2{\vartheta_W}$ measurement performed so  far  belongs  to  the  so-called  atomic  parity  violation (APV) experiment, using cesium atoms~\cite{Wood:1997zq, Guena:2004sq}.  This latter measurement depends on the value of $R_n(^{133}\text{Cs})$ that, at the time of Ref.~\cite{Dzuba:2012kx}, could only have been extrapolated from a compilation of antiprotonic atom x-rays data~\cite{PhysRevLett.87.082501}. A combination of COHERENT and APV data is thus highly beneficial to determine simultaneously in a model-independent way these two fundamental parameters, keeping their correlations into account.  

In this paper, we present improved measurements of the average  neutron  rms  radius  of $^{133}\text{Cs}$  and $^{127}\text{I}$  obtained analyzing the
updated COHERENT CsI data~\cite{Pershey:MCENNS20}. In combination with the APV experimental result, we derive the most precise measurement of $R_n$ of $^{133}\text{Cs}$ and $^{127}\text{I}$, disentangling for the first time the contributions of the two nuclei. Moreover, for the first time, we obtain a data-driven measurement of the low-energy weak mixing angle with a percent uncertainty,
independent of the value of the average neutron rms radius of $^{133}\text{Cs}$ and $^{127}\text{I}$
(that is allowed to vary freely in the analysis),
from a simultaneous fit of the COHERENT and APV experimental results.

The plan of the paper is as follows:
in Section~\ref{sec:COHERENT}
we introduce the \cenns cross section
and we describe the method of analysis of the COHERENT data;
in Section~\ref{sec:average}
we present the results on the average CsI neutron rms radius
obtained from the analysis of the
COHERENT \cenns data;
in Section~\ref{sec:APV}
we describe the APV data analysis;
in Section~\ref{sec:radii}
we present the results on the
$^{133}\text{Cs}$ and $^{127}\text{I}$ neutron radii
obtained from the combined analysis of
COHERENT \cenns and APV data;
in Section~\ref{sec:weak}
we discuss the determination of the weak mixing angle
from the combined analysis of
COHERENT \cenns and APV data;
finally, in Section~\ref{sec:conclusions},
we briefly summarize the results presented in the paper.

\section{COHERENT \cenns data analysis}
\label{sec:COHERENT}

\begin{table*}
\centering
\setlength{\tabcolsep}{0.4pt}
\begin{tabular}{l|cccccc|cccccc|cccccc|}
&
\multicolumn{6}{c|}{$^{127}\text{I}$}
&
\multicolumn{6}{c|}{$^{133}\text{Cs}$}
&
\multicolumn{6}{c|}{$^{208}\text{Pb}$}
\\
Model
&
$R_{p}^{\text{point}}$
&
$R_{p}$
&
$R_{n}^{\text{point}}$
&
$R_{n}$
&
$\Delta R_{np}^{\text{point}}$
&
$\Delta R_{np}$
&
$R_{p}^{\text{point}}$
&
$R_{p}$
&
$R_{n}^{\text{point}}$
&
$R_{n}$
&
$\Delta R_{np}^{\text{point}}$
&
$\Delta R_{np}$
&
$R_{p}^{\text{point}}$
&
$R_{p}$
&
$R_{n}^{\text{point}}$
&
$R_{n}$
&
$\Delta R_{np}^{\text{point}}$
&
$\Delta R_{np}$
\\
\hline
SHF SkI3 \cite{Reinhard:1995zz}
&
4.68
&
4.75
&
4.85
&
4.92
&
0.17
&
0.17
&
4.74
&
4.81
&
4.91
&
4.98
&
0.18
&
0.18
&
5.43
&
5.49
&
5.66
&
5.72
&
0.23
&
0.23
\\
SHF SkI4 \cite{Reinhard:1995zz}
&
4.67
&
4.74
&
4.81
&
4.88
&
0.14
&
0.14
&
4.73
&
4.80
&
4.88
&
4.95
&
0.15
&
0.14
&
5.43
&
5.49
&
5.61
&
5.67
&
0.18
&
0.18
\\
SHF Sly4 \cite{Chabanat:1997un}
&
4.71
&
4.78
&
4.84
&
4.91
&
0.13
&
0.13
&
4.78
&
4.85
&
4.90
&
4.98
&
0.13
&
0.13
&
5.46
&
5.53
&
5.62
&
5.69
&
0.16
&
0.16
\\
SHF Sly5 \cite{Chabanat:1997un}
&
4.70
&
4.77
&
4.83
&
4.90
&
0.13
&
0.13
&
4.77
&
4.84
&
4.90
&
4.97
&
0.13
&
0.13
&
5.45
&
5.52
&
5.62
&
5.68
&
0.16
&
0.16
\\
SHF Sly6 \cite{Chabanat:1997un}
&
4.70
&
4.77
&
4.83
&
4.90
&
0.13
&
0.13
&
4.77
&
4.84
&
4.89
&
4.97
&
0.13
&
0.13
&
5.46
&
5.52
&
5.62
&
5.68
&
0.16
&
0.16
\\
SHF Sly4d \cite{Kim-Otsuka-Bonche-1997}
&
4.71
&
4.79
&
4.84
&
4.91
&
0.13
&
0.12
&
4.78
&
4.85
&
4.90
&
4.97
&
0.12
&
0.12
&
5.48
&
5.54
&
5.65
&
5.71
&
0.17
&
0.17
\\
SHF SV-bas \cite{Klupfel:2008af}
&
4.68
&
4.76
&
4.80
&
4.88
&
0.12
&
0.12
&
4.74
&
4.82
&
4.87
&
4.94
&
0.13
&
0.12
&
5.44
&
5.51
&
5.60
&
5.66
&
0.15
&
0.15
\\
SHF UNEDF0 \cite{Kortelainen:2010hv}
&
4.69
&
4.76
&
4.83
&
4.91
&
0.14
&
0.14
&
4.76
&
4.83
&
4.92
&
4.99
&
0.16
&
0.15
&
5.46
&
5.52
&
5.65
&
5.71
&
0.19
&
0.19
\\
SHF UNEDF1 \cite{Kortelainen:2011ft}
&
4.68
&
4.76
&
4.83
&
4.91
&
0.15
&
0.15
&
4.76
&
4.83
&
4.90
&
4.98
&
0.15
&
0.15
&
5.46
&
5.52
&
5.64
&
5.70
&
0.18
&
0.17
\\
SHF SkM* \cite{Bartel:1982ed}
&
4.71
&
4.78
&
4.84
&
4.91
&
0.13
&
0.13
&
4.76
&
4.84
&
4.90
&
4.97
&
0.13
&
0.13
&
5.46
&
5.52
&
5.63
&
5.69
&
0.17
&
0.17
\\
SHF SkP \cite{Dobaczewski:1983zc}
&
4.72
&
4.80
&
4.84
&
4.91
&
0.12
&
0.12
&
4.79
&
4.86
&
4.91
&
4.98
&
0.12
&
0.12
&
5.48
&
5.54
&
5.62
&
5.68
&
0.15
&
0.14
\\
RMF DD-ME2 \cite{Niksic:2002yp}
&
4.67
&
4.75
&
4.82
&
4.89
&
0.15
&
0.15
&
4.74
&
4.81
&
4.89
&
4.96
&
0.15
&
0.15
&
5.46
&
5.52
&
5.65
&
5.71
&
0.19
&
0.19
\\
RMF DD-PC1 \cite{Niksic:2008vp}
&
4.68
&
4.75
&
4.83
&
4.90
&
0.15
&
0.15
&
4.74
&
4.82
&
4.90
&
4.97
&
0.16
&
0.15
&
5.45
&
5.52
&
5.65
&
5.71
&
0.20
&
0.20
\\
RMF NL1 \cite{Reinhard:1986qq}
&
4.70
&
4.78
&
4.94
&
5.01
&
0.23
&
0.23
&
4.76
&
4.84
&
5.01
&
5.08
&
0.25
&
0.24
&
5.48
&
5.55
&
5.80
&
5.86
&
0.32
&
0.31
\\
RMF NL3 \cite{Lalazissis:1996rd}
&
4.69
&
4.77
&
4.89
&
4.96
&
0.20
&
0.19
&
4.75
&
4.82
&
4.95
&
5.03
&
0.21
&
0.20
&
5.47
&
5.53
&
5.74
&
5.80
&
0.28
&
0.27
\\
RMF NL-Z2 \cite{Bender:1999yt}
&
4.73
&
4.80
&
4.94
&
5.01
&
0.21
&
0.21
&
4.79
&
4.86
&
5.01
&
5.08
&
0.22
&
0.22
&
5.52
&
5.58
&
5.81
&
5.87
&
0.29
&
0.29
\\
RMF NL-SH \cite{Sharma:1993it}
&
4.68
&
4.75
&
4.86
&
4.94
&
0.19
&
0.18
&
4.74
&
4.81
&
4.93
&
5.00
&
0.19
&
0.19
&
5.45
&
5.52
&
5.72
&
5.78
&
0.26
&
0.26
\\
\hline
\end{tabular}
\caption{ \label{tab:models}
Theoretical values in units of fermi of the rms
proton and neutron point and physical radii and the neutron skin of
$^{127}\text{I}$,
$^{133}\text{Cs}$,
and
$^{208}\text{Pb}$
obtained with
nonrelativistic Skyrme-Hartree-Fock (SHF)
and
relativistic mean field (RMF)
nuclear models.
}
\end{table*}

The SM \cenns differential cross section as a function of the true nuclear kinetic recoil energy $T'_\mathrm{nr}$,
considering a spin-zero nucleus $\mathcal{N}$ with $Z$ protons and $N$ neutrons, is given by~\cite{Drukier:1983gj,Barranco:2005yy,Patton:2012jr}
\begin{align}\nonumber
\dfrac{d\sigma_{\nu_{\ell}\text{-}\mathcal{N}}}{d T'_\mathrm{nr}}
(E,T'_\mathrm{nr})
\null & \null
= 
\dfrac{G_{\text{F}}^2 M}{\pi}
\left(
1 - \dfrac{M T'_\mathrm{nr}}{2 E^2}
\right)
\times
\\
\null & \null
\left[
g_{V}^{p}
Z
F_{Z}(|\vet{q}|^2)
+
g_{V}^{n}
N
F_{N}(|\vet{q}|^2)
\right]^2
,
\label{cs-std}
\end{align}
where $G_{\text{F}}$ is the Fermi constant, $\ell = e, \mu, \tau$ is the neutrino flavor, $E$ is the neutrino energy and $|\vec{q}| \simeq \sqrt{2 M T'_\mathrm{nr}}$ is the three-momentum transfer,
$M$ being the nuclear mass. As introduced in  Ref.~\cite{Cadeddu:2020lky}, we modify the tree-level values of the vector couplings 
$g_{V}^{p}=1/2- 2 \sin^2\!\vartheta^{\mathrm{SM}}_{W}=0.0229$
and
$g_{V}^{n}=-1/2$,
where $\sin^2\vartheta_{W}^{\mathrm{SM}}
=
0.23857 \pm 0.00005
\,
\text{\protect\cite{Tanabashi:2018oca}}$
is the low-energy value of the weak mixing angle, in order to take into account
radiative corrections in the $\overline{\text{MS}}$ scheme~\cite{Erler:2013xha}, namely
$g_{V}^{p}(\nu_{e})
=0.0401$,
$g_{V}^{p}(\nu_{\mu})
=
0.0318$, and
$g_{V}^{n}
=-0.5094$.
In Eq.~(\ref{cs-std}),
$F_{Z}(|\vet{q}|^2)$
and
$F_{N}(|\vet{q}|^2)$
are, respectively, the proton and neutron form factors.
They 
describe the loss of coherence for
$|\vet{q}| R_{p} \gtrsim 1$
and
$|\vet{q}| R_{n} \gtrsim 1$,
where $R_{p}$ is the rms radius of the proton distribution.
The three most popular parameterizations of the form factors are the
symmetrized Fermi~\cite{Piekarewicz:2016vbn},
Helm~\cite{Helm:1956zz}, and
Klein-Nystrand~\cite{Klein:1999qj},
that give practically identical results\footnote{
These parameterizations of the form factors depend on two parameters:
the rms radius and a parameter that quantifies the nuclear surface thickness.
For all the nuclear proton and neutron form factors,
we considered the standard surface thickness of 2.30 fm~\cite{Fricke:1995zz},
that is in agreement with the values extracted from measured charge distributions
of similar nuclei~\cite{Friedrich:1982esq}.
We verified that the results are practically independent of
small variations of the value of the surface thickness.}.

For the values of $R_{p}$, we correct the charge radii determined experimentally from muonic atom spectroscopy~\cite{Fricke:1995zz,Angeli:2013epw} as in Ref.~\cite{Cadeddu:2020lky}, obtaining
\begin{align}
\null & \null
R_{p}(^{133}\text{Cs})
=
4.821 \pm 0.005 \, \text{fm}
,
\label{RpCs}
\\
\null & \null
R_{p}(^{127}\text{I})
=
4.766 \pm 0.008 \, \text{fm}
.
\label{RpI}
\end{align}

For the neutron distribution there is only poor knowledge of $R_{n}(^{133}\text{Cs})$  and $R_{n}(^{127}\text{I})$
obtained in the analyses of the COHERENT 2017 data~\cite{Cadeddu:2017etk,Papoulias:2019lfi,Cadeddu:2018dux,Huang:2019ene,Papoulias_2020,Khan_2019,Cadeddu:2019eta}.
Plausible theoretical values
can be obtained using the
recent nuclear shell model (NSM) estimate of the corresponding neutron skins,
the differences between the neutron and the proton rms radii,
$0.27\, \text{fm}$ and $0.26\, \text{fm}$~\cite{Hoferichter_2020},
leading to
\begin{equation}
R_{n}^{\text{NSM}}(^{133}\text{Cs}) \simeq 5.09 \, \text{fm}
,
\quad
R_{n}^{\text{NSM}}(\mathrm{^{127}\text{I}}) \simeq 5.03 \, \text{fm}
.
\label{RnNSM}
\end{equation}
These values are slightly larger than those
in Table~\ref{tab:models},
that we obtained using
nonrelativistic Skyrme-Hartree-Fock (SHF)
and
relativistic mean-field (RMF)
nuclear models.
We calculated the physical proton and neutron radii
$R_{p,n}$
from the corresponding point-radii
$R_{p,n}^{\text{point}}$
given by the models
adding in quadrature the contribution of the
rms nucleon radius
$
\langle r_{N}^2 \rangle^{1/2}
\simeq
0.84 \, \text{fm}
$,
that is considered to be approximately equal for the proton and the neutron,
\begin{equation}
R_{p,n}^2
=
(R_{p,n}^{\text{point}})^2
+
\langle r_{N}^2 \rangle
.
\label{Rpn}
\end{equation}

The analysis of COHERENT data is performed in each nuclear recoil energy bin $i$ and time interval $j$ with the least-squares function
\begin{align}
\chi^2_{\mathrm{C}}
=
\null & \null
\sum_{i=2}^{9}
\sum_{j=1}^{11}
\left(
\dfrac
{
N_{ij}^{\text{exp}}
- \sum_{z=1}^{3}( 1 + \eta_{z} ) N_{ij}^{z}
}
{ \sigma_{ij} }
\right)^2
+ \sum_{z=1}^{3}
\left(
\dfrac{ \eta_{z} }{ \sigma_{z} }
\right)^2
,
\label{chi2coherent}
\end{align}
where $z=1,2,3$ stands for \cenns, Beam-Related Neutron (BRN) and Steady-State (SS) backgrounds.
$N_{ij}^{\text{exp}}$ is the experimental event number, $N_{ij}^{\text{\cenns}}$ is the predicted number of \cenns events in Eq.~\eqref{Nij},
$N_{ij}^{\text{BRN}}$ and $N_{ij}^{\text{SS}}$ are the estimated number of BRN and SS background events, respectively, and
$\sigma_{ij}$ is the statistical uncertainty, all taken from Ref.~\cite{Pershey:MCENNS20}.
The uncertainties of the $\eta_z$ nuisance parameters, which quantify
the systematic uncertainty of the signal rate, of the BRN and of the SS background rates, are $\sigma_{\text{\cenns}}
=
13\%$, $
\sigma_{\text{BRN}}
=
0.9\%$ and $
\sigma_{\text{SS}}
=
3\%$~\cite{Pershey:MCENNS20}.

We calculated the \cenns event number $N_{i}^{\text{\cenns}}$ in each nuclear recoil energy bin $i$ with
\begin{align}
N_{i}^{\text{\cenns}}
=
\null & \null
N_\text{CsI}
\int_{T_{\text{nr}}^{i}}^{T_{\text{nr}}^{i+1}}
dT_{\text{nr}}
\,
A(T_{\text{nr}})
\nonumber
\\
\null & \null
\times
\int_{0}^{T^{\prime\text{max}}_{\text{nr}}}
dT'_{\text{nr}}
\,
R(T_{\text{nr}},T'_{\text{nr}})
\int_{E_{\text{min}}(T'_{\text{nr}})}^{E_{\text{max}}}
dE
\nonumber
\\
\null & \null
\times
\sum_{\nu=\nu_e,\nu_\mu,\overline{\nu}_\mu}
\frac{dN_\nu}{dE}(E)
\,
\dfrac{d\sigma_{\nu\text{-}\text{CsI}}}{d T'_\text{nr}}(E,T'_{\text{nr}})
,
\label{Ni}
\end{align}
where
$T_{\text{nr}}$ is the reconstructed nuclear recoil energy,
$A(T_{\text{nr}})$ is the energy-dependent detector efficiency,
$T^{\prime\text{max}}_{\text{nr}} = 2 E_{\text{max}}^2 / M$,
$E_{\text{max}} = m_\mu/2 \sim 52.8$ MeV, $m_\mu$ being the muon mass, $E_{\text{min}}(T'_{\text{nr}}) = \sqrt{MT'_\text{nr}/2}$,
 $dN_\nu/dE$ is the neutrino flux integrated over the experiment lifetime and $N_\text{CsI}$ is the number of
$^{133}\text{Cs}$ and $^{127}\text{I}$ atoms in the detector. The latter is given by $N_\mathrm{A}\,M_{\mathrm{det}}/M_{\mathrm{CsI}}$, where $N_\mathrm{A}$ is the Avogadro number, $M_{\mathrm{det}}$ is the detector active mass equal to 14.6~kg and $M_{\mathrm{CsI}}$ is the molar mass of CsI. The neutrino flux from the spallation neutron source, $dN_\nu/dE$, is given by the sum of the prompt $\nu_\mu$ component and the delayed $\nu_e$ and $\bar{\nu}_\mu$ components, considering $8.48\times10^{-2}$ as the number of neutrinos per flavor
that are produced for each proton-on-target (POT).
A number of POT equal to $3.20 \times 10^{23}$
and a distance of $19.3 \, \text{m} $
between the source and the COHERENT detector are used.
The energy resolution function, $R(T_{\text{nr}},T'_{\text{nr}})$, is parameterized in terms of the number of photoelectrons (PE) following Ref.~\cite{Konovalov:MCENNS20}.
The number of PE is related to the nuclear recoil kinetic energy thanks to the light yield $13.348 \, N_{\text{PE}} / \text{keV}$~\cite{Konovalov:MCENNS20} and the quenching factor, $f_{\text{Q}}(T_{\text{nr}})$, that is parameterized as a fourth order polynomial as in Ref.~\cite{Konovalov:MCENNS20}.

In order to exploit also the arrival time information, we calculated the \cenns event number, $N_{ij}^{\text{\cenns}}$, in each nuclear recoil energy bin $i$ and time interval $j$ with
\begin{equation}
N_{ij}^{\text{\cenns}}
=
(N_{i}^{\text{\cenns}})_{\nu_{\mu}} P_{j}^{(\nu_{\mu})}
+
(N_{i}^{\text{\cenns}})_{\nu_{e},\bar\nu_{\mu}} P_{j}^{(\nu_{e},\bar\nu_{\mu})}
,
\label{Nij}
\end{equation}
where
$P_{j}^{(\nu_{\mu})}$ and $P_{j}^{(\nu_{e},\bar\nu_{\mu})}$
are obtained by integrating the arrival time distributions
in the corresponding time intervals with the time-dependent efficiency function ~\cite{Konovalov:MCENNS20,Pershey:MCENNS20}.

Using the SM inputs, the experimental values of
$R_{p}(^{133}\text{Cs})$
and
$R_{p}(^{127}\text{I})$
in Eqs.~\eqref{RpCs} and~\eqref{RpI},
and the
NSM values of
$R_{n}^{\text{EFT}}(^{133}\text{Cs})$
and
$R_{n}^{\text{EFT}}(\mathrm{^{127}\text{I}})$
in Eq.~\eqref{RnNSM}, the total number of predicted 
events is found to be $N^{\text{\cenns}}=311.8$.

\section{Average CsI neutron radius}
\label{sec:average}

We fitted the COHERENT CsI data to get information on the average neutron rms radius of $^{133}\text{Cs}$ and $^{127}\text{I}$, $R_n(\mathrm{CsI})$, obtaining\footnote{
We considered also a fit with equal $^{133}\text{Cs}$ and $^{127}\text{I}$
neutron skins,
which gave the almost equivalent result
$ R_{n}(\mathrm{Cs}) = 5.56 {}^{+0.45}_{-0.43} \, \text{fm}$
and
$ R_{n}(\mathrm{I}) = 5.51 {}^{+0.45}_{-0.43} \, \text{fm}$.
}
\begin{equation}
R_{n}(\mathrm{CsI})
=
5.55 \pm 0.44 \, \text{fm}
.
\label{Rn-fit}
\end{equation}
This result is almost a factor of 2.5 more precise than previous determinations~\cite{Cadeddu:2017etk} using the 2017 COHERENT dataset~\cite{Akimov:2017ade,Akimov:2018vzs}.

The average of the NSM expected values in Eq.~(\ref{RnNSM}),
$R_{n}^{\text{NSM}}(\text{CsI}) \simeq 5.06 \, \text{fm}$,
is compatible with the determination in Eq.~(\ref{Rn-fit})
at $1.1\sigma$.

The SHF and RMF predictions in Table~\ref{tab:models} give
$R_{n}^{\text{SHF}}(\text{CsI}) \simeq 4.92 - 4.95 \, \text{fm}$
and
$R_{n}^{\text{RMF}}(\text{CsI}) \simeq 4.93 - 5.05 \, \text{fm}$,
that differ from the value in Eq.~\eqref{Rn-fit} by about
$1.4\sigma$
and
$1.2\sigma$,
respectively.

Therefore,
the COHERENT CsI data still do not allow to exclude some nuclear models,
but tends to favor those that predict a relatively large value of $R_{n}$,
as the NSM, RMF NL1, and RMF NL-Z2.

\section{APV data analysis}
\label{sec:APV}

The COHERENT data do not allow us to disentangle the contributions of the two nuclei, but only to constrain their average. A separation of the two contributions can be achieved in combination with the low-energy measurement of the weak charge, $Q_W$, of $^{133}\text{Cs}$ in APV experiments, that is related to the  weak  mixing  angle  through the relation
\begin{align}
Q_W^{\mathrm{th}}(\sin^2\vartheta_W)
=
&
- 2 [ Z (g_{A V}^{e p}(\sin^2\vartheta_W) + 0.00005)
\nonumber
\\
&
+ N (g_{A V}^{e n} + 0.00006) ] 
\left( 1 - \dfrac{\alpha}{2 \pi} \right)
,
\label{QWSMthe}
\end{align}
where $\alpha$ is the fine-structure constant and the couplings of electrons to nucleons,
$g_{A V}^{e p}$ and $g_{A V}^{e n}$, using $\sin^2\vartheta_W^{\mathrm{SM}}$ and taking into account  radiative corrections in the SM~\cite{Zyla:2020zbs,Erler:2013xha,reverler}, are given by
(see Appendix~\ref{app:radiative})
\begin{align}
\null & \null
g_{A V,\text{SM}}^{e p}
=
2 g_{A V,\text{SM}}^{e u} + g_{A V,\text{SM}}^{e d}
=
- 0.0357
,
\label{gAVeppdg}
\\
\null & \null
g_{A V,\text{SM}}^{e n}
=
g_{A V,\text{SM}}^{e u} + 2 g_{A V,\text{SM}}^{e d}
=
0.495
,
\label{gAVenpdg}
\end{align}
where $g_{A V,\text{SM}}^{e u}
=
- 0.1888$ and $g_{A V,\text{SM}}^{e d}
=
0.3419$. These values give
$Q_{{W}}^{\text{SM}}
=
-73.23 \pm 0.01
\label{QWSMval}$.

Experimentally, the weak charge of a nucleus is extracted from the ratio of the parity violating amplitude, $E_{\text{PNC}}$, to the Stark vector transition polarizability, $\beta$, and by calculating theoretically $E_{\rm PNC}$ in terms of \QW
\begin{equation}
\label{QWeq}
Q_W= N \left( \dfrac{{\rm Im}\, E_{\rm PNC}}{\beta} \right)_{\rm exp.} 
\left( \dfrac{Q_W}{N\, {\rm Im}\, E_{\rm PNC}} \right)_{\rm th.} \beta_{\rm exp.+th.}\,,
\end{equation}
where $\beta_{\rm exp.+th.}$ and $(\mathrm{Im}\, E_{\rm PNC})_{\rm th.}$ are determined from atomic theory, and Im stands for imaginary part~\cite{Zyla:2020zbs}.
We use 
$({\rm Im}\, E_{\rm PNC}/{\beta})_{\rm exp} = (-3.0967 \pm 0.0107) \times 10^{-13} |e|/a_B^2$~\cite{Zyla:2020zbs}, where $a_B$ is the Bohr radius and $|e|$ is the absolute value of the electric charge, and $\beta_{\rm exp.+th.} = (27.064 \pm 0.033)\, a_B^3$~\cite{Zyla:2020zbs}.

For the imaginary part of $E_{\rm PNC}$ we use
$({\rm Im}\, E_{\rm PNC})_{\rm th.}^{\rm w.n.s.}=(0.8995\pm0.0040)\times10^{-11}|e|a_B \frac{Q_W}{N}$~\cite{Dzuba:2012kx}, where we subtracted the correction called ``neutron skin", introduced in Ref.~\cite{PhysRevA.65.012106} to take into account the difference between $R_n$ and $R_p$ that is not considered in the nominal atomic theory derivation. Here we remove this correction in order to be able to directly evaluate $R_n$ from a combined fit with the COHERENT data.
The neutron skin corrected value of the weak charge, $Q_W^\mathrm{n.s.}(R_n)$, is thus retrieved by summing to $({\rm Im}\, E_{\rm PNC})_{\rm th.}^{\rm w.n.s.}$ the correcting term $\delta E^\mathrm{n.s.}_\mathrm{PNC}(R_n)= \left[ (\mathrm{N}/Q_W)\left(1-(q_n(R_n)/q_p)\right) E_\mathrm{PNC}^\mathrm{w.n.s.} \right]$~\cite{Viatkina,Cadeddu:2019eta,Cadeddu:2018izq}.
The factors $q_p$ and $q_n$ incorporate the radial dependence of the electron axial transition matrix element by considering the proton and the neutron spatial distribution, respectively~\cite{PhysRevC.46.2587,Pollock1999,James_1999,Horowitz2001,Viatkina}.
Our calculation of $q_p$ and $q_n$ is described in Appendix~\ref{app:qpqn}.

\section{$^{133}\text{Cs}$ and $^{127}\text{I}$ neutron radii}
\label{sec:radii}

We performed the combined APV and COHERENT analysis with the least-squares function
\begin{align}
\chi^2
=
\chi^2_{\text{C}}
+
\left(
\dfrac{
Q_W^{\rm Cs\,n.s.}(R_n)
-
Q_W^{\mathrm{th}}(\sin^2\vartheta_W)
}{ \sigma_{\rm APV}(R_n,\sin^2\vartheta_W) }
\right)^2
\,,
\label{chicoherentapvRn}
\end{align} 
where the first term is defined in Eq.~(\ref{chi2coherent}) and the second term represents the contribution of the APV measurement, where $\sigma_{\rm APV}$ is the total uncertainty.

Considering that the COHERENT data depend separately on $R_n(^{133}\mathrm{Cs})$ and $R_n(^{127}\mathrm{I})$,
while APV depends only on $R_n(^{133}\mathrm{Cs})$,
we disentangled for the first time the two nuclear contributions.
Assuming $\sin^2\vartheta_W=\sin^2\vartheta_W^{\rm SM}$, we obtained
\begin{equation}
R_{n}(^{133}\mathrm{Cs})
=
5.27
{}^{+0.33}_{-0.33}\, \text{fm}
\,,\,
R_{n}(^{127}\mathrm{I})
=
5.9
{}^{+1.0}_{-0.9}\, \text{fm}
.
\label{RnCs-RnI-fit}
\end{equation}

\begin{figure}[!t]
\centering
\includegraphics*[width=\linewidth]{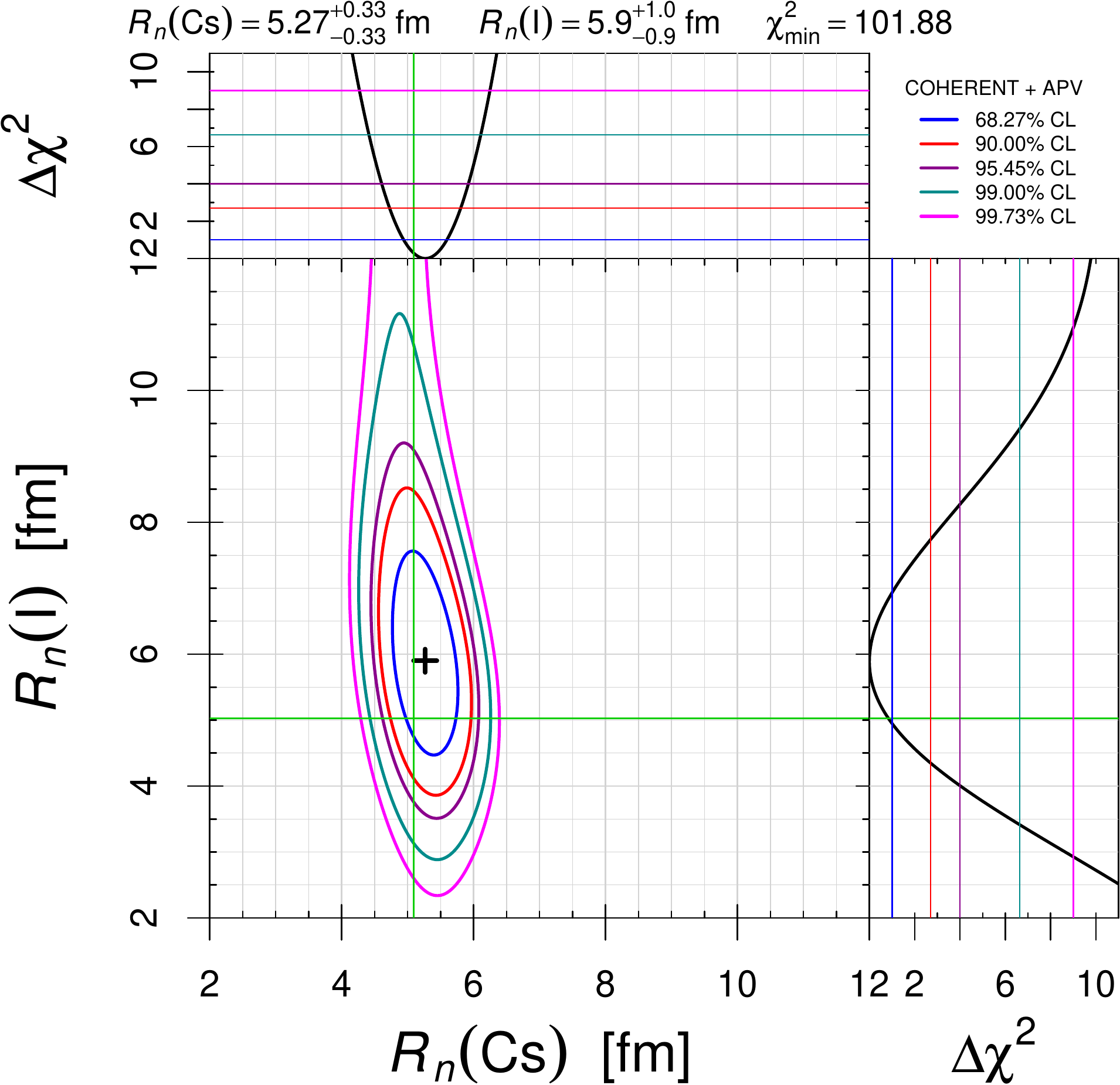}
\caption{ \label{fig:RnCsRnI} Contours at different CL of the allowed regions in the plane of $R_n(^{133}\mathrm{Cs})$ and $R_n(^{127}\mathrm{I})$, together with their marginalizations, obtained  from  the  combined  fit  of  the  COHERENT and APV data.
The green lines indicate the NSM expected values
$R_{n}^{\text{NSM}}(^{133}\text{Cs}) \simeq 5.09 \, \text{fm}$
and
$R_{n}^{\text{NSM}}(\mathrm{^{127}\text{I}}) \simeq 5.03 \, \text{fm}$.
}
\end{figure}

The contours at different confidence levels (CL) of the allowed regions in the plane of $R_n(^{133}\mathrm{Cs})$ and $R_n(^{127}\mathrm{I})$ are shown in~\autoref{fig:RnCsRnI},
from which one can see that
NSM expected values in Eq.~(\ref{RnNSM})
lie in the $1\sigma$ allowed region. Thanks to the combination with APV, $R_{n}(^{133}\mathrm{Cs})$ is well constrained and practically uncorrelated with $R_n(^{127}\mathrm{I})$.

The value in Eq.~(\ref{RnCs-RnI-fit}) represents the most precise determination of $R_{n}(\rm ^{133}Cs)$ and implies a
value of the neutron skin
\begin{equation}
\Delta R_{np}(\mathrm{^{133}Cs})
= R_{n}(\mathrm{^{133}Cs}) - R_{p}(\rm ^{133}Cs)=
0.45 
{}^{+0.33}_{-0.33}
\, \text{fm}
,
\label{Rnp-fit}
\end{equation}
that tends to be larger than the SHF and RMF
nuclear model predictions in Table~\ref{tab:models}.
\begin{figure}[!b]
\centering
\includegraphics*[width=\linewidth]{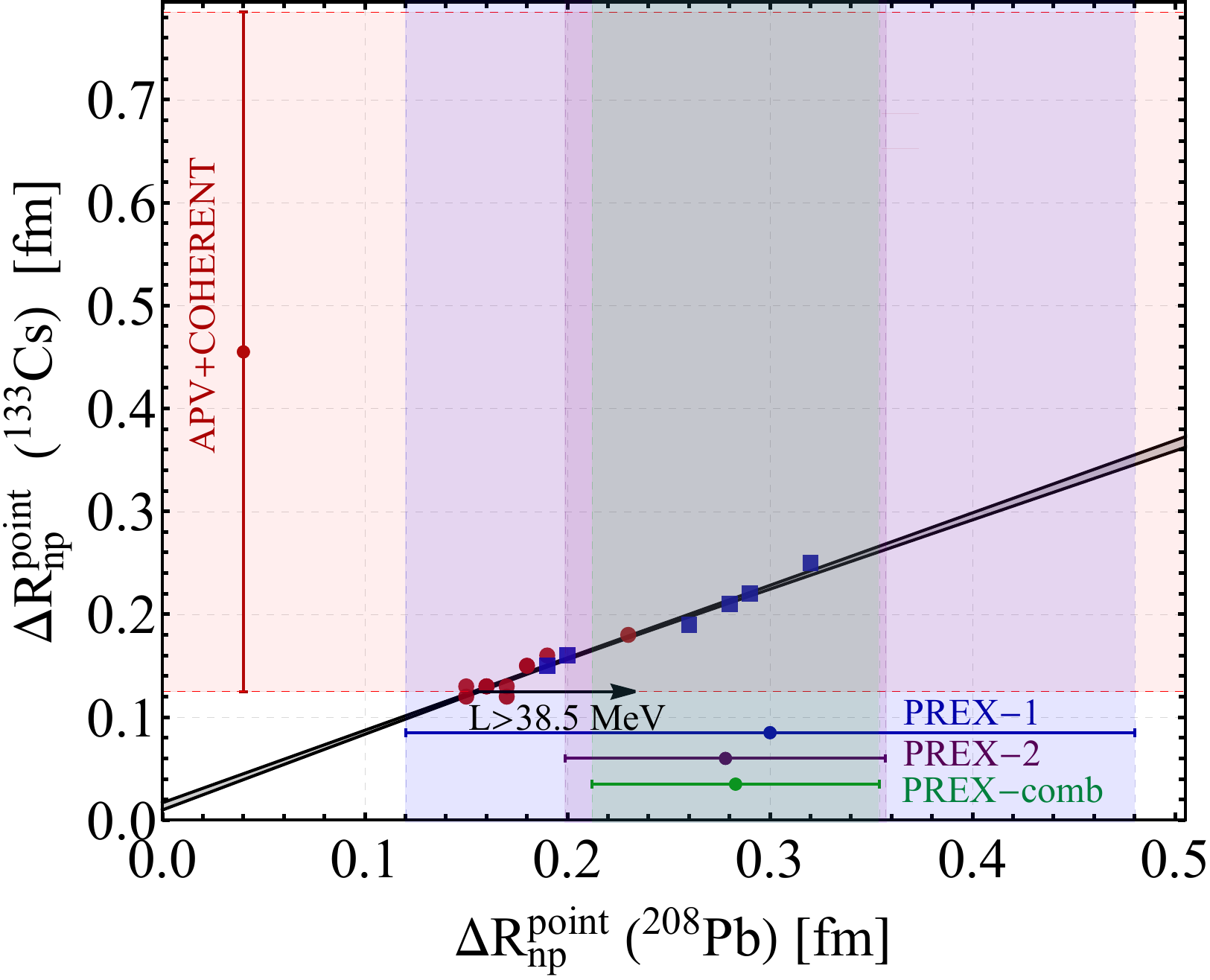}
\caption{ \label{fig:neutron_skin}
Point neutron skin predictions for $^{208}\text{Pb}$ and $^{133}\text{Cs}$ according to different models (blue squares and red circles, see text and Table~\ref{tab:models} for details). 
Constraints set by PREX-1~\cite{Abrahamyan_2012,Horowitz:2012tj}, PREX-2 and their combination~\cite{Adhikari:2021phr,PREXII} and the constraint derived in this work using COHERENT+APV data are also shown by the blue, purple, green, and light red bands, respectively.}
\end{figure}%
This value can be translated in terms of the proton and neutron point radii to allow a direct comparison with $\Delta R_{np}^{\mathrm{point}}$ measured with parity-violating electron scattering on $^{208}\text{Pb}$ in the PREX experiments~\cite{Horowitz:2012tj,Adhikari:2021phr, Abrahamyan_2012,PREXII}. The comparison is shown in~\autoref{fig:neutron_skin}, together with the neutron skin predictions
given in Table~\ref{tab:models},
that have been obtained 
with nonrelativistic Skyrme-Hartree-Fock models (red circles)
and
relativistic mean-field models (blue squares).
A clear model-independent linear correlation is present between the neutron skin of $^{208}\text{Pb}$ and $^{133}\text{Cs}$ within the nonrelativistic and relativistic models with different interactions. This has been already discussed in the literature~\cite{Yang:2019pbx, Zheng:2014nga, Sil:2005tg, PhysRevC.85.041302,Yue:2021yfx},
but here for the first time we are able to compare different experimental determinations of the neutron skin of two nuclei obtained exploiting three electroweak processes, namely atomic parity violation, \cenns, and parity-violating electron scattering.

The combination of the precise PREX results with the unique determination of $\Delta R_{np}(^{133}\text{Cs})$ from APV and COHERENT prefers models that predict large neutron skins. 
The neutron skin of a neutron-rich nucleus is the result of the competition  between  the Coulomb repulsion between the protons, the surface  tension, that decreases when the excess neutrons are pushed to the surface, and the symmetry  energy~\cite{Baldo:2016jhp}. The latter reflects the variation in binding energy of the nucleons as a function of the neutron to proton ratio. Its density dependence, that is a fundamental ingredient of the EOS, is expressed in terms of the slope parameter, $L$, that depends on the derivative of the symmetry energy with respect to density at saturation.

Theoretical calculations show a strong correlation~\cite{Zhang:2013wna, Furnstahl:2001un, PhysRevLett.106.252501, Warda:2009tc} between $\Delta R_{np}$ and $L$, namely larger neutron skins translate into larger values of $L$. Thus, an experimental measurement of $\Delta R_{np}$ represents the most reliable way to determine $L$, which in turn provides critical inputs to a wide range of problems in physics. Among others, it would greatly improve the  modeling of matter inside the cores of neutron stars~\cite{PhysRevLett.85.5296,Steiner:2004fi}, despite  a  difference  in  size with the nucleus  of  18  orders  of  magnitude. Specifically, given that $L$ is  directly  proportional  to  the  pressure  of  pure  neutron  matter  at  saturation density, larger values of $\Delta R_{np}$ imply a larger size of neutron stars~\cite{PhysRevLett.86.5647}. In~\autoref{fig:neutron_skin} we indicated the lower limit for $L$
suggested by the combined COHERENT and APV result, namely $L>38.5$~MeV.
Interestingly, these findings are not in contrast with laboratory experiments or astrophysical observations~\cite{Reed:2021nqk,PhysRevLett.111.162501,Yue:2021yfx}. Indeed, our bound is compatible with the constraints on the slope parameter $L$ derived in Ref.~\cite{PhysRevLett.125.202702,Tews_2017} from a combined analysis of a variety of experimental and theoretical approaches, comprising heavy ion collisions~\cite{PhysRevLett.102.122701}, neutron skin-thickness of tin isotopes~\cite{PhysRevC.82.024321}, giant dipole resonances~\cite{PhysRevC.77.061304}, the dipole polarizability of $^{208}\text{Pb}$~\cite{PhysRevLett.107.062502,PhysRevC.88.024316}, and nuclear masses~\cite{PhysRevC.82.024313}. All these constraints indicate an allowed region of $L$ corresponding to $40 \lesssim L \lesssim 65$.
However, the central value of the averaged PREX result as well as of the combined COHERENT and APV determination presented in this paper suggest rather large neutron skin-thicknesses that would imply a fairly stiff EOS at the typical densities found in atomic nuclei. This finding is in contrast with the current understanding of the neutron star parameters coming from the observation of gravitational waves from GW170817~\cite{Abbott_2017,Abbott_2018}.
If both are correct, it would imply the softening of the EOS at intermediate densities, followed by a stiffening at higher densities~\cite{Reed:2021nqk}, that may be indicative of a phase transition in the stellar core~\cite{PhysRevLett.120.172702}.

For completeness, using the result in Eq.~(\ref{RnCs-RnI-fit}) we are also able to measure for the first time the neutron skin of $^{127}$I, $\Delta R_{np}(\rm ^{127}I)={1.1}^{+1}_{-0.9}\,\text{fm}$, even though with large uncertainty.

\section{Weak mixing angle}
\label{sec:weak}

Leaving $\sin^2\vartheta_W$ free to vary in the $\chi^2$ in Eq.~(\ref{chicoherentapvRn}) and assuming $R_n(^{133}\mathrm{Cs}) \simeq R_n(^{127}\mathrm{I}) \simeq R_n(\mathrm{CsI})$, it is possible to constrain simultaneously $R_n(\mathrm{CsI})$ and $\sin^2\vartheta_W$.
In this analysis we assume that $\sin^2\vartheta_W$
has the same value at the momentum transfer scales of
COHERENT \cenns data
(about 100 MeV)
and the APV data
(about 2 MeV),
as in the SM prediction.
Therefore,
our analysis probes new physics beyond the SM
that can generate a deviation of $\sin^2\vartheta_W$ from the SM prediction
that is constant between about 2 and 100 MeV.

In this analysis we considered
$R_n(^{133}\mathrm{Cs}) \simeq R_n(^{127}\mathrm{I})$
because the data do not allow us to obtain separate information
on the two radii together with the weak mixing angle.
This choice is acceptable,
since the two radii are expected to have values that differ by less than
0.1 fm
(see Eq.~\eqref{RnNSM} and Table~\ref{tab:models}),
that is smaller than the uncertainties of the determinations
in Eq.~\eqref{RnCs-RnI-fit} of
the two radii assuming the SM value of $\sin^2\vartheta_W$.

\begin{figure}[!t]
\centering
\includegraphics*[width=\linewidth]{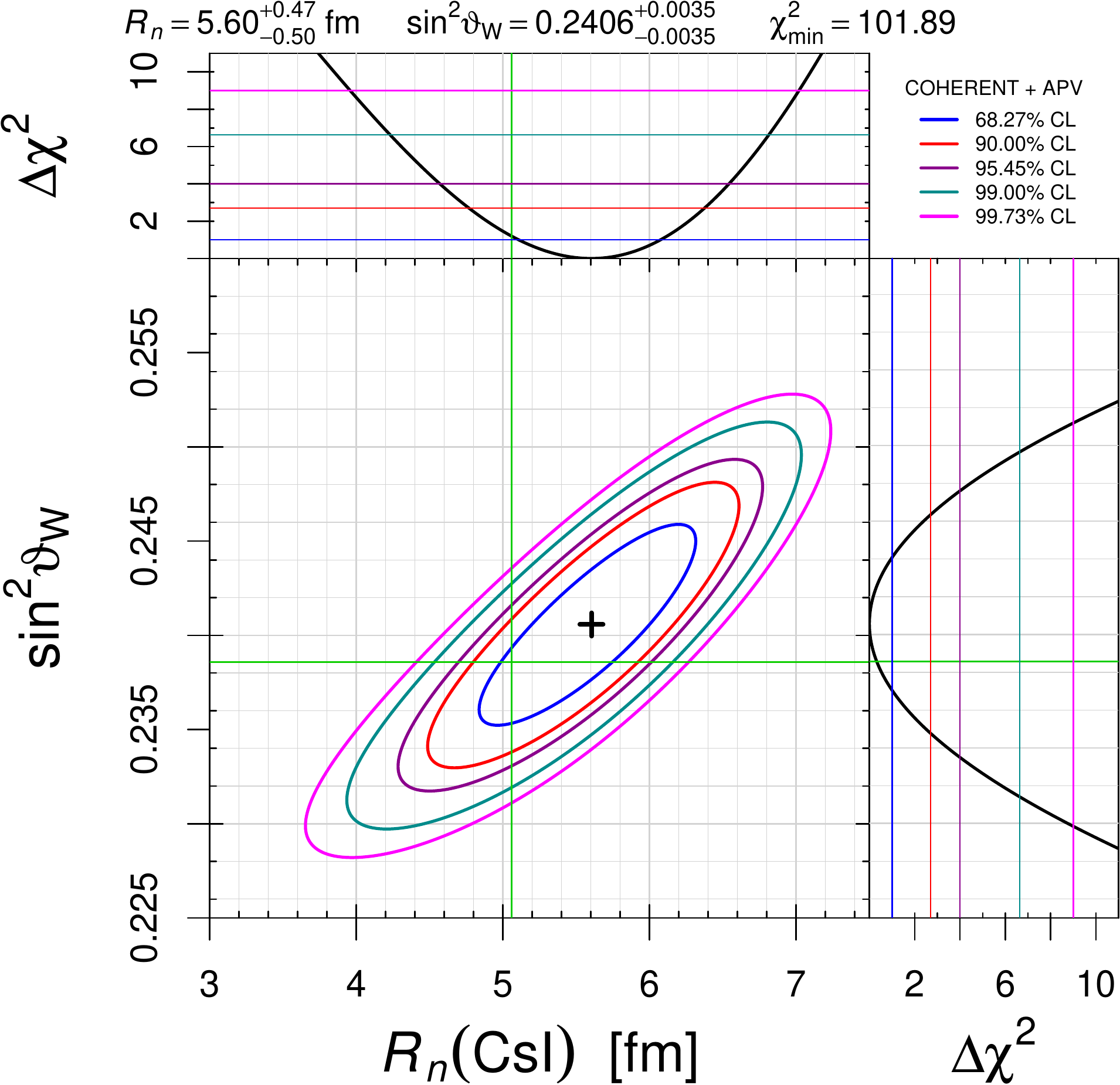}
\caption{ \label{fig:RnSin2} Contours at different CL of the allowed regions in the plane of  $R_n(^{133}\mathrm{Cs})$ and $\sin^2{\vartheta_W}$, together with their marginalizations, obtained  from  the  combined  fit  of  the  COHERENT and APV data.
The green lines indicate the
average of the NSM expected values in Eq.~(\ref{RnNSM}),
$R_{n}^{\text{NSM}}(\text{CsI}) \simeq 5.06 \, \text{fm}$,
and
$\sin^2\vartheta_W=\sin^2\vartheta_W^{\rm SM}\simeq0.23857$.}
\end{figure}

We obtained\footnote{
Considering a fit with equal $^{133}\text{Cs}$ and $^{127}\text{I}$
neutron skins,
we obtained the almost equivalent result
$ R_{n}(\mathrm{Cs}) = 5.63 {}^{+0.47}_{-0.50} \, \text{fm}$,
$ R_{n}(\mathrm{I}) = 5.58 {}^{+0.47}_{-0.50} \, \text{fm}$,
and
$\sin^2{\vartheta_W} = 0.2407 \pm 0.0035$.
}
\begin{equation}
R_{n}(\mathrm{CsI})
=
5.60
{}^{+0.47}_{-0.50}
\, \text{fm}
, \, 
\sin^2{\vartheta_W}
=
0.2406 \pm 0.0035
. 
\label{Rn-sin2-fit-APV}
\end{equation}
The contours at different CL in the plane of $R_n(\text{CsI})$ and  $\sin^2{\vartheta_W}$ are shown in~\autoref{fig:RnSin2}. 
One can see that the NSM expected value for
$R_{n}(\text{CsI})$
and the SM value of $\sin^2\vartheta_W$
lie in the $1\sigma$ allowed region.
The inclusion of the experimental input of $R_{n}(\text{CsI})$ has the effect of shifting the measurement of $\sin^2{\vartheta_W}$ towards larger values with respect to the Particle Data Group (PDG) APV value $\sin^2{\vartheta}_W^{\mathrm{PDG}}=0.2367\pm0.0018$~\cite{Zyla:2020zbs}, 
while keeping the uncertainty at the percent level.

\begin{figure}[!t]
\centering
\includegraphics*[width=\linewidth]{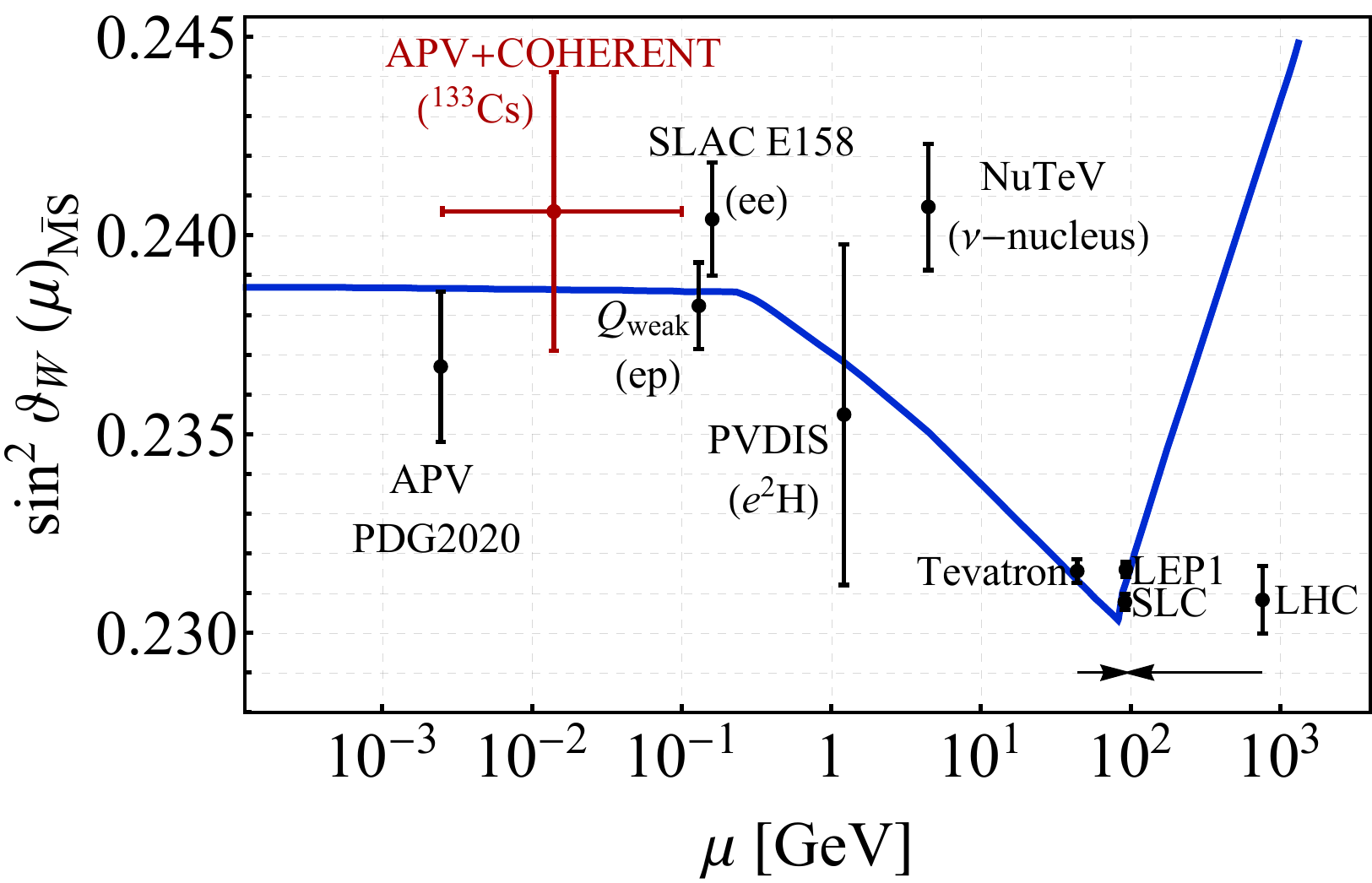}
\caption{ \label{fig:running}
Variation of $\sin^2 \vartheta_{\text{W}}$ with energy scale $\mu$. The SM prediction is shown as the solid curve, together with
experimental determinations in black~\cite{Tanabashi:2018oca,Wood:1997zq,Dzuba:2012kx,Anthony:2005pm,Anthony:2005pm,Wang:2014bba,Zeller:2001hh, Androic:2018kni}. The result derived in this paper is shown in red.}
\end{figure}

Our result is depicted by the red data point in~\autoref{fig:running}, where a summary of the weak mixing angle measurements as a function of the energy scale $\mu$ is shown along with the SM predicted running calculated in the $\overline{\text{MS}}$ scheme~\cite{Tanabashi:2018oca, Erler:2004in,Erler:2017knj}.

It is important to remark that, before this paper, the value of $R_n(^{133}\mathrm{Cs})$ used in the APV result was extrapolated from hadronic experiments using antiprotonic atoms~\cite{PhysRevLett.87.082501}, that are known to be affected, unlike electroweak measurements, by considerable model dependencies and uncontrolled approximations that may be underestimated in the nuclear uncertainty~\cite{Thiel:2019tkm}. Among others, antiprotonic atoms test the neutron distribution in the nuclear periphery, where the density drops exponentially, under the strong assumption that a two-parameter Fermi distribution can be safely used to 
extrapolate the information on the nuclear interior. Thus, it is legit to question if the uncertainty of the official APV result is realistic. On the contrary, the measurement of $\sin^2 \vartheta_{\text{W}}$ presented in this paper in Eq.~(\ref{Rn-sin2-fit-APV}) keeps into account the correlation with the value of $R_n$ determined simultaneously using two electroweak probes, that are known to be practically model independent.

In this regard, the precise determination of $R_n$ for different nuclei from electroweak measurements, as shown in this paper in Eq.~(\ref{RnCs-RnI-fit}) for $^{133}\mathrm{Cs}$, provides a valuable benchmark to calibrate the result of experiments involving hadronic probes, that are fundamental to map the large neutron skins of exotic nuclei.
In the future, the COHERENT program~\cite{akimov2018coherent} will include more detectors, each based on a different material allowing more determinations of $R_n$.  Besides more data that will be available using a single-phase liquid argon detector, that so far allowed a first constrain on $R_{n}(\text{Ar})$~\cite{Cadeddu:2020lky}, there will be two future experiments that are still being developed: a germanium detector, that is also the target used by the CONUS experiment~\cite{Bonet_2021}, and an array of NaI crystals. 

It is also important to note that the central value of the $\sin^2 \vartheta_{\text{W}}$ measurement presented in Eq.~(\ref{Rn-sin2-fit-APV}) is slightly larger with respect to the SM prediction. Combined with the other low-energy measurements, it could be interpreted in terms of a presence of a new dark boson~\cite{PhysRevLett.109.031802,Davoudiasl:2014kua, PhysRevD.92.055005,cadeddu2021muon}. Further measurements of $\sin^2 \vartheta_{\text{W}}$ in the low energy sector should come from the P2~\cite{Becker:2018ggl} and MOLLER~\cite{Benesch:2014bas} experiments, from the near DUNE detector~\cite{deGouvea:2019wav} and the exploitation of coherent elastic neutrino scattering in atoms~\cite{Cadeddu_2019} and nuclei~\cite{Cadeddu:2020lky,Fernandez_Moroni_2021,Ca_as_2018}.

\section{Conclusions}
\label{sec:conclusions}

In conclusion, in this paper we discussed the results on nuclear physics and on the low-energy electroweak mixing angle obtained from the analysis of the new COHERENT CsI data in combination with the atomic parity violation result in cesium. We  obtained the most precise measurement of the neutron  rms  radius and neutron-skin values  of $^{133}\text{Cs}$ and $^{127}\text{I}$, disentangling for the first time the two nuclear contributions. Moreover, for the first time, we derived a data-driven APV+COHERENT  measurement of the low-energy weak mixing angle with a percent uncertainty fully determined from electroweak processes
and independent of the average neutron  rms  radius of $\text{CsI}$
that was allowed to vary freely in the fit.

\begin{acknowledgments}
We would like to thank A. Konovalov and D. Pershey for the useful information provided for the analysis of the COHERENT data.
The work of C. Giunti  and C.A. Ternes is supported by the research grant "The Dark Universe: A Synergic Multimessenger Approach" number 2017X7X85K under the program PRIN 2017 funded by the Ministero dell'Istruzione, Universit\`a e della Ricerca (MIUR).
The work of Y.F. Li and Y.Y. Zhang is supported by the National Natural Science Foundation of China under Grant No.~12075255 and No.~11835013,
and by Beijing Natural Science Foundation under Grant No.~1192019.
Y.F. Li is also grateful for the support by the CAS Center for Excellence in Particle Physics (CCEPP).
\end{acknowledgments}

\appendix

\section{APV Weak Charge Calculation}
\label{app:radiative}

In order to determine the APV weak charge, $Q^{\mathrm{th}}_W$, it is necessary to study in detail the calculation of the couplings, taking into account the radiative corrections.
Following Refs.~\cite{reverler,Erler:2013xha,Marciano_Sirlin,Marciano_Sirlin_2} the lepton-fermion couplings are
\begin{eqnarray}
g_{AV}^{\ell f}&=&\rho \Big[ -\frac{1}{2}+2Q_f \hat{s}_0^2-2Q_f \varnothing_{\ell Z}+\boxtimes_{ZZ}^{\ell f}+\boxtimes_{\gamma Z}^{\ell f}\Big]\nonumber\\&-&2Q_f \varnothing_{\ell W}+\square_{WW},\ \ \ \ \mathrm{for}\ f=u, \label{gAVu}\\
g_{AV}^{\ell f}&=&\rho \Big[ \frac{1}{2}+2Q_f \hat{s}_0^2-2Q_f \varnothing_{\ell Z}+\boxtimes_{ZZ}^{\ell f}+\boxtimes_{\gamma Z}^{\ell f}\Big]\nonumber\\&-&2Q_f \varnothing_{\ell W}+\hourglass_{WW},\ \ \ \ \mathrm{for}\ f=d.\label{gAVd}
\end{eqnarray}
In these relations for up and down quarks, $\rho= 1.00063$ represents a low-energy correction for neutral-current processes and $Q_f$ is the fermion charge. Here $\hat{s}_0^2=\sin^2{\vartheta}_W^{\mathrm{SM}}$, which keeps the same value for $\mu< \mathcal{O}(0.1~\mathrm{GeV})$. The other corrections inserted in equations (\ref{gAVu})-(\ref{gAVd}) come from different contributions, such as electron charge radii ($\varnothing_{e W}$,$\varnothing_{e Z}$), EW box diagrams ($\boxtimes_{ZZ}^{\ell f}$, $\square_{WW}$, $\hourglass_{WW}$) and vacuum polarization of $\gamma Z$ diagrams ($\boxtimes_{\gamma Z}^{\ell f}$) \cite{reverler}.
They can be expressed as
\begin{subequations}
\begin{eqnarray}
\varnothing_{\ell W}&=&\frac{2\alpha}{9\pi},\label{philW}\\
\varnothing_{\ell Z}&=&\frac{\alpha}{6\pi}Q_\ell g_{VA}^{\ell \ell}\Big(\ln\frac{M_Z^2}{m_\ell^2}+\frac{1}{6}\Big),\label{philZ}\\
\boxtimes_{ZZ}^{\ell f}&=&-\frac{3\hat{\alpha}_Z}{16\pi \hat{s}_Z^2 \hat{c}_Z^2}\Big(g_{VA}^{\ell f} g_{VV}^{\ell f}+g_{AV}^{\ell f}g_{AA}^{\ell f}\Big)\times\nonumber\\&\null&\Big[ 1-\frac{\hat{\alpha}_s (M_Z)}{\pi}\Big],\label{boxZZ}\\
\boxtimes_{\gamma Z}^{\ell f}&=&\frac{3\hat{\alpha}_{fZ}}{2\pi} Q_f g_{VA}^{\ell f}\Big( \ln \frac{M_Z^2}{m_f^2}+\frac{3}{2}\Big),\label{boxgammaZ}\\
\square_{WW}&=&-\frac{\hat{\alpha}_Z}{2\pi \hat{s}^2_Z}\Big[1-\frac{\hat{\alpha}_s(M_W)}{2\pi} \Big],\label{squareWW}\\
\hourglass_{WW}&=&\frac{\hat{\alpha}_Z}{8\pi \hat{s}_Z^2}\Big[1+\frac{\hat{\alpha}_s(M_W)}{\pi}\Big]\label{hourglassWW}.
\end{eqnarray}
\end{subequations}
In the expressions above, $\ell$ indicates the lepton involved in the interaction (in our case $\ell=e$), while $f$ indicates the quarks (in our case $f=u,d$).\\
For the electromagnetic-running coupling we adopt the abbreviation $\hat{\alpha}_{ij}\equiv\hat{\alpha}(\sqrt{m_i M_j})$ and $\hat{\alpha}_Z\equiv \alpha(M_Z)$.
In particular, $\hat{\alpha}_{fZ}$, that is present in the $\boxtimes_{\gamma Z}^{\ell f}$ contribution in Eq.~(\ref{boxgammaZ}), is evaluated considering the quark masses equal to the proton one, and inside the logarithmic term the same value ($m_q=m_p$) is used. For the strong coupling, we use the values $\hat{\alpha}_s(M_{Z})=0.1185$~\cite{Zyla:2020zbs} and $\hat{\alpha}_s(M_W)=0.123$~\cite{alitti}.

Inside the correction diagrams in Eqs.~(\ref{philZ}),~(\ref{boxZZ}),~(\ref{boxgammaZ}), the neutral-current couplings enter at tree level and can be written as~\cite{Erler:2013xha} 
\begin{eqnarray}
g_V^f&\equiv& \sqrt{2}\frac{T_f^3-2Q_f\sin^2\vartheta_W(\mu)}{\cos\vartheta_W(\mu)},\\
g_A^f&\equiv& \sqrt{2}\frac{T_f^3}{\cos\vartheta_W(\mu)}.
\end{eqnarray}
Their products are defined as
\begin{equation}
    g_{\alpha \beta}^{\ell f}= \cos^2\vartheta_W(\mu)g_{\alpha}^{\ell}g_\beta^f \ \ \ \ \mathrm{for} \ \ \alpha,\beta = V,A.
\end{equation}
It is important to remark, as reported in Ref.~\cite{reverler}, that for the EW box corrections (Eqs.~(\ref{boxZZ}),~(\ref{squareWW}),~(\ref{hourglassWW})) the sine is evaluated at the value of the $Z$ mass, $\hat{s}_Z^2\equiv\sin^2\hat{\theta}_W(M_Z)=0.23121$  \cite{Zyla:2020zbs}, while in the $\boxtimes_{\gamma Z}^{\ell f}$ term (Eq.~(\ref{boxgammaZ})) the sine is evaluated at scale $\mu=\sqrt{m_p M_Z}$.
Finally, inside the $\varnothing_{\ell Z}$ term (Eq.~(\ref{philZ})) the coupling $g_{VA}^{\ell\ell}$ is obtained using the value $\sin^2\theta_W (\sqrt{m_\ell M_Z})$ as discussed in Ref.~\cite{reverler}.\\
In order to determine the couplings to the proton and to the neutron it is sufficient to use the fact that
\begin{eqnarray}
    g_{AV}^{ep}&=&2g_{AV}^{eu}+g_{AV}^{ed},\\
    g_{AV}^{en}&=&g_{AV}^{eu}+2g_{AV}^{ed}.
\end{eqnarray}\\
However, as pointed out in Refs. \cite{Erler:2013xha, reverler}, it is necessary to take into account also a correction relative to the $\boxtimes_{\gamma Z}^{\ell f}$ contribution, and this is obtained by adding to the proton and neutron couplings some small constants such that
\begin{eqnarray}
g_{AV}^{ep}\rightarrow g_{AV}^{ep}+0.00005,\\
g_{AV}^{en}\rightarrow g_{AV}^{en}+0.00006,
\end{eqnarray}
obtaining the theoretical expression for the APV weak charge written in Eq.~(7).

\section{Nuclear integrals calculation}
\label{app:qpqn}

The approach used to model the nuclear size and shape of the nucleus in APV experiments is based on Refs.~\cite{James_1999,Viatkina}, where the interaction matrix is proportional to the electroweak couplings to protons and neutrons
\begin{equation}
    \mathcal{M}\propto G_{\rm F} \widetilde{Q}_W.
\end{equation}
Here $G_{\rm F}$ is the Fermi constant and 
\begin{equation}
    \widetilde{Q}_W \equiv Z q_p (1-4\sin^2\vartheta_W)-Nq_n.
\end{equation}
This coupling depends on the integrals
\begin{equation}
    q_{p,n}=4\pi \int_0^\infty \rho_{p,n}(r) f(r) r^2\mathrm{d}r,
    \label{eq:qpn}
\end{equation}
where $\rho_{p,n}(r)$ are the proton and neutron densities in the
nucleus as functions of the radius $r$ and $f(r)$
is the matrix element of the electron axial current between the atomic
$s_{1/2}$ and $p_{1/2}$ wave functions inside the nucleus normalized to $f(0)=1$.
The function $f(r)$ can be expressed as a series in power of $(Z\alpha)$, and for most of the atoms of interest, in particular for $(Z\alpha)$ up to $\sim 0.7$, cutting off the series at $(Z\alpha)^2$ is more than adequate to fulfil the requirements of precision for the comparison with experimental observation. 
According to Eq. (13) of Ref. \cite{James_1999}, at order $(Z\alpha)^2$, for any nucleus, $f(r)$ is given by
\begin{align}
    f(r) &=& 1-2\int_0^r \frac{V(r')}{r'^2}\int_0^{r'}V(r'')r''^2 \mathrm{d}r'' \mathrm{d}r' \nonumber\\
    &&+ \left(\frac{1}{r}\int_0^r V(r')r'^2 \mathrm{d}r'\right)^2,
    \label{eq:f(r)}
\end{align}
where $V(r)$ represents the radial electric potential determined uniquely by the charge distribution $\rho_c(r)$ of the nucleus.
One can obtain the potential through the Poisson equation
\begin{equation}
  \frac{1}{r}\frac{d^2}{dr^2}[rV(r)]=-4\pi Z \alpha \rho_c(r),
\end{equation}
whose general solution is
\begin{equation}
    V(r)=4\pi Z \alpha \left[\frac{1}{r}\int_0^r \rho_c(r')r'^2\mathrm{d}r'+\int_r^\infty \rho_c(r')r'\mathrm{d}r'\right].
    \label{eq:potential_solution}
\end{equation}
At this point one has to choose how to parameterize the charge density in order to perform the calculation. 
The easiest choice is to imagine the nucleus as a sphere of radius $R_c$ and constant density
\begin{equation}
    \rho_c^{\mathrm{cd}}(r)=\frac{3}{4\pi R_c^3}\,\Theta(R_c-r),
\end{equation}
$\Theta(R_c-r)$ is the Heaviside function, and the potential, using Eq.~(\ref{eq:potential_solution}) turns out to be
\begin{equation}
    V^{\mathrm{cd}}(r) =\begin{cases} \frac{Z\alpha}{2R_c}\left(3-\frac{r^2}{R_c^2}\right) & \mbox{for} \  r<R_c  \\
     \frac{Z\alpha}{r} & \mbox{for} \ r>R_c \end{cases}.
\end{equation}
By using Eq. (\ref{eq:f(r)}), it is possible to derive the analytical form of $f^{\mathrm{cd}}(r)$ for $r<R_c$
\begin{align}
    f^{\mathrm{cd}}(r) &=  1-\frac{(Z\alpha)^2}{2}\left(\frac{r^2}{R_c^2}-\frac{r^4}{5R_c^4}+\frac{r^6}{75R_c^6}\right),
\end{align}
and for $r>R_c$
\begin{align}
    f^{\mathrm{cd}}(r) &= 1-\frac{(Z\alpha)^2}{2}\left(\frac{13}{30}+\frac{2R_c^2}{5r^2}-\frac{R_c^4}{50r^4}+2\ln\left(\frac{r}{R_c}\right) \right).
\end{align}
Using the above results and Eq.~(\ref{eq:qpn}), one can calculate the proton and neutron integrals. It is worth to notice that in the case of constant density, the integrals in Eq.~(\ref{eq:qpn}) have a cut-off at the value of the proton distribution radius $R_p$, and the neutron distribution radius $R_n$. Since both $R_p$ and $R_n$ are larger than $R_c$, one has to use both forms for $f(r)$, depending on the region of integration.
These considerations lead to
\begin{align}
q_{p,n}^{\mathrm{cd}}=1-(Z\alpha)^2
&
\left(
-\frac{7}{60} 
+\frac{3}{5}\frac{R_c^2}{R_{p,n}^2}
-\frac{16}{63}\frac{R_c^3}{R_{p,n}^3}
\right.
\nonumber \\ 
&
\left.
+\frac{3}{100}\frac{R_c^4}{R_{p,n}^4}+\ln{\frac{R_{p,n}}{R_c}}
\right).
\end{align}
Under the approximation $R_c \simeq R_p$ and for $R_n^2/R_p^2 -1 \ll 1$, it is possible to obtain the typically used forms of $q_{p,n}$
\begin{align}
q_p^{\mathrm{cd}}&\simeq 1-\frac{817}{3150}(Z\alpha)^2,\\
q_n^{\mathrm{cd}}&\simeq1-(Z\alpha)^2\left[\frac{817}{3150}+\frac{116}{525}\left(\frac{R_n^2}{R_p^2}-1\right)\right].
\end{align}
In this manuscript we performed the calculations considering the more accurate charge, proton and neutron distribution densities
that correspond to the form factors in the \cenns cross section.
Therefore,
we evaluated numerically the quantities in Eqs. (\ref{eq:qpn}), (\ref{eq:f(r)}), and (\ref{eq:potential_solution}). In practice,
we used the Helm parameterization \cite{Helm:1956zz} with $R_c(^{133}\mathrm{Cs})=4.8041$~fm and $R_p(^{133}\mathrm{Cs})=4.8212$~fm which, for reference, give as a result $q_p(^{133}\mathrm{Cs})=0.9567$.

\bibliographystyle{apsrev4-2}
\bibliography{biblio}

\begin{thebibliography}{117}%
\makeatletter
\providecommand \@ifxundefined [1]{%
\@ifx{#1\undefined}
}%
\providecommand \@ifnum [1]{%
\ifnum #1\expandafter \@firstoftwo
\else \expandafter \@secondoftwo
\fi
}%
\providecommand \@ifx [1]{%
\ifx #1\expandafter \@firstoftwo
\else \expandafter \@secondoftwo
\fi
}%
\providecommand \natexlab [1]{#1}%
\providecommand \enquote [1]{``#1''}%
\providecommand \bibnamefont [1]{#1}%
\providecommand \bibfnamefont [1]{#1}%
\providecommand \citenamefont [1]{#1}%
\providecommand \href@noop [0]{\@secondoftwo}%
\providecommand \href [0]{\begingroup \@sanitize@url \@href}%
\providecommand \@href[1]{\@@startlink{#1}\@@href}%
\providecommand \@@href[1]{\endgroup#1\@@endlink}%
\providecommand \@sanitize@url [0]{\catcode `\\12\catcode `\$12\catcode
`\&12\catcode `\#12\catcode `\^12\catcode `\_12\catcode `\%12\relax}%
\providecommand \@@startlink[1]{}%
\providecommand \@@endlink[0]{}%
\providecommand \url [0]{\begingroup\@sanitize@url \@url }%
\providecommand \@url [1]{\endgroup\@href {#1}{\urlprefix }}%
\providecommand \urlprefix [0]{URL }%
\providecommand \Eprint [0]{\href }%
\providecommand \doibase [0]{https://doi.org/}%
\providecommand \selectlanguage [0]{\@gobble}%
\providecommand \bibinfo [0]{\@secondoftwo}%
\providecommand \bibfield [0]{\@secondoftwo}%
\providecommand \translation [1]{[#1]}%
\providecommand \BibitemOpen [0]{}%
\providecommand \bibitemStop [0]{}%
\providecommand \bibitemNoStop [0]{.\EOS\space}%
\providecommand \EOS [0]{\spacefactor3000\relax}%
\providecommand \BibitemShut [1]{\csname bibitem#1\endcsname}%
\let\auto@bib@innerbib\@empty
\bibitem [{\citenamefont {Akimov}\ \emph {et~al.}(2017)\citenamefont {Akimov}
\emph {et~al.}}]{Akimov:2017ade}%
\BibitemOpen
\bibfield {author} {\bibinfo {author} {\bibfnamefont {D.}~\bibnamefont
{Akimov}} \emph {et~al.} (\bibinfo {collaboration} {COHERENT}),\ }\href
{https://doi.org/10.1126/science.aao0990} {\bibfield {journal} {\bibinfo
{journal} {Science}\ }\textbf {\bibinfo {volume} {357}},\ \bibinfo {pages}
{1123} (\bibinfo {year} {2017})},\ \Eprint
{https://arxiv.org/abs/arXiv:1708.01294} {arXiv:1708.01294 [nucl-ex]}
\BibitemShut {NoStop}%
\bibitem [{\citenamefont {Akimov}\ \emph
{et~al.}(2018{\natexlab{a}})\citenamefont {Akimov} \emph
{et~al.}}]{Akimov:2018vzs}%
\BibitemOpen
\bibfield {author} {\bibinfo {author} {\bibfnamefont {D.}~\bibnamefont
{Akimov}} \emph {et~al.} (\bibinfo {collaboration} {COHERENT}),\ }\href@noop
{} {\bibinfo {title} {{COHERENT Collaboration data release from the first
observation of coherent elastic neutrino-nucleus scattering}}} (\bibinfo
{year} {2018}{\natexlab{a}}),\ \Eprint
{https://arxiv.org/abs/arXiv:1804.09459} {arXiv:1804.09459 [nucl-ex]}
\BibitemShut {NoStop}%
\bibitem [{\citenamefont {Cadeddu}\ \emph
{et~al.}(2018{\natexlab{a}})\citenamefont {Cadeddu}, \citenamefont {Giunti},
\citenamefont {Li},\ and\ \citenamefont {Zhang}}]{Cadeddu:2017etk}%
\BibitemOpen
\bibfield {author} {\bibinfo {author} {\bibfnamefont {M.}~\bibnamefont
{Cadeddu}}, \bibinfo {author} {\bibfnamefont {C.}~\bibnamefont {Giunti}},
\bibinfo {author} {\bibfnamefont {Y.~F.}\ \bibnamefont {Li}},\ and\ \bibinfo
{author} {\bibfnamefont {Y.~Y.}\ \bibnamefont {Zhang}},\ }\href@noop {}
{\bibfield {journal} {\bibinfo {journal} {Phys.Rev.Lett.}\ }\textbf
{\bibinfo {volume} {120}},\ \bibinfo {pages} {072501} (\bibinfo {year}
{2018}{\natexlab{a}})},\ \Eprint {https://arxiv.org/abs/arXiv:1710.02730}
{arXiv:1710.02730 [hep-ph]} \BibitemShut {NoStop}%
\bibitem [{\citenamefont {Papoulias}\ \emph {et~al.}(2020)\citenamefont
{Papoulias}, \citenamefont {Kosmas}, \citenamefont {Sahu}, \citenamefont
{Kota},\ and\ \citenamefont {Hota}}]{Papoulias:2019lfi}%
\BibitemOpen
\bibfield {author} {\bibinfo {author} {\bibfnamefont {D.}~\bibnamefont
{Papoulias}}, \bibinfo {author} {\bibfnamefont {T.}~\bibnamefont {Kosmas}},
\bibinfo {author} {\bibfnamefont {R.}~\bibnamefont {Sahu}}, \bibinfo {author}
{\bibfnamefont {V.}~\bibnamefont {Kota}},\ and\ \bibinfo {author}
{\bibfnamefont {M.}~\bibnamefont {Hota}},\ }\href
{https://doi.org/10.1016/j.physletb.2019.135133} {\bibfield {journal}
{\bibinfo {journal} {Physics Letters B}\ }\textbf {\bibinfo {volume}
{800}},\ \bibinfo {pages} {135133} (\bibinfo {year} {2020})},\ \Eprint
{https://arxiv.org/abs/1903.03722} {arXiv:1903.03722 [hep-ph]} \BibitemShut
{NoStop}%
\bibitem [{\citenamefont {Coloma}\ \emph {et~al.}(2017)\citenamefont {Coloma},
\citenamefont {Gonzalez-Garcia}, \citenamefont {Maltoni},\ and\ \citenamefont
{Schwetz}}]{Coloma:2017ncl}%
\BibitemOpen
\bibfield {author} {\bibinfo {author} {\bibfnamefont {P.}~\bibnamefont
{Coloma}}, \bibinfo {author} {\bibfnamefont {M.~C.}\ \bibnamefont
{Gonzalez-Garcia}}, \bibinfo {author} {\bibfnamefont {M.}~\bibnamefont
{Maltoni}},\ and\ \bibinfo {author} {\bibfnamefont {T.}~\bibnamefont
{Schwetz}},\ }\href@noop {} {\bibfield {journal} {\bibinfo {journal}
{Phys.Rev.}\ }\textbf {\bibinfo {volume} {D96}},\ \bibinfo {pages} {115007}
(\bibinfo {year} {2017})},\ \Eprint {https://arxiv.org/abs/arXiv:1708.02899}
{arXiv:1708.02899 [hep-ph]} \BibitemShut {NoStop}%
\bibitem [{\citenamefont {Liao}\ and\ \citenamefont
{Marfatia}(2017)}]{Liao:2017uzy}%
\BibitemOpen
\bibfield {author} {\bibinfo {author} {\bibfnamefont {J.}~\bibnamefont
{Liao}}\ and\ \bibinfo {author} {\bibfnamefont {D.}~\bibnamefont
{Marfatia}},\ }\href {https://doi.org/10.1016/j.physletb.2017.10.046}
{\bibfield {journal} {\bibinfo {journal} {Phys. Lett. B}\ }\textbf
{\bibinfo {volume} {775}},\ \bibinfo {pages} {54} (\bibinfo {year} {2017})},\
\Eprint {https://arxiv.org/abs/1708.04255} {arXiv:1708.04255 [hep-ph]}
\BibitemShut {NoStop}%
\bibitem [{\citenamefont {Papoulias}\ and\ \citenamefont
{Kosmas}(2018)}]{Kosmas:2017tsq}%
\BibitemOpen
\bibfield {author} {\bibinfo {author} {\bibfnamefont {D.~K.}\ \bibnamefont
{Papoulias}}\ and\ \bibinfo {author} {\bibfnamefont {T.~S.}\ \bibnamefont
{Kosmas}},\ }\href@noop {} {\bibfield {journal} {\bibinfo {journal}
{Phys.Rev.}\ }\textbf {\bibinfo {volume} {D97}},\ \bibinfo {pages} {033003}
(\bibinfo {year} {2018})},\ \Eprint {https://arxiv.org/abs/arXiv:1711.09773}
{arXiv:1711.09773 [hep-ph]} \BibitemShut {NoStop}%
\bibitem [{\citenamefont {Denton}\ \emph {et~al.}(2018)\citenamefont {Denton},
\citenamefont {Farzan},\ and\ \citenamefont {Shoemaker}}]{Denton:2018xmq}%
\BibitemOpen
\bibfield {author} {\bibinfo {author} {\bibfnamefont {P.~B.}\ \bibnamefont
{Denton}}, \bibinfo {author} {\bibfnamefont {Y.}~\bibnamefont {Farzan}},\
and\ \bibinfo {author} {\bibfnamefont {I.~M.}\ \bibnamefont {Shoemaker}},\
}\href {https://doi.org/10.1007/JHEP07(2018)037} {\bibfield {journal}
{\bibinfo {journal} {JHEP}\ }\textbf {\bibinfo {volume} {07}},\ \bibinfo
{pages} {037}},\ \Eprint {https://arxiv.org/abs/1804.03660} {arXiv:1804.03660
[hep-ph]} \BibitemShut {NoStop}%
\bibitem [{\citenamefont {Aristizabal~Sierra}\ \emph
{et~al.}(2018)\citenamefont {Aristizabal~Sierra}, \citenamefont {De~Romeri},\
and\ \citenamefont {Rojas}}]{AristizabalSierra:2018eqm}%
\BibitemOpen
\bibfield {author} {\bibinfo {author} {\bibfnamefont {D.}~\bibnamefont
{Aristizabal~Sierra}}, \bibinfo {author} {\bibfnamefont {V.}~\bibnamefont
{De~Romeri}},\ and\ \bibinfo {author} {\bibfnamefont {N.}~\bibnamefont
{Rojas}},\ }\href@noop {} {\bibfield {journal} {\bibinfo {journal}
{Phys.Rev.}\ }\textbf {\bibinfo {volume} {D98}},\ \bibinfo {pages} {075018}
(\bibinfo {year} {2018})},\ \Eprint {https://arxiv.org/abs/arXiv:1806.07424}
{arXiv:1806.07424 [hep-ph]} \BibitemShut {NoStop}%
\bibitem [{\citenamefont {Cadeddu}\ \emph
{et~al.}(2018{\natexlab{b}})\citenamefont {Cadeddu}, \citenamefont {Giunti},
\citenamefont {Kouzakov}, \citenamefont {Li}, \citenamefont {Studenikin},\
and\ \citenamefont {Zhang}}]{Cadeddu:2018dux}%
\BibitemOpen
\bibfield {author} {\bibinfo {author} {\bibfnamefont {M.}~\bibnamefont
{Cadeddu}}, \bibinfo {author} {\bibfnamefont {C.}~\bibnamefont {Giunti}},
\bibinfo {author} {\bibfnamefont {K.}~\bibnamefont {Kouzakov}}, \bibinfo
{author} {\bibfnamefont {Y.~F.}\ \bibnamefont {Li}}, \bibinfo {author}
{\bibfnamefont {A.}~\bibnamefont {Studenikin}},\ and\ \bibinfo {author}
{\bibfnamefont {Y.~Y.}\ \bibnamefont {Zhang}},\ }\href@noop {} {\bibfield
{journal} {\bibinfo {journal} {Phys.Rev.}\ }\textbf {\bibinfo {volume}
{D98}},\ \bibinfo {pages} {113010} (\bibinfo {year} {2018}{\natexlab{b}})},\
\Eprint {https://arxiv.org/abs/arXiv:1810.05606} {arXiv:1810.05606 [hep-ph]}
\BibitemShut {NoStop}%
\bibitem [{\citenamefont {Dutta}\ \emph {et~al.}(2019)\citenamefont {Dutta},
\citenamefont {Liao}, \citenamefont {Sinha},\ and\ \citenamefont
{Strigari}}]{Dutta:2019eml}%
\BibitemOpen
\bibfield {author} {\bibinfo {author} {\bibfnamefont {B.}~\bibnamefont
{Dutta}}, \bibinfo {author} {\bibfnamefont {S.}~\bibnamefont {Liao}},
\bibinfo {author} {\bibfnamefont {S.}~\bibnamefont {Sinha}},\ and\ \bibinfo
{author} {\bibfnamefont {L.~E.}\ \bibnamefont {Strigari}},\ }\href@noop {}
{\bibfield {journal} {\bibinfo {journal} {Phys.Rev.Lett.}\ }\textbf
{\bibinfo {volume} {123}},\ \bibinfo {pages} {061801} (\bibinfo {year}
{2019})},\ \Eprint {https://arxiv.org/abs/arXiv:1903.10666} {arXiv:1903.10666
[hep-ph]} \BibitemShut {NoStop}%
\bibitem [{\citenamefont {Cadeddu}\ and\ \citenamefont
{Dordei}(2019)}]{Cadeddu:2018izq}%
\BibitemOpen
\bibfield {author} {\bibinfo {author} {\bibfnamefont {M.}~\bibnamefont
{Cadeddu}}\ and\ \bibinfo {author} {\bibfnamefont {F.}~\bibnamefont
{Dordei}},\ }\href {https://doi.org/10.1103/PhysRevD.99.033010} {\bibfield
{journal} {\bibinfo {journal} {Phys. Rev. D}\ }\textbf {\bibinfo {volume}
{99}},\ \bibinfo {pages} {033010} (\bibinfo {year} {2019})},\ \Eprint
{https://arxiv.org/abs/1808.10202} {arXiv:1808.10202 [hep-ph]} \BibitemShut
{NoStop}%
\bibitem [{\citenamefont {Dutta}\ \emph {et~al.}(2020)\citenamefont {Dutta},
\citenamefont {Kim}, \citenamefont {Liao}, \citenamefont {Park},
\citenamefont {Shin},\ and\ \citenamefont {Strigari}}]{Dutta:2019nbn}%
\BibitemOpen
\bibfield {author} {\bibinfo {author} {\bibfnamefont {B.}~\bibnamefont
{Dutta}}, \bibinfo {author} {\bibfnamefont {D.}~\bibnamefont {Kim}}, \bibinfo
{author} {\bibfnamefont {S.}~\bibnamefont {Liao}}, \bibinfo {author}
{\bibfnamefont {J.-C.}\ \bibnamefont {Park}}, \bibinfo {author}
{\bibfnamefont {S.}~\bibnamefont {Shin}},\ and\ \bibinfo {author}
{\bibfnamefont {L.~E.}\ \bibnamefont {Strigari}},\ }\href@noop {} {\bibfield
{journal} {\bibinfo {journal} {Phys.Rev.Lett.}\ }\textbf {\bibinfo {volume}
{124}},\ \bibinfo {pages} {121802} (\bibinfo {year} {2020})},\ \Eprint
{https://arxiv.org/abs/arXiv:1906.10745} {arXiv:1906.10745 [hep-ph]}
\BibitemShut {NoStop}%
\bibitem [{\citenamefont {Cadeddu}\ \emph
{et~al.}(2020{\natexlab{a}})\citenamefont {Cadeddu}, \citenamefont {Dordei},
\citenamefont {Giunti}, \citenamefont {Li},\ and\ \citenamefont
{Zhang}}]{Cadeddu:2019eta}%
\BibitemOpen
\bibfield {author} {\bibinfo {author} {\bibfnamefont {M.}~\bibnamefont
{Cadeddu}}, \bibinfo {author} {\bibfnamefont {F.}~\bibnamefont {Dordei}},
\bibinfo {author} {\bibfnamefont {C.}~\bibnamefont {Giunti}}, \bibinfo
{author} {\bibfnamefont {Y.~F.}\ \bibnamefont {Li}},\ and\ \bibinfo {author}
{\bibfnamefont {Y.~Y.}\ \bibnamefont {Zhang}},\ }\href
{https://doi.org/10.1103/PhysRevD.101.033004} {\bibfield {journal} {\bibinfo
{journal} {Phys. Rev.}\ }\textbf {\bibinfo {volume} {D101}},\ \bibinfo
{pages} {033004} (\bibinfo {year} {2020}{\natexlab{a}})},\ \Eprint
{https://arxiv.org/abs/1908.06045} {arXiv:1908.06045 [hep-ph]} \BibitemShut
{NoStop}%
\bibitem [{\citenamefont {Papoulias}(2020)}]{Papoulias_2020}%
\BibitemOpen
\bibfield {author} {\bibinfo {author} {\bibfnamefont {D.}~\bibnamefont
{Papoulias}},\ }\bibfield {journal} {\bibinfo {journal} {Physical Review
D}\ }\textbf {\bibinfo {volume} {102}},\ \href
{https://doi.org/10.1103/physrevd.102.113004} {10.1103/physrevd.102.113004}
(\bibinfo {year} {2020})\BibitemShut {NoStop}%
\bibitem [{\citenamefont {Khan}\ and\ \citenamefont
{Rodejohann}(2019)}]{Khan_2019}%
\BibitemOpen
\bibfield {author} {\bibinfo {author} {\bibfnamefont {A.~N.}\ \bibnamefont
{Khan}}\ and\ \bibinfo {author} {\bibfnamefont {W.}~\bibnamefont
{Rodejohann}},\ }\bibfield {journal} {\bibinfo {journal} {Physical Review
D}\ }\textbf {\bibinfo {volume} {100}},\ \href
{https://doi.org/10.1103/physrevd.100.113003} {10.1103/physrevd.100.113003}
(\bibinfo {year} {2019})\BibitemShut {NoStop}%
\bibitem [{\citenamefont {Cadeddu}\ \emph
{et~al.}(2021{\natexlab{a}})\citenamefont {Cadeddu}, \citenamefont
{Cargioli}, \citenamefont {Dordei}, \citenamefont {Giunti}, \citenamefont
{Li}, \citenamefont {Picciau},\ and\ \citenamefont
{Zhang}}]{Cadeddu:2020nbr}%
\BibitemOpen
\bibfield {author} {\bibinfo {author} {\bibfnamefont {M.}~\bibnamefont
{Cadeddu}}, \bibinfo {author} {\bibfnamefont {N.}~\bibnamefont {Cargioli}},
\bibinfo {author} {\bibfnamefont {F.}~\bibnamefont {Dordei}}, \bibinfo
{author} {\bibfnamefont {C.}~\bibnamefont {Giunti}}, \bibinfo {author}
{\bibfnamefont {Y.~F.}\ \bibnamefont {Li}}, \bibinfo {author} {\bibfnamefont
{E.}~\bibnamefont {Picciau}},\ and\ \bibinfo {author} {\bibfnamefont {Y.~Y.}\
\bibnamefont {Zhang}},\ }\href {https://doi.org/10.1007/JHEP01(2021)116}
{\bibfield {journal} {\bibinfo {journal} {JHEP}\ }\textbf {\bibinfo
{volume} {01}},\ \bibinfo {pages} {116}},\ \Eprint
{https://arxiv.org/abs/2008.05022} {arXiv:2008.05022 [hep-ph]} \BibitemShut
{NoStop}%
\bibitem [{\citenamefont {Akimov}\ \emph {et~al.}(2021)\citenamefont {Akimov}
\emph {et~al.}}]{Akimov:2020pdx}%
\BibitemOpen
\bibfield {author} {\bibinfo {author} {\bibfnamefont {D.}~\bibnamefont
{Akimov}} \emph {et~al.} (\bibinfo {collaboration} {COHERENT}),\ }\href
{https://doi.org/10.1103/PhysRevLett.126.012002} {\bibfield {journal}
{\bibinfo {journal} {Phys. Rev. Lett.}\ }\textbf {\bibinfo {volume} {126}},\
\bibinfo {pages} {012002} (\bibinfo {year} {2021})},\ \Eprint
{https://arxiv.org/abs/2003.10630} {arXiv:2003.10630 [nucl-ex]} \BibitemShut
{NoStop}%
\bibitem [{\citenamefont {Cadeddu}\ \emph
{et~al.}(2020{\natexlab{b}})\citenamefont {Cadeddu}, \citenamefont {Dordei},
\citenamefont {Giunti}, \citenamefont {Li}, \citenamefont {Picciau},\ and\
\citenamefont {Zhang}}]{Cadeddu:2020lky}%
\BibitemOpen
\bibfield {author} {\bibinfo {author} {\bibfnamefont {M.}~\bibnamefont
{Cadeddu}}, \bibinfo {author} {\bibfnamefont {F.}~\bibnamefont {Dordei}},
\bibinfo {author} {\bibfnamefont {C.}~\bibnamefont {Giunti}}, \bibinfo
{author} {\bibfnamefont {Y.}~\bibnamefont {Li}}, \bibinfo {author}
{\bibfnamefont {E.}~\bibnamefont {Picciau}},\ and\ \bibinfo {author}
{\bibfnamefont {Y.}~\bibnamefont {Zhang}},\ }\href
{https://doi.org/10.1103/PhysRevD.102.015030} {\bibfield {journal} {\bibinfo
{journal} {Phys. Rev. D}\ }\textbf {\bibinfo {volume} {102}},\ \bibinfo
{pages} {015030} (\bibinfo {year} {2020}{\natexlab{b}})},\ \Eprint
{https://arxiv.org/abs/2005.01645} {arXiv:2005.01645 [hep-ph]} \BibitemShut
{NoStop}%
\bibitem [{\citenamefont {Miranda}\ \emph {et~al.}(2020)\citenamefont
{Miranda}, \citenamefont {Papoulias}, \citenamefont {Garcia}, \citenamefont
{Sanders}, \citenamefont {Tórtola},\ and\ \citenamefont
{Valle}}]{Miranda_2020}%
\BibitemOpen
\bibfield {author} {\bibinfo {author} {\bibfnamefont {O.}~\bibnamefont
{Miranda}}, \bibinfo {author} {\bibfnamefont {D.}~\bibnamefont {Papoulias}},
\bibinfo {author} {\bibfnamefont {G.~S.}\ \bibnamefont {Garcia}}, \bibinfo
{author} {\bibfnamefont {O.}~\bibnamefont {Sanders}}, \bibinfo {author}
{\bibfnamefont {M.}~\bibnamefont {Tórtola}},\ and\ \bibinfo {author}
{\bibfnamefont {J.}~\bibnamefont {Valle}},\ }\bibfield {journal} {\bibinfo
{journal} {Journal of High Energy Physics}\ }\textbf {\bibinfo {volume}
{2020}},\ \href {https://doi.org/10.1007/jhep05(2020)130}
{10.1007/jhep05(2020)130} (\bibinfo {year} {2020})\BibitemShut {NoStop}%
\bibitem [{\citenamefont {Pershey}(2020)}]{Pershey:MCENNS20}%
\BibitemOpen
\bibfield {author} {\bibinfo {author} {\bibfnamefont {D.}~\bibnamefont
{Pershey}},\ }\href@noop {} {\bibinfo {title} {New coherent results}}
(\bibinfo {year} {2020}),\ \bibinfo {note} {talk presented at {Magnificent
CE$\nu$NS 2020, 16-20 November 2020}}\BibitemShut {NoStop}%
\bibitem [{\citenamefont {Huang}\ and\ \citenamefont
{Chen}(2019)}]{Huang:2019ene}%
\BibitemOpen
\bibfield {author} {\bibinfo {author} {\bibfnamefont {X.-R.}\ \bibnamefont
{Huang}}\ and\ \bibinfo {author} {\bibfnamefont {L.-W.}\ \bibnamefont
{Chen}},\ }\href {https://doi.org/10.1103/PhysRevD.100.071301} {\bibfield
{journal} {\bibinfo {journal} {Phys. Rev.}\ }\textbf {\bibinfo {volume}
{D100}},\ \bibinfo {pages} {071301} (\bibinfo {year} {2019})},\ \Eprint
{https://arxiv.org/abs/1902.07625} {arXiv:1902.07625 [hep-ph]} \BibitemShut
{NoStop}%
\bibitem [{\citenamefont {Lattimer}\ and\ \citenamefont
{Prakash}(2004)}]{Lattimer:2004pg}%
\BibitemOpen
\bibfield {author} {\bibinfo {author} {\bibfnamefont {J.~M.}\ \bibnamefont
{Lattimer}}\ and\ \bibinfo {author} {\bibfnamefont {M.}~\bibnamefont
{Prakash}},\ }\href {https://doi.org/10.1126/science.1090720} {\bibfield
{journal} {\bibinfo {journal} {Science}\ }\textbf {\bibinfo {volume}
{304}},\ \bibinfo {pages} {536} (\bibinfo {year} {2004})},\ \Eprint
{https://arxiv.org/abs/astro-ph/0405262} {arXiv:astro-ph/0405262}
\BibitemShut {NoStop}%
\bibitem [{\citenamefont {Steiner}\ \emph {et~al.}(2005)\citenamefont
{Steiner}, \citenamefont {Prakash}, \citenamefont {Lattimer},\ and\
\citenamefont {Ellis}}]{Steiner:2004fi}%
\BibitemOpen
\bibfield {author} {\bibinfo {author} {\bibfnamefont {A.~W.}\ \bibnamefont
{Steiner}}, \bibinfo {author} {\bibfnamefont {M.}~\bibnamefont {Prakash}},
\bibinfo {author} {\bibfnamefont {J.~M.}\ \bibnamefont {Lattimer}},\ and\
\bibinfo {author} {\bibfnamefont {P.~J.}\ \bibnamefont {Ellis}},\ }\href
{https://doi.org/10.1016/j.physrep.2005.02.004} {\bibfield {journal}
{\bibinfo {journal} {Phys. Rept.}\ }\textbf {\bibinfo {volume} {411}},\
\bibinfo {pages} {325} (\bibinfo {year} {2005})},\ \Eprint
{https://arxiv.org/abs/nucl-th/0410066} {arXiv:nucl-th/0410066} \BibitemShut
{NoStop}%
\bibitem [{\citenamefont {Alex~Brown}(2000)}]{PhysRevLett.85.5296}%
\BibitemOpen
\bibfield {author} {\bibinfo {author} {\bibfnamefont {B.}~\bibnamefont
{Alex~Brown}},\ }\href {https://doi.org/10.1103/PhysRevLett.85.5296}
{\bibfield {journal} {\bibinfo {journal} {Phys. Rev. Lett.}\ }\textbf
{\bibinfo {volume} {85}},\ \bibinfo {pages} {5296} (\bibinfo {year}
{2000})}\BibitemShut {NoStop}%
\bibitem [{\citenamefont {Typel}\ and\ \citenamefont
{Brown}(2001)}]{PhysRevC.64.027302}%
\BibitemOpen
\bibfield {author} {\bibinfo {author} {\bibfnamefont {S.}~\bibnamefont
{Typel}}\ and\ \bibinfo {author} {\bibfnamefont {B.~A.}\ \bibnamefont
{Brown}},\ }\href {https://doi.org/10.1103/PhysRevC.64.027302} {\bibfield
{journal} {\bibinfo {journal} {Phys. Rev. C}\ }\textbf {\bibinfo {volume}
{64}},\ \bibinfo {pages} {027302} (\bibinfo {year} {2001})}\BibitemShut
{NoStop}%
\bibitem [{\citenamefont {Fricke}\ \emph
{et~al.}(1995{\natexlab{a}})\citenamefont {Fricke}, \citenamefont
{Bernhardt}, \citenamefont {Heilig}, \citenamefont {Schaller}, \citenamefont
{Schellenberg}, \citenamefont {Shera},\ and\ \citenamefont
{de~Jager}}]{Fricke}%
\BibitemOpen
\bibfield {author} {\bibinfo {author} {\bibfnamefont {G.}~\bibnamefont
{Fricke}}, \bibinfo {author} {\bibfnamefont {C.}~\bibnamefont {Bernhardt}},
\bibinfo {author} {\bibfnamefont {K.}~\bibnamefont {Heilig}}, \bibinfo
{author} {\bibfnamefont {L.}~\bibnamefont {Schaller}}, \bibinfo {author}
{\bibfnamefont {L.}~\bibnamefont {Schellenberg}}, \bibinfo {author}
{\bibfnamefont {E.}~\bibnamefont {Shera}},\ and\ \bibinfo {author}
{\bibfnamefont {C.}~\bibnamefont {de~Jager}},\ }\href
{https://doi.org/10.1006/adnd.1995.1007} {\bibfield {journal} {\bibinfo
{journal} {Atom. Data Nucl. Data Tabl.}\ }\textbf {\bibinfo {volume} {60}},\
\bibinfo {pages} {177} (\bibinfo {year} {1995}{\natexlab{a}})}\BibitemShut
{NoStop}%
\bibitem [{\citenamefont {Angeli}\ and\ \citenamefont
{Marinova}(2013{\natexlab{a}})}]{Angeli}%
\BibitemOpen
\bibfield {author} {\bibinfo {author} {\bibfnamefont {I.}~\bibnamefont
{Angeli}}\ and\ \bibinfo {author} {\bibfnamefont {K.~P.}\ \bibnamefont
{Marinova}},\ }\href {https://doi.org/10.1016/J.ADT.2011.12.006} {\bibfield
{journal} {\bibinfo {journal} {Atom. Data Nucl. Data Tabl.}\ }\textbf
{\bibinfo {volume} {99}},\ \bibinfo {pages} {69} (\bibinfo {year}
{2013}{\natexlab{a}})}\BibitemShut {NoStop}%
\bibitem [{\citenamefont {Thiel}\ \emph {et~al.}(2019)\citenamefont {Thiel},
\citenamefont {Sfienti}, \citenamefont {Piekarewicz}, \citenamefont
{Horowitz},\ and\ \citenamefont {Vanderhaeghen}}]{Thiel:2019tkm}%
\BibitemOpen
\bibfield {author} {\bibinfo {author} {\bibfnamefont {M.}~\bibnamefont
{Thiel}}, \bibinfo {author} {\bibfnamefont {C.}~\bibnamefont {Sfienti}},
\bibinfo {author} {\bibfnamefont {J.}~\bibnamefont {Piekarewicz}}, \bibinfo
{author} {\bibfnamefont {C.~J.}\ \bibnamefont {Horowitz}},\ and\ \bibinfo
{author} {\bibfnamefont {M.}~\bibnamefont {Vanderhaeghen}},\ }\href
{https://doi.org/10.1088/1361-6471/ab2c6d} {\bibfield {journal} {\bibinfo
{journal} {J. Phys. G}\ }\textbf {\bibinfo {volume} {46}},\ \bibinfo {pages}
{093003} (\bibinfo {year} {2019})},\ \Eprint
{https://arxiv.org/abs/1904.12269} {arXiv:1904.12269 [nucl-ex]} \BibitemShut
{NoStop}%
\bibitem [{\citenamefont {Wood}\ \emph {et~al.}(1997)\citenamefont {Wood},
\citenamefont {Bennett}, \citenamefont {Cho}, \citenamefont {Masterson},
\citenamefont {Roberts}, \citenamefont {Tanner},\ and\ \citenamefont
{Wieman}}]{Wood:1997zq}%
\BibitemOpen
\bibfield {author} {\bibinfo {author} {\bibfnamefont {C.~S.}\ \bibnamefont
{Wood}}, \bibinfo {author} {\bibfnamefont {S.~C.}\ \bibnamefont {Bennett}},
\bibinfo {author} {\bibfnamefont {D.}~\bibnamefont {Cho}}, \bibinfo {author}
{\bibfnamefont {B.~P.}\ \bibnamefont {Masterson}}, \bibinfo {author}
{\bibfnamefont {J.~L.}\ \bibnamefont {Roberts}}, \bibinfo {author}
{\bibfnamefont {C.~E.}\ \bibnamefont {Tanner}},\ and\ \bibinfo {author}
{\bibfnamefont {C.~E.}\ \bibnamefont {Wieman}},\ }\href
{https://doi.org/10.1126/science.275.5307.1759} {\bibfield {journal}
{\bibinfo {journal} {Science}\ }\textbf {\bibinfo {volume} {275}},\ \bibinfo
{pages} {1759} (\bibinfo {year} {1997})}\BibitemShut {NoStop}%
\bibitem [{\citenamefont {Guena}\ \emph {et~al.}(2005)\citenamefont {Guena},
\citenamefont {Lintz},\ and\ \citenamefont {Bouchiat}}]{Guena:2004sq}%
\BibitemOpen
\bibfield {author} {\bibinfo {author} {\bibfnamefont {J.}~\bibnamefont
{Guena}}, \bibinfo {author} {\bibfnamefont {M.}~\bibnamefont {Lintz}},\ and\
\bibinfo {author} {\bibfnamefont {M.~A.}\ \bibnamefont {Bouchiat}},\ }\href
{https://doi.org/10.1103/PhysRevA.71.042108} {\bibfield {journal} {\bibinfo
{journal} {Phys. Rev. A}\ }\textbf {\bibinfo {volume} {71}},\ \bibinfo
{pages} {042108} (\bibinfo {year} {2005})},\ \Eprint
{https://arxiv.org/abs/physics/0412017} {arXiv:physics/0412017} \BibitemShut
{NoStop}%
\bibitem [{\citenamefont {Dzuba}\ \emph {et~al.}(2012)\citenamefont {Dzuba},
\citenamefont {Berengut}, \citenamefont {Flambaum},\ and\ \citenamefont
{Roberts}}]{Dzuba:2012kx}%
\BibitemOpen
\bibfield {author} {\bibinfo {author} {\bibfnamefont {V.~A.}\ \bibnamefont
{Dzuba}}, \bibinfo {author} {\bibfnamefont {J.~C.}\ \bibnamefont {Berengut}},
\bibinfo {author} {\bibfnamefont {V.~V.}\ \bibnamefont {Flambaum}},\ and\
\bibinfo {author} {\bibfnamefont {B.}~\bibnamefont {Roberts}},\ }\href
{https://doi.org/10.1103/PhysRevLett.109.203003} {\bibfield {journal}
{\bibinfo {journal} {Phys. Rev. Lett.}\ }\textbf {\bibinfo {volume} {109}},\
\bibinfo {pages} {203003} (\bibinfo {year} {2012})},\ \Eprint
{https://arxiv.org/abs/1207.5864} {arXiv:1207.5864 [hep-ph]} \BibitemShut
{NoStop}%
\bibitem [{\citenamefont {Trzci\ifmmode~\acute{n}\else \'{n}\fi{}ska}\ \emph
{et~al.}(2001)\citenamefont {Trzci\ifmmode~\acute{n}\else \'{n}\fi{}ska},
\citenamefont {Jastrz\ifmmode~\mbox{\c{e}}\else \c{e}\fi{}bski},
\citenamefont {Lubi\ifmmode~\acute{n}\else \'{n}\fi{}ski}, \citenamefont
{Hartmann}, \citenamefont {Schmidt}, \citenamefont {von Egidy},\ and\
\citenamefont {K\l{}os}}]{PhysRevLett.87.082501}%
\BibitemOpen
\bibfield {author} {\bibinfo {author} {\bibfnamefont {A.}~\bibnamefont
{Trzci\ifmmode~\acute{n}\else \'{n}\fi{}ska}}, \bibinfo {author}
{\bibfnamefont {J.}~\bibnamefont {Jastrz\ifmmode~\mbox{\c{e}}\else
\c{e}\fi{}bski}}, \bibinfo {author} {\bibfnamefont {P.}~\bibnamefont
{Lubi\ifmmode~\acute{n}\else \'{n}\fi{}ski}}, \bibinfo {author}
{\bibfnamefont {F.~J.}\ \bibnamefont {Hartmann}}, \bibinfo {author}
{\bibfnamefont {R.}~\bibnamefont {Schmidt}}, \bibinfo {author} {\bibfnamefont
{T.}~\bibnamefont {von Egidy}},\ and\ \bibinfo {author} {\bibfnamefont
{B.}~\bibnamefont {K\l{}os}},\ }\href
{https://doi.org/10.1103/PhysRevLett.87.082501} {\bibfield {journal}
{\bibinfo {journal} {Phys. Rev. Lett.}\ }\textbf {\bibinfo {volume} {87}},\
\bibinfo {pages} {082501} (\bibinfo {year} {2001})}\BibitemShut {NoStop}%
\bibitem [{\citenamefont {Reinhard}\ and\ \citenamefont
{Flocard}(1995)}]{Reinhard:1995zz}%
\BibitemOpen
\bibfield {author} {\bibinfo {author} {\bibfnamefont {P.~G.}\ \bibnamefont
{Reinhard}}\ and\ \bibinfo {author} {\bibfnamefont {H.}~\bibnamefont
{Flocard}},\ }\href {https://doi.org/10.1016/0375-9474(94)00770-N} {\bibfield
{journal} {\bibinfo {journal} {Nucl. Phys.}\ }\textbf {\bibinfo {volume}
{A584}},\ \bibinfo {pages} {467} (\bibinfo {year} {1995})}\BibitemShut
{NoStop}%
\bibitem [{\citenamefont {Chabanat}\ \emph {et~al.}(1998)\citenamefont
{Chabanat}, \citenamefont {Bonche}, \citenamefont {Haensel}, \citenamefont
{Meyer},\ and\ \citenamefont {Schaeffer}}]{Chabanat:1997un}%
\BibitemOpen
\bibfield {author} {\bibinfo {author} {\bibfnamefont {E.}~\bibnamefont
{Chabanat}}, \bibinfo {author} {\bibfnamefont {P.}~\bibnamefont {Bonche}},
\bibinfo {author} {\bibfnamefont {P.}~\bibnamefont {Haensel}}, \bibinfo
{author} {\bibfnamefont {J.}~\bibnamefont {Meyer}},\ and\ \bibinfo {author}
{\bibfnamefont {R.}~\bibnamefont {Schaeffer}},\ }\href
{https://doi.org/10.1016/S0375-9474(98)00570-3,
10.1016/S0375-9474(98)00180-8} {\bibfield {journal} {\bibinfo {journal}
{Nucl. Phys.}\ }\textbf {\bibinfo {volume} {A635}},\ \bibinfo {pages} {231}
(\bibinfo {year} {1998})}\BibitemShut {NoStop}%
\bibitem [{\citenamefont {{Kim}}\ \emph {et~al.}(1997)\citenamefont {{Kim}},
\citenamefont {{Otsuka}},\ and\ \citenamefont
{{Bonche}}}]{Kim-Otsuka-Bonche-1997}%
\BibitemOpen
\bibfield {author} {\bibinfo {author} {\bibfnamefont {K.-H.}\ \bibnamefont
{{Kim}}}, \bibinfo {author} {\bibfnamefont {T.}~\bibnamefont {{Otsuka}}},\
and\ \bibinfo {author} {\bibfnamefont {P.}~\bibnamefont {{Bonche}}},\ }\href
{https://doi.org/10.1088/0954-3899/23/10/014} {\bibfield {journal} {\bibinfo
{journal} {Journal of Physics G Nuclear Physics}\ }\textbf {\bibinfo
{volume} {23}},\ \bibinfo {pages} {1267} (\bibinfo {year}
{1997})}\BibitemShut {NoStop}%
\bibitem [{\citenamefont {Klupfel}\ \emph {et~al.}(2009)\citenamefont
{Klupfel}, \citenamefont {Reinhard}, \citenamefont {Burvenich},\ and\
\citenamefont {Maruhn}}]{Klupfel:2008af}%
\BibitemOpen
\bibfield {author} {\bibinfo {author} {\bibfnamefont {P.}~\bibnamefont
{Klupfel}}, \bibinfo {author} {\bibfnamefont {P.~G.}\ \bibnamefont
{Reinhard}}, \bibinfo {author} {\bibfnamefont {T.~J.}\ \bibnamefont
{Burvenich}},\ and\ \bibinfo {author} {\bibfnamefont {J.~A.}\ \bibnamefont
{Maruhn}},\ }\href {https://doi.org/10.1103/PhysRevC.79.034310} {\bibfield
{journal} {\bibinfo {journal} {Phys. Rev.}\ }\textbf {\bibinfo {volume}
{C79}},\ \bibinfo {pages} {034310} (\bibinfo {year} {2009})},\ \Eprint
{https://arxiv.org/abs/arXiv:0804.3385} {arXiv:0804.3385 [nucl-th]}
\BibitemShut {NoStop}%
\bibitem [{\citenamefont {Kortelainen}\ \emph
{et~al.}(2010{\natexlab{a}})\citenamefont {Kortelainen}, \citenamefont
{Lesinski}, \citenamefont {More}, \citenamefont {Nazarewicz}, \citenamefont
{Sarich}, \citenamefont {Schunck}, \citenamefont {Stoitsov},\ and\
\citenamefont {Wild}}]{Kortelainen:2010hv}%
\BibitemOpen
\bibfield {author} {\bibinfo {author} {\bibfnamefont {M.}~\bibnamefont
{Kortelainen}}, \bibinfo {author} {\bibfnamefont {T.}~\bibnamefont
{Lesinski}}, \bibinfo {author} {\bibfnamefont {J.}~\bibnamefont {More}},
\bibinfo {author} {\bibfnamefont {W.}~\bibnamefont {Nazarewicz}}, \bibinfo
{author} {\bibfnamefont {J.}~\bibnamefont {Sarich}}, \bibinfo {author}
{\bibfnamefont {N.}~\bibnamefont {Schunck}}, \bibinfo {author} {\bibfnamefont
{M.~V.}\ \bibnamefont {Stoitsov}},\ and\ \bibinfo {author} {\bibfnamefont
{S.}~\bibnamefont {Wild}},\ }\href
{https://doi.org/10.1103/PhysRevC.82.024313} {\bibfield {journal} {\bibinfo
{journal} {Phys. Rev.}\ }\textbf {\bibinfo {volume} {C82}},\ \bibinfo {pages}
{024313} (\bibinfo {year} {2010}{\natexlab{a}})},\ \Eprint
{https://arxiv.org/abs/arXiv:1005.5145} {arXiv:1005.5145 [nucl-th]}
\BibitemShut {NoStop}%
\bibitem [{\citenamefont {Kortelainen}\ \emph {et~al.}(2012)\citenamefont
{Kortelainen}, \citenamefont {McDonnell}, \citenamefont {Nazarewicz},
\citenamefont {Reinhard}, \citenamefont {Sarich}, \citenamefont {Schunck},
\citenamefont {Stoitsov},\ and\ \citenamefont {Wild}}]{Kortelainen:2011ft}%
\BibitemOpen
\bibfield {author} {\bibinfo {author} {\bibfnamefont {M.}~\bibnamefont
{Kortelainen}}, \bibinfo {author} {\bibfnamefont {J.}~\bibnamefont
{McDonnell}}, \bibinfo {author} {\bibfnamefont {W.}~\bibnamefont
{Nazarewicz}}, \bibinfo {author} {\bibfnamefont {P.~G.}\ \bibnamefont
{Reinhard}}, \bibinfo {author} {\bibfnamefont {J.}~\bibnamefont {Sarich}},
\bibinfo {author} {\bibfnamefont {N.}~\bibnamefont {Schunck}}, \bibinfo
{author} {\bibfnamefont {M.~V.}\ \bibnamefont {Stoitsov}},\ and\ \bibinfo
{author} {\bibfnamefont {S.~M.}\ \bibnamefont {Wild}},\ }\href
{https://doi.org/10.1103/PhysRevC.85.024304} {\bibfield {journal} {\bibinfo
{journal} {Phys. Rev.}\ }\textbf {\bibinfo {volume} {C85}},\ \bibinfo {pages}
{024304} (\bibinfo {year} {2012})},\ \Eprint
{https://arxiv.org/abs/arXiv:1111.4344} {arXiv:1111.4344 [nucl-th]}
\BibitemShut {NoStop}%
\bibitem [{\citenamefont {Bartel}\ \emph {et~al.}(1982)\citenamefont {Bartel},
\citenamefont {Quentin}, \citenamefont {Brack}, \citenamefont {Guet},\ and\
\citenamefont {Hakansson}}]{Bartel:1982ed}%
\BibitemOpen
\bibfield {author} {\bibinfo {author} {\bibfnamefont {J.}~\bibnamefont
{Bartel}}, \bibinfo {author} {\bibfnamefont {P.}~\bibnamefont {Quentin}},
\bibinfo {author} {\bibfnamefont {M.}~\bibnamefont {Brack}}, \bibinfo
{author} {\bibfnamefont {C.}~\bibnamefont {Guet}},\ and\ \bibinfo {author}
{\bibfnamefont {H.~B.}\ \bibnamefont {Hakansson}},\ }\href
{https://doi.org/10.1016/0375-9474(82)90403-1} {\bibfield {journal}
{\bibinfo {journal} {Nucl. Phys.}\ }\textbf {\bibinfo {volume} {A386}},\
\bibinfo {pages} {79} (\bibinfo {year} {1982})}\BibitemShut {NoStop}%
\bibitem [{\citenamefont {Dobaczewski}\ \emph {et~al.}(1984)\citenamefont
{Dobaczewski}, \citenamefont {Flocard},\ and\ \citenamefont
{Treiner}}]{Dobaczewski:1983zc}%
\BibitemOpen
\bibfield {author} {\bibinfo {author} {\bibfnamefont {J.}~\bibnamefont
{Dobaczewski}}, \bibinfo {author} {\bibfnamefont {H.}~\bibnamefont
{Flocard}},\ and\ \bibinfo {author} {\bibfnamefont {J.}~\bibnamefont
{Treiner}},\ }\href {https://doi.org/10.1016/0375-9474(84)90433-0} {\bibfield
{journal} {\bibinfo {journal} {Nucl. Phys.}\ }\textbf {\bibinfo {volume}
{A422}},\ \bibinfo {pages} {103} (\bibinfo {year} {1984})}\BibitemShut
{NoStop}%
\bibitem [{\citenamefont {Niksic}\ \emph {et~al.}(2002)\citenamefont {Niksic},
\citenamefont {Vretenar}, \citenamefont {Finelli},\ and\ \citenamefont
{Ring}}]{Niksic:2002yp}%
\BibitemOpen
\bibfield {author} {\bibinfo {author} {\bibfnamefont {T.}~\bibnamefont
{Niksic}}, \bibinfo {author} {\bibfnamefont {D.}~\bibnamefont {Vretenar}},
\bibinfo {author} {\bibfnamefont {P.}~\bibnamefont {Finelli}},\ and\ \bibinfo
{author} {\bibfnamefont {P.}~\bibnamefont {Ring}},\ }\href
{https://doi.org/10.1103/PhysRevC.66.024306} {\bibfield {journal} {\bibinfo
{journal} {Phys. Rev.}\ }\textbf {\bibinfo {volume} {C66}},\ \bibinfo {pages}
{024306} (\bibinfo {year} {2002})},\ \Eprint
{https://arxiv.org/abs/nucl-th/0205009} {nucl-th/0205009 [nucl-th]}
\BibitemShut {NoStop}%
\bibitem [{\citenamefont {Niksic}\ \emph {et~al.}(2008)\citenamefont {Niksic},
\citenamefont {Vretenar},\ and\ \citenamefont {Ring}}]{Niksic:2008vp}%
\BibitemOpen
\bibfield {author} {\bibinfo {author} {\bibfnamefont {T.}~\bibnamefont
{Niksic}}, \bibinfo {author} {\bibfnamefont {D.}~\bibnamefont {Vretenar}},\
and\ \bibinfo {author} {\bibfnamefont {P.}~\bibnamefont {Ring}},\ }\href
{https://doi.org/10.1103/PhysRevC.78.034318} {\bibfield {journal} {\bibinfo
{journal} {Phys. Rev.}\ }\textbf {\bibinfo {volume} {C78}},\ \bibinfo {pages}
{034318} (\bibinfo {year} {2008})},\ \Eprint
{https://arxiv.org/abs/arXiv:0809.1375} {arXiv:0809.1375 [nucl-th]}
\BibitemShut {NoStop}%
\bibitem [{\citenamefont {Reinhard}\ \emph {et~al.}(1986)\citenamefont
{Reinhard}, \citenamefont {Rufa}, \citenamefont {Maruhn}, \citenamefont
{Greiner},\ and\ \citenamefont {Friedrich}}]{Reinhard:1986qq}%
\BibitemOpen
\bibfield {author} {\bibinfo {author} {\bibfnamefont {P.~G.}\ \bibnamefont
{Reinhard}}, \bibinfo {author} {\bibfnamefont {M.}~\bibnamefont {Rufa}},
\bibinfo {author} {\bibfnamefont {J.}~\bibnamefont {Maruhn}}, \bibinfo
{author} {\bibfnamefont {W.}~\bibnamefont {Greiner}},\ and\ \bibinfo {author}
{\bibfnamefont {J.}~\bibnamefont {Friedrich}},\ }\href@noop {} {\bibfield
{journal} {\bibinfo {journal} {Z. Phys. A}\ }\textbf {\bibinfo {volume}
{323}},\ \bibinfo {pages} {13} (\bibinfo {year} {1986})}\BibitemShut
{NoStop}%
\bibitem [{\citenamefont {Lalazissis}\ \emph {et~al.}(1997)\citenamefont
{Lalazissis}, \citenamefont {Konig},\ and\ \citenamefont
{Ring}}]{Lalazissis:1996rd}%
\BibitemOpen
\bibfield {author} {\bibinfo {author} {\bibfnamefont {G.~A.}\ \bibnamefont
{Lalazissis}}, \bibinfo {author} {\bibfnamefont {J.}~\bibnamefont {Konig}},\
and\ \bibinfo {author} {\bibfnamefont {P.}~\bibnamefont {Ring}},\ }\href
{https://doi.org/10.1103/PhysRevC.55.540} {\bibfield {journal} {\bibinfo
{journal} {Phys. Rev.}\ }\textbf {\bibinfo {volume} {C55}},\ \bibinfo {pages}
{540} (\bibinfo {year} {1997})},\ \Eprint
{https://arxiv.org/abs/nucl-th/9607039} {nucl-th/9607039 [nucl-th]}
\BibitemShut {NoStop}%
\bibitem [{\citenamefont {Bender}\ \emph {et~al.}(1999)\citenamefont {Bender},
\citenamefont {Rutz}, \citenamefont {Reinhard}, \citenamefont {Maruhn},\ and\
\citenamefont {Greiner}}]{Bender:1999yt}%
\BibitemOpen
\bibfield {author} {\bibinfo {author} {\bibfnamefont {M.}~\bibnamefont
{Bender}}, \bibinfo {author} {\bibfnamefont {K.}~\bibnamefont {Rutz}},
\bibinfo {author} {\bibfnamefont {P.~G.}\ \bibnamefont {Reinhard}}, \bibinfo
{author} {\bibfnamefont {J.~A.}\ \bibnamefont {Maruhn}},\ and\ \bibinfo
{author} {\bibfnamefont {W.}~\bibnamefont {Greiner}},\ }\href
{https://doi.org/10.1103/PhysRevC.60.034304} {\bibfield {journal} {\bibinfo
{journal} {Phys. Rev.}\ }\textbf {\bibinfo {volume} {C60}},\ \bibinfo {pages}
{034304} (\bibinfo {year} {1999})},\ \Eprint
{https://arxiv.org/abs/nucl-th/9906030} {nucl-th/9906030 [nucl-th]}
\BibitemShut {NoStop}%
\bibitem [{\citenamefont {Sharma}\ \emph {et~al.}(1993)\citenamefont {Sharma},
\citenamefont {Nagarajan},\ and\ \citenamefont {Ring}}]{Sharma:1993it}%
\BibitemOpen
\bibfield {author} {\bibinfo {author} {\bibfnamefont {M.~M.}\ \bibnamefont
{Sharma}}, \bibinfo {author} {\bibfnamefont {M.~A.}\ \bibnamefont
{Nagarajan}},\ and\ \bibinfo {author} {\bibfnamefont {P.}~\bibnamefont
{Ring}},\ }\href {https://doi.org/10.1016/0370-2693(93)90970-S} {\bibfield
{journal} {\bibinfo {journal} {Phys. Lett.}\ }\textbf {\bibinfo {volume}
{B312}},\ \bibinfo {pages} {377} (\bibinfo {year} {1993})}\BibitemShut
{NoStop}%
\bibitem [{\citenamefont {Drukier}\ and\ \citenamefont
{Stodolsky}(1984)}]{Drukier:1983gj}%
\BibitemOpen
\bibfield {author} {\bibinfo {author} {\bibfnamefont {A.}~\bibnamefont
{Drukier}}\ and\ \bibinfo {author} {\bibfnamefont {L.}~\bibnamefont
{Stodolsky}},\ }\href {https://doi.org/10.1103/PhysRevD.30.2295} {\bibfield
{journal} {\bibinfo {journal} {Phys. Rev.}\ }\textbf {\bibinfo {volume}
{D30}},\ \bibinfo {pages} {2295} (\bibinfo {year} {1984})}\BibitemShut
{NoStop}%
\bibitem [{\citenamefont {Barranco}\ \emph {et~al.}(2005)\citenamefont
{Barranco}, \citenamefont {Miranda},\ and\ \citenamefont
{Rashba}}]{Barranco:2005yy}%
\BibitemOpen
\bibfield {author} {\bibinfo {author} {\bibfnamefont {J.}~\bibnamefont
{Barranco}}, \bibinfo {author} {\bibfnamefont {O.~G.}\ \bibnamefont
{Miranda}},\ and\ \bibinfo {author} {\bibfnamefont {T.~I.}\ \bibnamefont
{Rashba}},\ }\href {https://doi.org/10.1088/1126-6708/2005/12/021} {\bibfield
{journal} {\bibinfo {journal} {JHEP}\ }\textbf {\bibinfo {volume} {12}},\
\bibinfo {pages} {021}},\ \Eprint {https://arxiv.org/abs/hep-ph/0508299}
{arXiv:hep-ph/0508299 [hep-ph]} \BibitemShut {NoStop}%
\bibitem [{\citenamefont {Patton}\ \emph {et~al.}(2012)\citenamefont {Patton},
\citenamefont {Engel}, \citenamefont {McLaughlin},\ and\ \citenamefont
{Schunck}}]{Patton:2012jr}%
\BibitemOpen
\bibfield {author} {\bibinfo {author} {\bibfnamefont {K.}~\bibnamefont
{Patton}}, \bibinfo {author} {\bibfnamefont {J.}~\bibnamefont {Engel}},
\bibinfo {author} {\bibfnamefont {G.~C.}\ \bibnamefont {McLaughlin}},\ and\
\bibinfo {author} {\bibfnamefont {N.}~\bibnamefont {Schunck}},\ }\href
{https://doi.org/10.1103/PhysRevC.86.024612} {\bibfield {journal} {\bibinfo
{journal} {Phys. Rev.}\ }\textbf {\bibinfo {volume} {C86}},\ \bibinfo {pages}
{024612} (\bibinfo {year} {2012})},\ \Eprint
{https://arxiv.org/abs/1207.0693} {arXiv:1207.0693 [nucl-th]} \BibitemShut
{NoStop}%
\bibitem [{\citenamefont {Tanabashi}\ \emph {et~al.}(2018)\citenamefont
{Tanabashi} \emph {et~al.}}]{Tanabashi:2018oca}%
\BibitemOpen
\bibfield {author} {\bibinfo {author} {\bibfnamefont {M.}~\bibnamefont
{Tanabashi}} \emph {et~al.} (\bibinfo {collaboration} {Particle Data
Group}),\ }\href {https://doi.org/10.1103/PhysRevD.98.030001} {\bibfield
{journal} {\bibinfo {journal} {Phys. Rev.}\ }\textbf {\bibinfo {volume}
{D98}},\ \bibinfo {pages} {030001} (\bibinfo {year} {2018})}\BibitemShut
{NoStop}%
\bibitem [{\citenamefont {Erler}\ and\ \citenamefont
{Su}(2013)}]{Erler:2013xha}%
\BibitemOpen
\bibfield {author} {\bibinfo {author} {\bibfnamefont {J.}~\bibnamefont
{Erler}}\ and\ \bibinfo {author} {\bibfnamefont {S.}~\bibnamefont {Su}},\
}\href {https://doi.org/10.1016/j.ppnp.2013.03.004} {\bibfield {journal}
{\bibinfo {journal} {Prog. Part. Nucl. Phys.}\ }\textbf {\bibinfo {volume}
{71}},\ \bibinfo {pages} {119} (\bibinfo {year} {2013})},\ \Eprint
{https://arxiv.org/abs/1303.5522} {arXiv:1303.5522 [hep-ph]} \BibitemShut
{NoStop}%
\bibitem [{\citenamefont {Piekarewicz}\ \emph {et~al.}(2016)\citenamefont
{Piekarewicz}, \citenamefont {Linero}, \citenamefont {Giuliani},\ and\
\citenamefont {Chicken}}]{Piekarewicz:2016vbn}%
\BibitemOpen
\bibfield {author} {\bibinfo {author} {\bibfnamefont {J.}~\bibnamefont
{Piekarewicz}}, \bibinfo {author} {\bibfnamefont {A.~R.}\ \bibnamefont
{Linero}}, \bibinfo {author} {\bibfnamefont {P.}~\bibnamefont {Giuliani}},\
and\ \bibinfo {author} {\bibfnamefont {E.}~\bibnamefont {Chicken}},\ }\href
{https://doi.org/10.1103/PhysRevC.94.034316} {\bibfield {journal} {\bibinfo
{journal} {Phys. Rev.}\ }\textbf {\bibinfo {volume} {C94}},\ \bibinfo {pages}
{034316} (\bibinfo {year} {2016})},\ \Eprint
{https://arxiv.org/abs/1604.07799} {arXiv:1604.07799 [nucl-th]} \BibitemShut
{NoStop}%
\bibitem [{\citenamefont {Helm}(1956)}]{Helm:1956zz}%
\BibitemOpen
\bibfield {author} {\bibinfo {author} {\bibfnamefont {R.~H.}\ \bibnamefont
{Helm}},\ }\href {https://doi.org/10.1103/PhysRev.104.1466} {\bibfield
{journal} {\bibinfo {journal} {Phys. Rev.}\ }\textbf {\bibinfo {volume}
{104}},\ \bibinfo {pages} {1466} (\bibinfo {year} {1956})}\BibitemShut
{NoStop}%
\bibitem [{\citenamefont {Klein}\ and\ \citenamefont
{Nystrand}(1999)}]{Klein:1999qj}%
\BibitemOpen
\bibfield {author} {\bibinfo {author} {\bibfnamefont {S.}~\bibnamefont
{Klein}}\ and\ \bibinfo {author} {\bibfnamefont {J.}~\bibnamefont
{Nystrand}},\ }\href {https://doi.org/10.1103/PhysRevC.60.014903} {\bibfield
{journal} {\bibinfo {journal} {Phys. Rev.}\ }\textbf {\bibinfo {volume}
{C60}},\ \bibinfo {pages} {014903} (\bibinfo {year} {1999})},\ \Eprint
{https://arxiv.org/abs/hep-ph/9902259} {arXiv:hep-ph/9902259 [hep-ph]}
\BibitemShut {NoStop}%
\bibitem [{\citenamefont {Fricke}\ \emph
{et~al.}(1995{\natexlab{b}})\citenamefont {Fricke}, \citenamefont
{Bernhardt}, \citenamefont {Heilig}, \citenamefont {Schaller}, \citenamefont
{Schellenberg}, \citenamefont {Shera},\ and\ \citenamefont
{de~Jager}}]{Fricke:1995zz}%
\BibitemOpen
\bibfield {author} {\bibinfo {author} {\bibfnamefont {G.}~\bibnamefont
{Fricke}}, \bibinfo {author} {\bibfnamefont {C.}~\bibnamefont {Bernhardt}},
\bibinfo {author} {\bibfnamefont {K.}~\bibnamefont {Heilig}}, \bibinfo
{author} {\bibfnamefont {L.~A.}\ \bibnamefont {Schaller}}, \bibinfo {author}
{\bibfnamefont {L.}~\bibnamefont {Schellenberg}}, \bibinfo {author}
{\bibfnamefont {E.~B.}\ \bibnamefont {Shera}},\ and\ \bibinfo {author}
{\bibfnamefont {C.~W.}\ \bibnamefont {de~Jager}},\ }\href
{https://doi.org/10.1006/adnd.1995.1007} {\bibfield {journal} {\bibinfo
{journal} {Atom. Data Nucl. Data Tabl.}\ }\textbf {\bibinfo {volume} {60}},\
\bibinfo {pages} {177} (\bibinfo {year} {1995}{\natexlab{b}})}\BibitemShut
{NoStop}%
\bibitem [{\citenamefont {Friedrich}\ and\ \citenamefont
{Voegler}(1982)}]{Friedrich:1982esq}%
\BibitemOpen
\bibfield {author} {\bibinfo {author} {\bibfnamefont {J.}~\bibnamefont
{Friedrich}}\ and\ \bibinfo {author} {\bibfnamefont {N.}~\bibnamefont
{Voegler}},\ }\href {https://doi.org/10.1016/0375-9474(82)90147-6} {\bibfield
{journal} {\bibinfo {journal} {Nucl. Phys. A}\ }\textbf {\bibinfo {volume}
{373}},\ \bibinfo {pages} {192} (\bibinfo {year} {1982})}\BibitemShut
{NoStop}%
\bibitem [{\citenamefont {Angeli}\ and\ \citenamefont
{Marinova}(2013{\natexlab{b}})}]{Angeli:2013epw}%
\BibitemOpen
\bibfield {author} {\bibinfo {author} {\bibfnamefont {I.}~\bibnamefont
{Angeli}}\ and\ \bibinfo {author} {\bibfnamefont {K.~P.}\ \bibnamefont
{Marinova}},\ }\href {https://doi.org/10.1016/j.adt.2011.12.006} {\bibfield
{journal} {\bibinfo {journal} {Atom. Data Nucl. Data Tabl.}\ }\textbf
{\bibinfo {volume} {99}},\ \bibinfo {pages} {69} (\bibinfo {year}
{2013}{\natexlab{b}})}\BibitemShut {NoStop}%
\bibitem [{\citenamefont {Hoferichter}\ \emph {et~al.}(2020)\citenamefont
{Hoferichter}, \citenamefont {Menéndez},\ and\ \citenamefont
{Schwenk}}]{Hoferichter_2020}%
\BibitemOpen
\bibfield {author} {\bibinfo {author} {\bibfnamefont {M.}~\bibnamefont
{Hoferichter}}, \bibinfo {author} {\bibfnamefont {J.}~\bibnamefont
{Menéndez}},\ and\ \bibinfo {author} {\bibfnamefont {A.}~\bibnamefont
{Schwenk}},\ }\bibfield {journal} {\bibinfo {journal} {Physical Review D}\
}\textbf {\bibinfo {volume} {102}},\ \href
{https://doi.org/10.1103/physrevd.102.074018} {10.1103/physrevd.102.074018}
(\bibinfo {year} {2020})\BibitemShut {NoStop}%
\bibitem [{\citenamefont {Konovalov}(2020)}]{Konovalov:MCENNS20}%
\BibitemOpen
\bibfield {author} {\bibinfo {author} {\bibfnamefont {A.}~\bibnamefont
{Konovalov}},\ }\href@noop {} {\bibinfo {title} {{COHERENT}}} (\bibinfo
{year} {2020}),\ \bibinfo {note} {talk presented at {Magnificent CE$\nu$NS
2020, 16-20 November 2020}}\BibitemShut {NoStop}%
\bibitem [{\citenamefont {Zyla}\ \emph {et~al.}(2020)\citenamefont {Zyla} \emph
{et~al.}}]{Zyla:2020zbs}%
\BibitemOpen
\bibfield {author} {\bibinfo {author} {\bibfnamefont {P.}~\bibnamefont
{Zyla}} \emph {et~al.} (\bibinfo {collaboration} {Particle Data Group}),\
}\href {https://doi.org/10.1093/ptep/ptaa104} {\bibfield {journal} {\bibinfo
{journal} {PTEP}\ }\textbf {\bibinfo {volume} {2020}},\ \bibinfo {pages}
{083C01} (\bibinfo {year} {2020})}\BibitemShut {NoStop}%
\bibitem [{\citenamefont {Erler}\ \emph {et~al.}(2014)\citenamefont {Erler},
\citenamefont {Horowitz}, \citenamefont {Mantry},\ and\ \citenamefont
{Souder}}]{reverler}%
\BibitemOpen
\bibfield {author} {\bibinfo {author} {\bibfnamefont {J.}~\bibnamefont
{Erler}}, \bibinfo {author} {\bibfnamefont {C.~J.}\ \bibnamefont {Horowitz}},
\bibinfo {author} {\bibfnamefont {S.}~\bibnamefont {Mantry}},\ and\ \bibinfo
{author} {\bibfnamefont {P.~A.}\ \bibnamefont {Souder}},\ }\href
{https://doi.org/10.1146/annurev-nucl-102313-025520} {\bibfield {journal}
{\bibinfo {journal} {Ann. Rev. Nucl. Part. Sci.}\ }\textbf {\bibinfo
{volume} {64}},\ \bibinfo {pages} {269} (\bibinfo {year} {2014})},\ \Eprint
{https://arxiv.org/abs/1401.6199} {arXiv:1401.6199 [hep-ph]} \BibitemShut
{NoStop}%
\bibitem [{\citenamefont {Derevianko}(2001)}]{PhysRevA.65.012106}%
\BibitemOpen
\bibfield {author} {\bibinfo {author} {\bibfnamefont {A.}~\bibnamefont
{Derevianko}},\ }\href {https://doi.org/10.1103/PhysRevA.65.012106}
{\bibfield {journal} {\bibinfo {journal} {Phys. Rev. A}\ }\textbf {\bibinfo
{volume} {65}},\ \bibinfo {pages} {012106} (\bibinfo {year}
{2001})}\BibitemShut {NoStop}%
\bibitem [{\citenamefont {Viatkina}\ \emph {et~al.}(2019)\citenamefont
{Viatkina}, \citenamefont {Antypas}, \citenamefont {Kozlov}, \citenamefont
{Budker},\ and\ \citenamefont {Flambaum}}]{Viatkina}%
\BibitemOpen
\bibfield {author} {\bibinfo {author} {\bibfnamefont {A.~V.}\ \bibnamefont
{Viatkina}}, \bibinfo {author} {\bibfnamefont {D.}~\bibnamefont {Antypas}},
\bibinfo {author} {\bibfnamefont {M.~G.}\ \bibnamefont {Kozlov}}, \bibinfo
{author} {\bibfnamefont {D.}~\bibnamefont {Budker}},\ and\ \bibinfo {author}
{\bibfnamefont {V.~V.}\ \bibnamefont {Flambaum}},\ }\href
{https://doi.org/10.1103/PhysRevC.100.034318} {\bibfield {journal} {\bibinfo
{journal} {Phys. Rev. C}\ }\textbf {\bibinfo {volume} {100}},\ \bibinfo
{pages} {034318} (\bibinfo {year} {2019})}\BibitemShut {NoStop}%
\bibitem [{\citenamefont {Pollock}\ \emph {et~al.}(1992)\citenamefont
{Pollock}, \citenamefont {Fortson},\ and\ \citenamefont
{Wilets}}]{PhysRevC.46.2587}%
\BibitemOpen
\bibfield {author} {\bibinfo {author} {\bibfnamefont {S.~J.}\ \bibnamefont
{Pollock}}, \bibinfo {author} {\bibfnamefont {E.~N.}\ \bibnamefont
{Fortson}},\ and\ \bibinfo {author} {\bibfnamefont {L.}~\bibnamefont
{Wilets}},\ }\href {https://doi.org/10.1103/PhysRevC.46.2587} {\bibfield
{journal} {\bibinfo {journal} {Phys. Rev. C}\ }\textbf {\bibinfo {volume}
{46}},\ \bibinfo {pages} {2587} (\bibinfo {year} {1992})}\BibitemShut
{NoStop}%
\bibitem [{\citenamefont {Pollock}\ and\ \citenamefont
{Welliver}(1999)}]{Pollock1999}%
\BibitemOpen
\bibfield {author} {\bibinfo {author} {\bibfnamefont {S.}~\bibnamefont
{Pollock}}\ and\ \bibinfo {author} {\bibfnamefont {M.}~\bibnamefont
{Welliver}},\ }\href {https://doi.org/10.1016/s0370-2693(99)00987-9}
{\bibfield {journal} {\bibinfo {journal} {Physics Letters B}\ }\textbf
{\bibinfo {volume} {464}},\ \bibinfo {pages} {177–182} (\bibinfo {year}
{1999})}\BibitemShut {NoStop}%
\bibitem [{\citenamefont {James}\ and\ \citenamefont
{Sandars}(1999)}]{James_1999}%
\BibitemOpen
\bibfield {author} {\bibinfo {author} {\bibfnamefont {J.}~\bibnamefont
{James}}\ and\ \bibinfo {author} {\bibfnamefont {P.~G.~H.}\ \bibnamefont
{Sandars}},\ }\href {https://doi.org/10.1088/0953-4075/32/14/301} {\bibfield
{journal} {\bibinfo {journal} {Journal of Physics B: Atomic, Molecular and
Optical Physics}\ }\textbf {\bibinfo {volume} {32}},\ \bibinfo {pages} {3295}
(\bibinfo {year} {1999})}\BibitemShut {NoStop}%
\bibitem [{\citenamefont {Horowitz}\ \emph {et~al.}(2001)\citenamefont
{Horowitz}, \citenamefont {Pollock}, \citenamefont {Souder},\ and\
\citenamefont {Michaels}}]{Horowitz2001}%
\BibitemOpen
\bibfield {author} {\bibinfo {author} {\bibfnamefont {C.~J.}\ \bibnamefont
{Horowitz}}, \bibinfo {author} {\bibfnamefont {S.~J.}\ \bibnamefont
{Pollock}}, \bibinfo {author} {\bibfnamefont {P.~A.}\ \bibnamefont
{Souder}},\ and\ \bibinfo {author} {\bibfnamefont {R.}~\bibnamefont
{Michaels}},\ }\bibfield {journal} {\bibinfo {journal} {Physical Review C}\
}\textbf {\bibinfo {volume} {63}},\ \href
{https://doi.org/10.1103/physrevc.63.025501} {10.1103/physrevc.63.025501}
(\bibinfo {year} {2001})\BibitemShut {NoStop}%
\bibitem [{\citenamefont {Abrahamyan}\ \emph {et~al.}(2012)\citenamefont
{Abrahamyan}, \citenamefont {Ahmed}, \citenamefont {Albataineh},
\citenamefont {Aniol}, \citenamefont {Armstrong}, \citenamefont {Armstrong},
\citenamefont {Averett}, \citenamefont {Babineau}, \citenamefont {Barbieri},
\citenamefont {Bellini},\ and\ \citenamefont {et~al.}}]{Abrahamyan_2012}%
\BibitemOpen
\bibfield {author} {\bibinfo {author} {\bibfnamefont {S.}~\bibnamefont
{Abrahamyan}}, \bibinfo {author} {\bibfnamefont {Z.}~\bibnamefont {Ahmed}},
\bibinfo {author} {\bibfnamefont {H.}~\bibnamefont {Albataineh}}, \bibinfo
{author} {\bibfnamefont {K.}~\bibnamefont {Aniol}}, \bibinfo {author}
{\bibfnamefont {D.~S.}\ \bibnamefont {Armstrong}}, \bibinfo {author}
{\bibfnamefont {W.}~\bibnamefont {Armstrong}}, \bibinfo {author}
{\bibfnamefont {T.}~\bibnamefont {Averett}}, \bibinfo {author} {\bibfnamefont
{B.}~\bibnamefont {Babineau}}, \bibinfo {author} {\bibfnamefont
{A.}~\bibnamefont {Barbieri}}, \bibinfo {author} {\bibfnamefont
{V.}~\bibnamefont {Bellini}},\ and\ \bibinfo {author} {\bibnamefont
{et~al.}},\ }\bibfield {journal} {\bibinfo {journal} {Physical Review
Letters}\ }\textbf {\bibinfo {volume} {108}},\ \href
{https://doi.org/10.1103/physrevlett.108.112502}
{10.1103/physrevlett.108.112502} (\bibinfo {year} {2012})\BibitemShut
{NoStop}%
\bibitem [{\citenamefont {Horowitz}\ \emph {et~al.}(2012)\citenamefont
{Horowitz} \emph {et~al.}}]{Horowitz:2012tj}%
\BibitemOpen
\bibfield {author} {\bibinfo {author} {\bibfnamefont {C.~J.}\ \bibnamefont
{Horowitz}} \emph {et~al.},\ }\href
{https://doi.org/10.1103/PhysRevC.85.032501} {\bibfield {journal} {\bibinfo
{journal} {Phys. Rev. C}\ }\textbf {\bibinfo {volume} {85}},\ \bibinfo
{pages} {032501} (\bibinfo {year} {2012})},\ \Eprint
{https://arxiv.org/abs/1202.1468} {arXiv:1202.1468 [nucl-ex]} \BibitemShut
{NoStop}%
\bibitem [{\citenamefont {Adhikari}\ \emph {et~al.}(2021)\citenamefont
{Adhikari} \emph {et~al.}}]{Adhikari:2021phr}%
\BibitemOpen
\bibfield {author} {\bibinfo {author} {\bibfnamefont {D.}~\bibnamefont
{Adhikari}} \emph {et~al.},\ }\href@noop {} {\bibinfo {title} {{An Accurate
Determination of the Neutron Skin Thickness of $^{208}$Pb through
Parity-Violation in Electron Scattering}}} (\bibinfo {year} {2021}),\ \Eprint
{https://arxiv.org/abs/2102.10767} {arXiv:2102.10767 [nucl-ex]} \BibitemShut
{NoStop}%
\bibitem [{\citenamefont {Reed}(2020)}]{PREXII}%
\BibitemOpen
\bibfield {author} {\bibinfo {author} {\bibfnamefont {B.}~\bibnamefont
{Reed}},\ }\href
{https://indico.cern.ch/event/943069/contributions/4105294/attachments/2145999/3617186/CEvNS_2020.pdf}
{\bibfield {journal} {\bibinfo {journal} {Presentation on behalf of the
PREX-II Collaboration at the Magnificent CEvNS 2020 workshop}\ } (\bibinfo
{year} {2020})}\BibitemShut {NoStop}%
\bibitem [{\citenamefont {Yang}\ \emph {et~al.}(2019)\citenamefont {Yang},
\citenamefont {Hernandez},\ and\ \citenamefont {Piekarewicz}}]{Yang:2019pbx}%
\BibitemOpen
\bibfield {author} {\bibinfo {author} {\bibfnamefont {J.}~\bibnamefont
{Yang}}, \bibinfo {author} {\bibfnamefont {J.~A.}\ \bibnamefont
{Hernandez}},\ and\ \bibinfo {author} {\bibfnamefont {J.}~\bibnamefont
{Piekarewicz}},\ }\href {https://doi.org/10.1103/PhysRevC.100.054301}
{\bibfield {journal} {\bibinfo {journal} {Phys. Rev. C}\ }\textbf {\bibinfo
{volume} {100}},\ \bibinfo {pages} {054301} (\bibinfo {year} {2019})},\
\Eprint {https://arxiv.org/abs/1908.10939} {arXiv:1908.10939 [nucl-th]}
\BibitemShut {NoStop}%
\bibitem [{\citenamefont {Zheng}\ \emph {et~al.}(2014)\citenamefont {Zheng},
\citenamefont {Zhang},\ and\ \citenamefont {Chen}}]{Zheng:2014nga}%
\BibitemOpen
\bibfield {author} {\bibinfo {author} {\bibfnamefont {H.}~\bibnamefont
{Zheng}}, \bibinfo {author} {\bibfnamefont {Z.}~\bibnamefont {Zhang}},\ and\
\bibinfo {author} {\bibfnamefont {L.-W.}\ \bibnamefont {Chen}},\ }\href
{https://doi.org/10.1088/1475-7516/2014/08/011} {\bibfield {journal}
{\bibinfo {journal} {JCAP}\ }\textbf {\bibinfo {volume} {08}},\ \bibinfo
{pages} {011}},\ \Eprint {https://arxiv.org/abs/1403.5134} {arXiv:1403.5134
[nucl-th]} \BibitemShut {NoStop}%
\bibitem [{\citenamefont {Sil}\ \emph {et~al.}(2005)\citenamefont {Sil},
\citenamefont {Centelles}, \citenamefont {Vinas},\ and\ \citenamefont
{Piekarewicz}}]{Sil:2005tg}%
\BibitemOpen
\bibfield {author} {\bibinfo {author} {\bibfnamefont {T.}~\bibnamefont
{Sil}}, \bibinfo {author} {\bibfnamefont {M.}~\bibnamefont {Centelles}},
\bibinfo {author} {\bibfnamefont {X.}~\bibnamefont {Vinas}},\ and\ \bibinfo
{author} {\bibfnamefont {J.}~\bibnamefont {Piekarewicz}},\ }\href
{https://doi.org/10.1103/PhysRevC.71.045502} {\bibfield {journal} {\bibinfo
{journal} {Phys. Rev. C}\ }\textbf {\bibinfo {volume} {71}},\ \bibinfo
{pages} {045502} (\bibinfo {year} {2005})},\ \Eprint
{https://arxiv.org/abs/nucl-th/0501014} {arXiv:nucl-th/0501014} \BibitemShut
{NoStop}%
\bibitem [{\citenamefont {Piekarewicz}\ \emph {et~al.}(2012)\citenamefont
{Piekarewicz}, \citenamefont {Agrawal}, \citenamefont {Col\`o}, \citenamefont
{Nazarewicz}, \citenamefont {Paar}, \citenamefont {Reinhard}, \citenamefont
{Roca-Maza},\ and\ \citenamefont {Vretenar}}]{PhysRevC.85.041302}%
\BibitemOpen
\bibfield {author} {\bibinfo {author} {\bibfnamefont {J.}~\bibnamefont
{Piekarewicz}}, \bibinfo {author} {\bibfnamefont {B.~K.}\ \bibnamefont
{Agrawal}}, \bibinfo {author} {\bibfnamefont {G.}~\bibnamefont {Col\`o}},
\bibinfo {author} {\bibfnamefont {W.}~\bibnamefont {Nazarewicz}}, \bibinfo
{author} {\bibfnamefont {N.}~\bibnamefont {Paar}}, \bibinfo {author}
{\bibfnamefont {P.-G.}\ \bibnamefont {Reinhard}}, \bibinfo {author}
{\bibfnamefont {X.}~\bibnamefont {Roca-Maza}},\ and\ \bibinfo {author}
{\bibfnamefont {D.}~\bibnamefont {Vretenar}},\ }\href
{https://doi.org/10.1103/PhysRevC.85.041302} {\bibfield {journal} {\bibinfo
{journal} {Phys. Rev. C}\ }\textbf {\bibinfo {volume} {85}},\ \bibinfo
{pages} {041302} (\bibinfo {year} {2012})}\BibitemShut {NoStop}%
\bibitem [{\citenamefont {Yue}\ \emph {et~al.}(2021)\citenamefont {Yue},
\citenamefont {Chen}, \citenamefont {Zhang},\ and\ \citenamefont
{Zhou}}]{Yue:2021yfx}%
\BibitemOpen
\bibfield {author} {\bibinfo {author} {\bibfnamefont {T.-G.}\ \bibnamefont
{Yue}}, \bibinfo {author} {\bibfnamefont {L.-W.}\ \bibnamefont {Chen}},
\bibinfo {author} {\bibfnamefont {Z.}~\bibnamefont {Zhang}},\ and\ \bibinfo
{author} {\bibfnamefont {Y.}~\bibnamefont {Zhou}},\ }\href@noop {} {\bibinfo
{title} {{Constraints on the Symmetry Energy from PREX-II in the
Multimessenger Era}}} (\bibinfo {year} {2021}),\ \Eprint
{https://arxiv.org/abs/2102.05267} {arXiv:2102.05267 [nucl-th]} \BibitemShut
{NoStop}%
\bibitem [{\citenamefont {Baldo}\ and\ \citenamefont
{Burgio}(2016)}]{Baldo:2016jhp}%
\BibitemOpen
\bibfield {author} {\bibinfo {author} {\bibfnamefont {M.}~\bibnamefont
{Baldo}}\ and\ \bibinfo {author} {\bibfnamefont {G.~F.}\ \bibnamefont
{Burgio}},\ }\href {https://doi.org/10.1016/j.ppnp.2016.06.006} {\bibfield
{journal} {\bibinfo {journal} {Prog. Part. Nucl. Phys.}\ }\textbf {\bibinfo
{volume} {91}},\ \bibinfo {pages} {203} (\bibinfo {year} {2016})},\ \Eprint
{https://arxiv.org/abs/1606.08838} {arXiv:1606.08838 [nucl-th]} \BibitemShut
{NoStop}%
\bibitem [{\citenamefont {Zhang}\ and\ \citenamefont
{Chen}(2013)}]{Zhang:2013wna}%
\BibitemOpen
\bibfield {author} {\bibinfo {author} {\bibfnamefont {Z.}~\bibnamefont
{Zhang}}\ and\ \bibinfo {author} {\bibfnamefont {L.-W.}\ \bibnamefont
{Chen}},\ }\href {https://doi.org/10.1016/j.physletb.2013.08.002} {\bibfield
{journal} {\bibinfo {journal} {Phys. Lett. B}\ }\textbf {\bibinfo {volume}
{726}},\ \bibinfo {pages} {234} (\bibinfo {year} {2013})},\ \Eprint
{https://arxiv.org/abs/1302.5327} {arXiv:1302.5327 [nucl-th]} \BibitemShut
{NoStop}%
\bibitem [{\citenamefont {Furnstahl}(2002)}]{Furnstahl:2001un}%
\BibitemOpen
\bibfield {author} {\bibinfo {author} {\bibfnamefont {R.~J.}\ \bibnamefont
{Furnstahl}},\ }\href {https://doi.org/10.1016/S0375-9474(02)00867-9}
{\bibfield {journal} {\bibinfo {journal} {Nucl. Phys. A}\ }\textbf
{\bibinfo {volume} {706}},\ \bibinfo {pages} {85} (\bibinfo {year} {2002})},\
\Eprint {https://arxiv.org/abs/nucl-th/0112085} {arXiv:nucl-th/0112085}
\BibitemShut {NoStop}%
\bibitem [{\citenamefont {Roca-Maza}\ \emph {et~al.}(2011)\citenamefont
{Roca-Maza}, \citenamefont {Centelles}, \citenamefont {Vi\~nas},\ and\
\citenamefont {Warda}}]{PhysRevLett.106.252501}%
\BibitemOpen
\bibfield {author} {\bibinfo {author} {\bibfnamefont {X.}~\bibnamefont
{Roca-Maza}}, \bibinfo {author} {\bibfnamefont {M.}~\bibnamefont
{Centelles}}, \bibinfo {author} {\bibfnamefont {X.}~\bibnamefont {Vi\~nas}},\
and\ \bibinfo {author} {\bibfnamefont {M.}~\bibnamefont {Warda}},\ }\href
{https://doi.org/10.1103/PhysRevLett.106.252501} {\bibfield {journal}
{\bibinfo {journal} {Phys. Rev. Lett.}\ }\textbf {\bibinfo {volume} {106}},\
\bibinfo {pages} {252501} (\bibinfo {year} {2011})}\BibitemShut {NoStop}%
\bibitem [{\citenamefont {Warda}\ \emph {et~al.}(2009)\citenamefont {Warda},
\citenamefont {Vinas}, \citenamefont {Roca-Maza},\ and\ \citenamefont
{Centelles}}]{Warda:2009tc}%
\BibitemOpen
\bibfield {author} {\bibinfo {author} {\bibfnamefont {M.}~\bibnamefont
{Warda}}, \bibinfo {author} {\bibfnamefont {X.}~\bibnamefont {Vinas}},
\bibinfo {author} {\bibfnamefont {X.}~\bibnamefont {Roca-Maza}},\ and\
\bibinfo {author} {\bibfnamefont {M.}~\bibnamefont {Centelles}},\ }\href
{https://doi.org/10.1103/PhysRevC.80.024316} {\bibfield {journal} {\bibinfo
{journal} {Phys. Rev. C}\ }\textbf {\bibinfo {volume} {80}},\ \bibinfo
{pages} {024316} (\bibinfo {year} {2009})},\ \Eprint
{https://arxiv.org/abs/0906.0932} {arXiv:0906.0932 [nucl-th]} \BibitemShut
{NoStop}%
\bibitem [{\citenamefont {Horowitz}\ and\ \citenamefont
{Piekarewicz}(2001)}]{PhysRevLett.86.5647}%
\BibitemOpen
\bibfield {author} {\bibinfo {author} {\bibfnamefont {C.~J.}\ \bibnamefont
{Horowitz}}\ and\ \bibinfo {author} {\bibfnamefont {J.}~\bibnamefont
{Piekarewicz}},\ }\href {https://doi.org/10.1103/PhysRevLett.86.5647}
{\bibfield {journal} {\bibinfo {journal} {Phys. Rev. Lett.}\ }\textbf
{\bibinfo {volume} {86}},\ \bibinfo {pages} {5647} (\bibinfo {year}
{2001})}\BibitemShut {NoStop}%
\bibitem [{\citenamefont {Reed}\ \emph {et~al.}(2021)\citenamefont {Reed},
\citenamefont {Fattoyev}, \citenamefont {Horowitz},\ and\ \citenamefont
{Piekarewicz}}]{Reed:2021nqk}%
\BibitemOpen
\bibfield {author} {\bibinfo {author} {\bibfnamefont {B.~T.}\ \bibnamefont
{Reed}}, \bibinfo {author} {\bibfnamefont {F.~J.}\ \bibnamefont {Fattoyev}},
\bibinfo {author} {\bibfnamefont {C.~J.}\ \bibnamefont {Horowitz}},\ and\
\bibinfo {author} {\bibfnamefont {J.}~\bibnamefont {Piekarewicz}},\
}\href@noop {} {\bibinfo {title} {Implications of {PREX-II} on the equation
of state of neutron-rich matter}} (\bibinfo {year} {2021}),\ \Eprint
{https://arxiv.org/abs/2101.03193} {arXiv:2101.03193 [nucl-th]} \BibitemShut
{NoStop}%
\bibitem [{\citenamefont {Fattoyev}\ and\ \citenamefont
{Piekarewicz}(2013)}]{PhysRevLett.111.162501}%
\BibitemOpen
\bibfield {author} {\bibinfo {author} {\bibfnamefont {F.~J.}\ \bibnamefont
{Fattoyev}}\ and\ \bibinfo {author} {\bibfnamefont {J.}~\bibnamefont
{Piekarewicz}},\ }\href {https://doi.org/10.1103/PhysRevLett.111.162501}
{\bibfield {journal} {\bibinfo {journal} {Phys. Rev. Lett.}\ }\textbf
{\bibinfo {volume} {111}},\ \bibinfo {pages} {162501} (\bibinfo {year}
{2013})}\BibitemShut {NoStop}%
\bibitem [{\citenamefont {Drischler}\ \emph {et~al.}(2020)\citenamefont
{Drischler}, \citenamefont {Furnstahl}, \citenamefont {Melendez},\ and\
\citenamefont {Phillips}}]{PhysRevLett.125.202702}%
\BibitemOpen
\bibfield {author} {\bibinfo {author} {\bibfnamefont {C.}~\bibnamefont
{Drischler}}, \bibinfo {author} {\bibfnamefont {R.~J.}\ \bibnamefont
{Furnstahl}}, \bibinfo {author} {\bibfnamefont {J.~A.}\ \bibnamefont
{Melendez}},\ and\ \bibinfo {author} {\bibfnamefont {D.~R.}\ \bibnamefont
{Phillips}},\ }\href {https://doi.org/10.1103/PhysRevLett.125.202702}
{\bibfield {journal} {\bibinfo {journal} {Phys. Rev. Lett.}\ }\textbf
{\bibinfo {volume} {125}},\ \bibinfo {pages} {202702} (\bibinfo {year}
{2020})}\BibitemShut {NoStop}%
\bibitem [{\citenamefont {Tews}\ \emph {et~al.}(2017)\citenamefont {Tews},
\citenamefont {Lattimer}, \citenamefont {Ohnishi},\ and\ \citenamefont
{Kolomeitsev}}]{Tews_2017}%
\BibitemOpen
\bibfield {author} {\bibinfo {author} {\bibfnamefont {I.}~\bibnamefont
{Tews}}, \bibinfo {author} {\bibfnamefont {J.~M.}\ \bibnamefont {Lattimer}},
\bibinfo {author} {\bibfnamefont {A.}~\bibnamefont {Ohnishi}},\ and\ \bibinfo
{author} {\bibfnamefont {E.~E.}\ \bibnamefont {Kolomeitsev}},\ }\href
{https://doi.org/10.3847/1538-4357/aa8db9} {\bibfield {journal} {\bibinfo
{journal} {The Astrophysical Journal}\ }\textbf {\bibinfo {volume} {848}},\
\bibinfo {pages} {105} (\bibinfo {year} {2017})}\BibitemShut {NoStop}%
\bibitem [{\citenamefont {Tsang}\ \emph {et~al.}(2009)\citenamefont {Tsang},
\citenamefont {Zhang}, \citenamefont {Danielewicz}, \citenamefont {Famiano},
\citenamefont {Li}, \citenamefont {Lynch},\ and\ \citenamefont
{Steiner}}]{PhysRevLett.102.122701}%
\BibitemOpen
\bibfield {author} {\bibinfo {author} {\bibfnamefont {M.~B.}\ \bibnamefont
{Tsang}}, \bibinfo {author} {\bibfnamefont {Y.}~\bibnamefont {Zhang}},
\bibinfo {author} {\bibfnamefont {P.}~\bibnamefont {Danielewicz}}, \bibinfo
{author} {\bibfnamefont {M.}~\bibnamefont {Famiano}}, \bibinfo {author}
{\bibfnamefont {Z.}~\bibnamefont {Li}}, \bibinfo {author} {\bibfnamefont
{W.~G.}\ \bibnamefont {Lynch}},\ and\ \bibinfo {author} {\bibfnamefont
{A.~W.}\ \bibnamefont {Steiner}},\ }\href
{https://doi.org/10.1103/PhysRevLett.102.122701} {\bibfield {journal}
{\bibinfo {journal} {Phys. Rev. Lett.}\ }\textbf {\bibinfo {volume} {102}},\
\bibinfo {pages} {122701} (\bibinfo {year} {2009})}\BibitemShut {NoStop}%
\bibitem [{\citenamefont {Chen}\ \emph {et~al.}(2010)\citenamefont {Chen},
\citenamefont {Ko}, \citenamefont {Li},\ and\ \citenamefont
{Xu}}]{PhysRevC.82.024321}%
\BibitemOpen
\bibfield {author} {\bibinfo {author} {\bibfnamefont {L.-W.}\ \bibnamefont
{Chen}}, \bibinfo {author} {\bibfnamefont {C.~M.}\ \bibnamefont {Ko}},
\bibinfo {author} {\bibfnamefont {B.-A.}\ \bibnamefont {Li}},\ and\ \bibinfo
{author} {\bibfnamefont {J.}~\bibnamefont {Xu}},\ }\href
{https://doi.org/10.1103/PhysRevC.82.024321} {\bibfield {journal} {\bibinfo
{journal} {Phys. Rev. C}\ }\textbf {\bibinfo {volume} {82}},\ \bibinfo
{pages} {024321} (\bibinfo {year} {2010})}\BibitemShut {NoStop}%
\bibitem [{\citenamefont {Trippa}\ \emph {et~al.}(2008)\citenamefont {Trippa},
\citenamefont {Col\`o},\ and\ \citenamefont {Vigezzi}}]{PhysRevC.77.061304}%
\BibitemOpen
\bibfield {author} {\bibinfo {author} {\bibfnamefont {L.}~\bibnamefont
{Trippa}}, \bibinfo {author} {\bibfnamefont {G.}~\bibnamefont {Col\`o}},\
and\ \bibinfo {author} {\bibfnamefont {E.}~\bibnamefont {Vigezzi}},\ }\href
{https://doi.org/10.1103/PhysRevC.77.061304} {\bibfield {journal} {\bibinfo
{journal} {Phys. Rev. C}\ }\textbf {\bibinfo {volume} {77}},\ \bibinfo
{pages} {061304} (\bibinfo {year} {2008})}\BibitemShut {NoStop}%
\bibitem [{\citenamefont {Tamii}\ \emph {et~al.}(2011)\citenamefont {Tamii},
\citenamefont {Poltoratska}, \citenamefont {von Neumann-Cosel}, \citenamefont
{Fujita}, \citenamefont {Adachi}, \citenamefont {Bertulani}, \citenamefont
{Carter}, \citenamefont {Dozono}, \citenamefont {Fujita}, \citenamefont
{Fujita}, \citenamefont {Hatanaka}, \citenamefont {Ishikawa}, \citenamefont
{Itoh}, \citenamefont {Kawabata}, \citenamefont {Kalmykov}, \citenamefont
{Krumbholz}, \citenamefont {Litvinova}, \citenamefont {Matsubara},
\citenamefont {Nakanishi}, \citenamefont {Neveling}, \citenamefont {Okamura},
\citenamefont {Ong}, \citenamefont {\"Ozel-Tashenov}, \citenamefont
{Ponomarev}, \citenamefont {Richter}, \citenamefont {Rubio}, \citenamefont
{Sakaguchi}, \citenamefont {Sakemi}, \citenamefont {Sasamoto}, \citenamefont
{Shimbara}, \citenamefont {Shimizu}, \citenamefont {Smit}, \citenamefont
{Suzuki}, \citenamefont {Tameshige}, \citenamefont {Wambach}, \citenamefont
{Yamada}, \citenamefont {Yosoi},\ and\ \citenamefont
{Zenihiro}}]{PhysRevLett.107.062502}%
\BibitemOpen
\bibfield {author} {\bibinfo {author} {\bibfnamefont {A.}~\bibnamefont
{Tamii}}, \bibinfo {author} {\bibfnamefont {I.}~\bibnamefont {Poltoratska}},
\bibinfo {author} {\bibfnamefont {P.}~\bibnamefont {von Neumann-Cosel}},
\bibinfo {author} {\bibfnamefont {Y.}~\bibnamefont {Fujita}}, \bibinfo
{author} {\bibfnamefont {T.}~\bibnamefont {Adachi}}, \bibinfo {author}
{\bibfnamefont {C.~A.}\ \bibnamefont {Bertulani}}, \bibinfo {author}
{\bibfnamefont {J.}~\bibnamefont {Carter}}, \bibinfo {author} {\bibfnamefont
{M.}~\bibnamefont {Dozono}}, \bibinfo {author} {\bibfnamefont
{H.}~\bibnamefont {Fujita}}, \bibinfo {author} {\bibfnamefont
{K.}~\bibnamefont {Fujita}}, \bibinfo {author} {\bibfnamefont
{K.}~\bibnamefont {Hatanaka}}, \bibinfo {author} {\bibfnamefont
{D.}~\bibnamefont {Ishikawa}}, \bibinfo {author} {\bibfnamefont
{M.}~\bibnamefont {Itoh}}, \bibinfo {author} {\bibfnamefont {T.}~\bibnamefont
{Kawabata}}, \bibinfo {author} {\bibfnamefont {Y.}~\bibnamefont {Kalmykov}},
\bibinfo {author} {\bibfnamefont {A.~M.}\ \bibnamefont {Krumbholz}}, \bibinfo
{author} {\bibfnamefont {E.}~\bibnamefont {Litvinova}}, \bibinfo {author}
{\bibfnamefont {H.}~\bibnamefont {Matsubara}}, \bibinfo {author}
{\bibfnamefont {K.}~\bibnamefont {Nakanishi}}, \bibinfo {author}
{\bibfnamefont {R.}~\bibnamefont {Neveling}}, \bibinfo {author}
{\bibfnamefont {H.}~\bibnamefont {Okamura}}, \bibinfo {author} {\bibfnamefont
{H.~J.}\ \bibnamefont {Ong}}, \bibinfo {author} {\bibfnamefont
{B.}~\bibnamefont {\"Ozel-Tashenov}}, \bibinfo {author} {\bibfnamefont
{V.~Y.}\ \bibnamefont {Ponomarev}}, \bibinfo {author} {\bibfnamefont
{A.}~\bibnamefont {Richter}}, \bibinfo {author} {\bibfnamefont
{B.}~\bibnamefont {Rubio}}, \bibinfo {author} {\bibfnamefont
{H.}~\bibnamefont {Sakaguchi}}, \bibinfo {author} {\bibfnamefont
{Y.}~\bibnamefont {Sakemi}}, \bibinfo {author} {\bibfnamefont
{Y.}~\bibnamefont {Sasamoto}}, \bibinfo {author} {\bibfnamefont
{Y.}~\bibnamefont {Shimbara}}, \bibinfo {author} {\bibfnamefont
{Y.}~\bibnamefont {Shimizu}}, \bibinfo {author} {\bibfnamefont {F.~D.}\
\bibnamefont {Smit}}, \bibinfo {author} {\bibfnamefont {T.}~\bibnamefont
{Suzuki}}, \bibinfo {author} {\bibfnamefont {Y.}~\bibnamefont {Tameshige}},
\bibinfo {author} {\bibfnamefont {J.}~\bibnamefont {Wambach}}, \bibinfo
{author} {\bibfnamefont {R.}~\bibnamefont {Yamada}}, \bibinfo {author}
{\bibfnamefont {M.}~\bibnamefont {Yosoi}},\ and\ \bibinfo {author}
{\bibfnamefont {J.}~\bibnamefont {Zenihiro}},\ }\href
{https://doi.org/10.1103/PhysRevLett.107.062502} {\bibfield {journal}
{\bibinfo {journal} {Phys. Rev. Lett.}\ }\textbf {\bibinfo {volume} {107}},\
\bibinfo {pages} {062502} (\bibinfo {year} {2011})}\BibitemShut {NoStop}%
\bibitem [{\citenamefont {Roca-Maza}\ \emph {et~al.}(2013)\citenamefont
{Roca-Maza}, \citenamefont {Brenna}, \citenamefont {Col\`o}, \citenamefont
{Centelles}, \citenamefont {Vi\~nas}, \citenamefont {Agrawal}, \citenamefont
{Paar}, \citenamefont {Vretenar},\ and\ \citenamefont
{Piekarewicz}}]{PhysRevC.88.024316}%
\BibitemOpen
\bibfield {author} {\bibinfo {author} {\bibfnamefont {X.}~\bibnamefont
{Roca-Maza}}, \bibinfo {author} {\bibfnamefont {M.}~\bibnamefont {Brenna}},
\bibinfo {author} {\bibfnamefont {G.}~\bibnamefont {Col\`o}}, \bibinfo
{author} {\bibfnamefont {M.}~\bibnamefont {Centelles}}, \bibinfo {author}
{\bibfnamefont {X.}~\bibnamefont {Vi\~nas}}, \bibinfo {author} {\bibfnamefont
{B.~K.}\ \bibnamefont {Agrawal}}, \bibinfo {author} {\bibfnamefont
{N.}~\bibnamefont {Paar}}, \bibinfo {author} {\bibfnamefont {D.}~\bibnamefont
{Vretenar}},\ and\ \bibinfo {author} {\bibfnamefont {J.}~\bibnamefont
{Piekarewicz}},\ }\href {https://doi.org/10.1103/PhysRevC.88.024316}
{\bibfield {journal} {\bibinfo {journal} {Phys. Rev. C}\ }\textbf {\bibinfo
{volume} {88}},\ \bibinfo {pages} {024316} (\bibinfo {year}
{2013})}\BibitemShut {NoStop}%
\bibitem [{\citenamefont {Kortelainen}\ \emph
{et~al.}(2010{\natexlab{b}})\citenamefont {Kortelainen}, \citenamefont
{Lesinski}, \citenamefont {Mor\'e}, \citenamefont {Nazarewicz}, \citenamefont
{Sarich}, \citenamefont {Schunck}, \citenamefont {Stoitsov},\ and\
\citenamefont {Wild}}]{PhysRevC.82.024313}%
\BibitemOpen
\bibfield {author} {\bibinfo {author} {\bibfnamefont {M.}~\bibnamefont
{Kortelainen}}, \bibinfo {author} {\bibfnamefont {T.}~\bibnamefont
{Lesinski}}, \bibinfo {author} {\bibfnamefont {J.}~\bibnamefont {Mor\'e}},
\bibinfo {author} {\bibfnamefont {W.}~\bibnamefont {Nazarewicz}}, \bibinfo
{author} {\bibfnamefont {J.}~\bibnamefont {Sarich}}, \bibinfo {author}
{\bibfnamefont {N.}~\bibnamefont {Schunck}}, \bibinfo {author} {\bibfnamefont
{M.~V.}\ \bibnamefont {Stoitsov}},\ and\ \bibinfo {author} {\bibfnamefont
{S.}~\bibnamefont {Wild}},\ }\href
{https://doi.org/10.1103/PhysRevC.82.024313} {\bibfield {journal} {\bibinfo
{journal} {Phys. Rev. C}\ }\textbf {\bibinfo {volume} {82}},\ \bibinfo
{pages} {024313} (\bibinfo {year} {2010}{\natexlab{b}})}\BibitemShut
{NoStop}%
\bibitem [{\citenamefont {Abbott}\ \emph {et~al.}(2017)\citenamefont {Abbott},
\citenamefont {Abbott}, \citenamefont {Abbott}, \citenamefont {Acernese},
\citenamefont {Ackley}, \citenamefont {Adams}, \citenamefont {Adams},
\citenamefont {Addesso}, \citenamefont {Adhikari}, \citenamefont {Adya},\
and\ \citenamefont {et~al.}}]{Abbott_2017}%
\BibitemOpen
\bibfield {author} {\bibinfo {author} {\bibfnamefont {B.}~\bibnamefont
{Abbott}}, \bibinfo {author} {\bibfnamefont {R.}~\bibnamefont {Abbott}},
\bibinfo {author} {\bibfnamefont {T.}~\bibnamefont {Abbott}}, \bibinfo
{author} {\bibfnamefont {F.}~\bibnamefont {Acernese}}, \bibinfo {author}
{\bibfnamefont {K.}~\bibnamefont {Ackley}}, \bibinfo {author} {\bibfnamefont
{C.}~\bibnamefont {Adams}}, \bibinfo {author} {\bibfnamefont
{T.}~\bibnamefont {Adams}}, \bibinfo {author} {\bibfnamefont
{P.}~\bibnamefont {Addesso}}, \bibinfo {author} {\bibfnamefont
{R.}~\bibnamefont {Adhikari}}, \bibinfo {author} {\bibfnamefont
{V.}~\bibnamefont {Adya}},\ and\ \bibinfo {author} {\bibnamefont {et~al.}},\
}\bibfield {journal} {\bibinfo {journal} {Physical Review Letters}\
}\textbf {\bibinfo {volume} {119}},\ \href
{https://doi.org/10.1103/physrevlett.119.161101}
{10.1103/physrevlett.119.161101} (\bibinfo {year} {2017})\BibitemShut
{NoStop}%
\bibitem [{\citenamefont {Abbott}\ \emph {et~al.}(2018)\citenamefont {Abbott},
\citenamefont {Abbott}, \citenamefont {Abbott}, \citenamefont {Acernese},
\citenamefont {Ackley}, \citenamefont {Adams}, \citenamefont {Adams},
\citenamefont {Addesso}, \citenamefont {Adhikari}, \citenamefont {Adya},\
and\ \citenamefont {et~al.}}]{Abbott_2018}%
\BibitemOpen
\bibfield {author} {\bibinfo {author} {\bibfnamefont {B.}~\bibnamefont
{Abbott}}, \bibinfo {author} {\bibfnamefont {R.}~\bibnamefont {Abbott}},
\bibinfo {author} {\bibfnamefont {T.}~\bibnamefont {Abbott}}, \bibinfo
{author} {\bibfnamefont {F.}~\bibnamefont {Acernese}}, \bibinfo {author}
{\bibfnamefont {K.}~\bibnamefont {Ackley}}, \bibinfo {author} {\bibfnamefont
{C.}~\bibnamefont {Adams}}, \bibinfo {author} {\bibfnamefont
{T.}~\bibnamefont {Adams}}, \bibinfo {author} {\bibfnamefont
{P.}~\bibnamefont {Addesso}}, \bibinfo {author} {\bibfnamefont
{R.}~\bibnamefont {Adhikari}}, \bibinfo {author} {\bibfnamefont
{V.}~\bibnamefont {Adya}},\ and\ \bibinfo {author} {\bibnamefont {et~al.}},\
}\bibfield {journal} {\bibinfo {journal} {Physical Review Letters}\
}\textbf {\bibinfo {volume} {121}},\ \href
{https://doi.org/10.1103/physrevlett.121.161101}
{10.1103/physrevlett.121.161101} (\bibinfo {year} {2018})\BibitemShut
{NoStop}%
\bibitem [{\citenamefont {Fattoyev}\ \emph {et~al.}(2018)\citenamefont
{Fattoyev}, \citenamefont {Piekarewicz},\ and\ \citenamefont
{Horowitz}}]{PhysRevLett.120.172702}%
\BibitemOpen
\bibfield {author} {\bibinfo {author} {\bibfnamefont {F.~J.}\ \bibnamefont
{Fattoyev}}, \bibinfo {author} {\bibfnamefont {J.}~\bibnamefont
{Piekarewicz}},\ and\ \bibinfo {author} {\bibfnamefont {C.~J.}\ \bibnamefont
{Horowitz}},\ }\href {https://doi.org/10.1103/PhysRevLett.120.172702}
{\bibfield {journal} {\bibinfo {journal} {Phys. Rev. Lett.}\ }\textbf
{\bibinfo {volume} {120}},\ \bibinfo {pages} {172702} (\bibinfo {year}
{2018})}\BibitemShut {NoStop}%
\bibitem [{\citenamefont {Anthony}\ \emph {et~al.}(2005)\citenamefont {Anthony}
\emph {et~al.}}]{Anthony:2005pm}%
\BibitemOpen
\bibfield {author} {\bibinfo {author} {\bibfnamefont {P.~L.}\ \bibnamefont
{Anthony}} \emph {et~al.} (\bibinfo {collaboration} {SLAC E158}),\ }\href
{https://doi.org/10.1103/PhysRevLett.95.081601} {\bibfield {journal}
{\bibinfo {journal} {Phys. Rev. Lett.}\ }\textbf {\bibinfo {volume} {95}},\
\bibinfo {pages} {081601} (\bibinfo {year} {2005})},\ \Eprint
{https://arxiv.org/abs/hep-ex/0504049} {hep-ex/0504049 [hep-ex]} \BibitemShut
{NoStop}%
\bibitem [{\citenamefont {Wang}\ \emph {et~al.}(2014)\citenamefont {Wang} \emph
{et~al.}}]{Wang:2014bba}%
\BibitemOpen
\bibfield {author} {\bibinfo {author} {\bibfnamefont {D.}~\bibnamefont
{Wang}} \emph {et~al.} (\bibinfo {collaboration} {PVDIS}),\ }\href
{https://doi.org/10.1038/nature12964} {\bibfield {journal} {\bibinfo
{journal} {Nature}\ }\textbf {\bibinfo {volume} {506}},\ \bibinfo {pages}
{67} (\bibinfo {year} {2014})}\BibitemShut {NoStop}%
\bibitem [{\citenamefont {Zeller}\ \emph {et~al.}(2002)\citenamefont {Zeller}
\emph {et~al.}}]{Zeller:2001hh}%
\BibitemOpen
\bibfield {author} {\bibinfo {author} {\bibfnamefont {G.~P.}\ \bibnamefont
{Zeller}} \emph {et~al.} (\bibinfo {collaboration} {NuTeV}),\ }\href@noop {}
{\bibfield {journal} {\bibinfo {journal} {Phys. Rev. Lett.}\ }\textbf
{\bibinfo {volume} {88}},\ \bibinfo {pages} {091802} (\bibinfo {year}
{2002})},\ \Eprint {https://arxiv.org/abs/hep-ex/0110059} {hep-ex/0110059}
\BibitemShut {NoStop}%
\bibitem [{\citenamefont {Androic}\ \emph {et~al.}(2018)\citenamefont {Androic}
\emph {et~al.}}]{Androic:2018kni}%
\BibitemOpen
\bibfield {author} {\bibinfo {author} {\bibfnamefont {D.}~\bibnamefont
{Androic}} \emph {et~al.} (\bibinfo {collaboration} {Qweak}),\ }\href
{https://doi.org/10.1038/s41586-018-0096-0} {\bibfield {journal} {\bibinfo
{journal} {Nature}\ }\textbf {\bibinfo {volume} {557}},\ \bibinfo {pages}
{207} (\bibinfo {year} {2018})}\BibitemShut {NoStop}%
\bibitem [{\citenamefont {Erler}\ and\ \citenamefont
{Ramsey-Musolf}(2005)}]{Erler:2004in}%
\BibitemOpen
\bibfield {author} {\bibinfo {author} {\bibfnamefont {J.}~\bibnamefont
{Erler}}\ and\ \bibinfo {author} {\bibfnamefont {M.~J.}\ \bibnamefont
{Ramsey-Musolf}},\ }\href {https://doi.org/10.1103/PhysRevD.72.073003}
{\bibfield {journal} {\bibinfo {journal} {Phys. Rev.}\ }\textbf {\bibinfo
{volume} {D72}},\ \bibinfo {pages} {073003} (\bibinfo {year} {2005})},\
\Eprint {https://arxiv.org/abs/hep-ph/0409169} {arXiv:hep-ph/0409169
[hep-ph]} \BibitemShut {NoStop}%
\bibitem [{\citenamefont {Erler}\ and\ \citenamefont
{Ferro-Hernández}(2018)}]{Erler:2017knj}%
\BibitemOpen
\bibfield {author} {\bibinfo {author} {\bibfnamefont {J.}~\bibnamefont
{Erler}}\ and\ \bibinfo {author} {\bibfnamefont {R.}~\bibnamefont
{Ferro-Hernández}},\ }\href {http://dx.doi.org/10.1007/JHEP03(2018)196}
{\bibfield {journal} {\bibinfo {journal} {JHEP}\ }\textbf {\bibinfo
{volume} {03}},\ \bibinfo {pages} {196}},\ \Eprint
{https://arxiv.org/abs/1712.09146} {arXiv:1712.09146 [hep-ph]} \BibitemShut
{NoStop}%
\bibitem [{\citenamefont {Akimov}\ \emph
{et~al.}(2018{\natexlab{b}})\citenamefont {Akimov} \emph
{et~al.}}]{akimov2018coherent}%
\BibitemOpen
\bibfield {author} {\bibinfo {author} {\bibfnamefont {D.}~\bibnamefont
{Akimov}} \emph {et~al.} (\bibinfo {collaboration} {COHERENT
collaboration}),\ }\href@noop {} {\bibinfo {title} {Coherent 2018 at the
spallation neutron source}} (\bibinfo {year} {2018}{\natexlab{b}}),\ \Eprint
{https://arxiv.org/abs/1803.09183} {arXiv:1803.09183 [physics.ins-det]}
\BibitemShut {NoStop}%
\bibitem [{\citenamefont {Bonet}\ \emph {et~al.}(2021)\citenamefont {Bonet},
\citenamefont {Bonhomme}, \citenamefont {Buck}, \citenamefont {Fülber},
\citenamefont {Hakenmüller}, \citenamefont {Heusser}, \citenamefont {Hugle},
\citenamefont {Lindner}, \citenamefont {Maneschg}, \citenamefont {Rink},\
and\ \citenamefont {et~al.}}]{Bonet_2021}%
\BibitemOpen
\bibfield {author} {\bibinfo {author} {\bibfnamefont {H.}~\bibnamefont
{Bonet}}, \bibinfo {author} {\bibfnamefont {A.}~\bibnamefont {Bonhomme}},
\bibinfo {author} {\bibfnamefont {C.}~\bibnamefont {Buck}}, \bibinfo {author}
{\bibfnamefont {K.}~\bibnamefont {Fülber}}, \bibinfo {author} {\bibfnamefont
{J.}~\bibnamefont {Hakenmüller}}, \bibinfo {author} {\bibfnamefont
{G.}~\bibnamefont {Heusser}}, \bibinfo {author} {\bibfnamefont
{T.}~\bibnamefont {Hugle}}, \bibinfo {author} {\bibfnamefont
{M.}~\bibnamefont {Lindner}}, \bibinfo {author} {\bibfnamefont
{W.}~\bibnamefont {Maneschg}}, \bibinfo {author} {\bibfnamefont
{T.}~\bibnamefont {Rink}},\ and\ \bibinfo {author} {\bibnamefont {et~al.}},\
}\bibfield {journal} {\bibinfo {journal} {Physical Review Letters}\
}\textbf {\bibinfo {volume} {126}},\ \href
{https://doi.org/10.1103/physrevlett.126.041804}
{10.1103/physrevlett.126.041804} (\bibinfo {year} {2021})\BibitemShut
{NoStop}%
\bibitem [{\citenamefont {Davoudiasl}\ \emph {et~al.}(2012)\citenamefont
{Davoudiasl}, \citenamefont {Lee},\ and\ \citenamefont
{Marciano}}]{PhysRevLett.109.031802}%
\BibitemOpen
\bibfield {author} {\bibinfo {author} {\bibfnamefont {H.}~\bibnamefont
{Davoudiasl}}, \bibinfo {author} {\bibfnamefont {H.-S.}\ \bibnamefont
{Lee}},\ and\ \bibinfo {author} {\bibfnamefont {W.~J.}\ \bibnamefont
{Marciano}},\ }\href {https://doi.org/10.1103/PhysRevLett.109.031802}
{\bibfield {journal} {\bibinfo {journal} {Phys. Rev. Lett.}\ }\textbf
{\bibinfo {volume} {109}},\ \bibinfo {pages} {031802} (\bibinfo {year}
{2012})}\BibitemShut {NoStop}%
\bibitem [{\citenamefont {Davoudiasl}\ \emph {et~al.}(2014)\citenamefont
{Davoudiasl}, \citenamefont {Lee},\ and\ \citenamefont
{Marciano}}]{Davoudiasl:2014kua}%
\BibitemOpen
\bibfield {author} {\bibinfo {author} {\bibfnamefont {H.}~\bibnamefont
{Davoudiasl}}, \bibinfo {author} {\bibfnamefont {H.-S.}\ \bibnamefont
{Lee}},\ and\ \bibinfo {author} {\bibfnamefont {W.~J.}\ \bibnamefont
{Marciano}},\ }\href {https://doi.org/10.1103/PhysRevD.89.095006} {\bibfield
{journal} {\bibinfo {journal} {Phys. Rev. D}\ }\textbf {\bibinfo {volume}
{89}},\ \bibinfo {pages} {095006} (\bibinfo {year} {2014})},\ \Eprint
{https://arxiv.org/abs/1402.3620} {arXiv:1402.3620 [hep-ph]} \BibitemShut
{NoStop}%
\bibitem [{\citenamefont {Davoudiasl}\ \emph {et~al.}(2015)\citenamefont
{Davoudiasl}, \citenamefont {Lee},\ and\ \citenamefont
{Marciano}}]{PhysRevD.92.055005}%
\BibitemOpen
\bibfield {author} {\bibinfo {author} {\bibfnamefont {H.}~\bibnamefont
{Davoudiasl}}, \bibinfo {author} {\bibfnamefont {H.-S.}\ \bibnamefont
{Lee}},\ and\ \bibinfo {author} {\bibfnamefont {W.~J.}\ \bibnamefont
{Marciano}},\ }\href {https://doi.org/10.1103/PhysRevD.92.055005} {\bibfield
{journal} {\bibinfo {journal} {Phys. Rev. D}\ }\textbf {\bibinfo {volume}
{92}},\ \bibinfo {pages} {055005} (\bibinfo {year} {2015})}\BibitemShut
{NoStop}%
\bibitem [{\citenamefont {Cadeddu}\ \emph
{et~al.}(2021{\natexlab{b}})\citenamefont {Cadeddu}, \citenamefont
{Cargioli}, \citenamefont {Dordei}, \citenamefont {Giunti},\ and\
\citenamefont {Picciau}}]{cadeddu2021muon}%
\BibitemOpen
\bibfield {author} {\bibinfo {author} {\bibfnamefont {M.}~\bibnamefont
{Cadeddu}}, \bibinfo {author} {\bibfnamefont {N.}~\bibnamefont {Cargioli}},
\bibinfo {author} {\bibfnamefont {F.}~\bibnamefont {Dordei}}, \bibinfo
{author} {\bibfnamefont {C.}~\bibnamefont {Giunti}},\ and\ \bibinfo {author}
{\bibfnamefont {E.}~\bibnamefont {Picciau}},\ }\href@noop {} {\bibinfo
{title} {Muon and electron g-2, proton and cesium weak charges implications
on dark $\mathbf{Z_d}$ models}} (\bibinfo {year} {2021}{\natexlab{b}}),\
\Eprint {https://arxiv.org/abs/2104.03280} {arXiv:2104.03280 [hep-ph]}
\BibitemShut {NoStop}%
\bibitem [{\citenamefont {Becker}\ \emph {et~al.}(2018)\citenamefont {Becker}
\emph {et~al.}}]{Becker:2018ggl}%
\BibitemOpen
\bibfield {author} {\bibinfo {author} {\bibfnamefont {D.}~\bibnamefont
{Becker}} \emph {et~al.},\ }\href
{https://doi.org/10.1140/epja/i2018-12611-6} {\bibfield {journal} {\bibinfo
{journal} {Eur. Phys. J. A}\ }\textbf {\bibinfo {volume} {54}},\ \bibinfo
{pages} {208} (\bibinfo {year} {2018})},\ \Eprint
{https://arxiv.org/abs/1802.04759} {arXiv:1802.04759 [nucl-ex]} \BibitemShut
{NoStop}%
\bibitem [{\citenamefont {Benesch}\ \emph {et~al.}(2014)\citenamefont {Benesch}
\emph {et~al.}}]{Benesch:2014bas}%
\BibitemOpen
\bibfield {author} {\bibinfo {author} {\bibfnamefont {J.}~\bibnamefont
{Benesch}} \emph {et~al.} (\bibinfo {collaboration} {MOLLER}),\ }\href@noop
{} {\bibinfo {title} {{The MOLLER Experiment: An Ultra-Precise Measurement of
the Weak Mixing Angle Using M{\textbackslash{}o}ller Scattering}}} (\bibinfo
{year} {2014}),\ \Eprint {https://arxiv.org/abs/1411.4088} {arXiv:1411.4088
[nucl-ex]} \BibitemShut {NoStop}%
\bibitem [{\citenamefont {de~Gouvea}\ \emph {et~al.}(2020)\citenamefont
{de~Gouvea}, \citenamefont {Machado}, \citenamefont {Perez-Gonzalez},\ and\
\citenamefont {Tabrizi}}]{deGouvea:2019wav}%
\BibitemOpen
\bibfield {author} {\bibinfo {author} {\bibfnamefont {A.}~\bibnamefont
{de~Gouvea}}, \bibinfo {author} {\bibfnamefont {P.~A.~N.}\ \bibnamefont
{Machado}}, \bibinfo {author} {\bibfnamefont {Y.~F.}\ \bibnamefont
{Perez-Gonzalez}},\ and\ \bibinfo {author} {\bibfnamefont {Z.}~\bibnamefont
{Tabrizi}},\ }\href {https://doi.org/10.1103/PhysRevLett.125.051803}
{\bibfield {journal} {\bibinfo {journal} {Phys. Rev. Lett.}\ }\textbf
{\bibinfo {volume} {125}},\ \bibinfo {pages} {051803} (\bibinfo {year}
{2020})},\ \Eprint {https://arxiv.org/abs/1912.06658} {arXiv:1912.06658
[hep-ph]} \BibitemShut {NoStop}%
\bibitem [{\citenamefont {Cadeddu}\ \emph {et~al.}(2019)\citenamefont
{Cadeddu}, \citenamefont {Dordei}, \citenamefont {Giunti}, \citenamefont
{Kouzakov}, \citenamefont {Picciau},\ and\ \citenamefont
{Studenikin}}]{Cadeddu_2019}%
\BibitemOpen
\bibfield {author} {\bibinfo {author} {\bibfnamefont {M.}~\bibnamefont
{Cadeddu}}, \bibinfo {author} {\bibfnamefont {F.}~\bibnamefont {Dordei}},
\bibinfo {author} {\bibfnamefont {C.}~\bibnamefont {Giunti}}, \bibinfo
{author} {\bibfnamefont {K.}~\bibnamefont {Kouzakov}}, \bibinfo {author}
{\bibfnamefont {E.}~\bibnamefont {Picciau}},\ and\ \bibinfo {author}
{\bibfnamefont {A.}~\bibnamefont {Studenikin}},\ }\bibfield {journal}
{\bibinfo {journal} {Physical Review D}\ }\textbf {\bibinfo {volume}
{100}},\ \href {https://doi.org/10.1103/physrevd.100.073014}
{10.1103/physrevd.100.073014} (\bibinfo {year} {2019})\BibitemShut {NoStop}%
\bibitem [{\citenamefont {Fernandez-Moroni}\ \emph {et~al.}(2021)\citenamefont
{Fernandez-Moroni}, \citenamefont {Machado}, \citenamefont {Martinez-Soler},
\citenamefont {Perez-Gonzalez}, \citenamefont {Rodrigues},\ and\
\citenamefont {Rosauro-Alcaraz}}]{Fernandez_Moroni_2021}%
\BibitemOpen
\bibfield {author} {\bibinfo {author} {\bibfnamefont {G.}~\bibnamefont
{Fernandez-Moroni}}, \bibinfo {author} {\bibfnamefont {P.~A.~N.}\
\bibnamefont {Machado}}, \bibinfo {author} {\bibfnamefont {I.}~\bibnamefont
{Martinez-Soler}}, \bibinfo {author} {\bibfnamefont {Y.~F.}\ \bibnamefont
{Perez-Gonzalez}}, \bibinfo {author} {\bibfnamefont {D.}~\bibnamefont
{Rodrigues}},\ and\ \bibinfo {author} {\bibfnamefont {S.}~\bibnamefont
{Rosauro-Alcaraz}},\ }\bibfield {journal} {\bibinfo {journal} {Journal of
High Energy Physics}\ }\textbf {\bibinfo {volume} {2021}},\ \href
{https://doi.org/10.1007/jhep03(2021)186} {10.1007/jhep03(2021)186} (\bibinfo
{year} {2021})\BibitemShut {NoStop}%
\bibitem [{\citenamefont {Cañas}\ \emph {et~al.}(2018)\citenamefont {Cañas},
\citenamefont {Garcés}, \citenamefont {Miranda},\ and\ \citenamefont
{Parada}}]{Ca_as_2018}%
\BibitemOpen
\bibfield {author} {\bibinfo {author} {\bibfnamefont {B.}~\bibnamefont
{Cañas}}, \bibinfo {author} {\bibfnamefont {E.}~\bibnamefont {Garcés}},
\bibinfo {author} {\bibfnamefont {O.}~\bibnamefont {Miranda}},\ and\ \bibinfo
{author} {\bibfnamefont {A.}~\bibnamefont {Parada}},\ }\href
{https://doi.org/10.1016/j.physletb.2018.07.049} {\bibfield {journal}
{\bibinfo {journal} {Physics Letters B}\ }\textbf {\bibinfo {volume}
{784}},\ \bibinfo {pages} {159–162} (\bibinfo {year} {2018})}\BibitemShut
{NoStop}%
\bibitem [{\citenamefont {Marciano}\ and\ \citenamefont
{Sirlin}(1983)}]{Marciano_Sirlin}%
\BibitemOpen
\bibfield {author} {\bibinfo {author} {\bibfnamefont {W.~J.}\ \bibnamefont
{Marciano}}\ and\ \bibinfo {author} {\bibfnamefont {A.}~\bibnamefont
{Sirlin}},\ }\href {https://doi.org/10.1103/PhysRevD.27.552} {\bibfield
{journal} {\bibinfo {journal} {Phys. Rev. D}\ }\textbf {\bibinfo {volume}
{27}},\ \bibinfo {pages} {552} (\bibinfo {year} {1983})}\BibitemShut
{NoStop}%
\bibitem [{\citenamefont {Marciano}\ and\ \citenamefont
{Sirlin}(1984)}]{Marciano_Sirlin_2}%
\BibitemOpen
\bibfield {author} {\bibinfo {author} {\bibfnamefont {W.~J.}\ \bibnamefont
{Marciano}}\ and\ \bibinfo {author} {\bibfnamefont {A.}~\bibnamefont
{Sirlin}},\ }\href {https://doi.org/10.1103/PhysRevD.29.75} {\bibfield
{journal} {\bibinfo {journal} {Phys. Rev. D}\ }\textbf {\bibinfo {volume}
{29}},\ \bibinfo {pages} {75} (\bibinfo {year} {1984})}\BibitemShut {NoStop}%
\bibitem [{\citenamefont {Alitti}\ \emph {et~al.}(1991)\citenamefont {Alitti},
\citenamefont {Ambrosini}, \citenamefont {Ansari}, \citenamefont {Autiero},
\citenamefont {Bareyre}, \citenamefont {Bertram}, \citenamefont {Blaylock},
\citenamefont {Bonamy}, \citenamefont {Bonesini}, \citenamefont {Borer},
\citenamefont {Bourliaud}, \citenamefont {Buskulic}, \citenamefont {Carboni},
\citenamefont {Cavalli}, \citenamefont {Cavasinni}, \citenamefont {Cenci},
\citenamefont {Chollet}, \citenamefont {Conta}, \citenamefont {Costa},\ and\
\citenamefont {Iacopini}}]{alitti}%
\BibitemOpen
\bibfield {author} {\bibinfo {author} {\bibfnamefont {J.}~\bibnamefont
{Alitti}}, \bibinfo {author} {\bibfnamefont {G.}~\bibnamefont {Ambrosini}},
\bibinfo {author} {\bibfnamefont {R.}~\bibnamefont {Ansari}}, \bibinfo
{author} {\bibfnamefont {D.}~\bibnamefont {Autiero}}, \bibinfo {author}
{\bibfnamefont {P.}~\bibnamefont {Bareyre}}, \bibinfo {author} {\bibfnamefont
{I.}~\bibnamefont {Bertram}}, \bibinfo {author} {\bibfnamefont
{G.}~\bibnamefont {Blaylock}}, \bibinfo {author} {\bibfnamefont
{P.}~\bibnamefont {Bonamy}}, \bibinfo {author} {\bibfnamefont
{M.}~\bibnamefont {Bonesini}}, \bibinfo {author} {\bibfnamefont
{K.}~\bibnamefont {Borer}}, \bibinfo {author} {\bibfnamefont
{M.}~\bibnamefont {Bourliaud}}, \bibinfo {author} {\bibfnamefont
{D.}~\bibnamefont {Buskulic}}, \bibinfo {author} {\bibfnamefont
{G.}~\bibnamefont {Carboni}}, \bibinfo {author} {\bibfnamefont
{D.}~\bibnamefont {Cavalli}}, \bibinfo {author} {\bibfnamefont
{V.}~\bibnamefont {Cavasinni}}, \bibinfo {author} {\bibfnamefont
{P.}~\bibnamefont {Cenci}}, \bibinfo {author} {\bibfnamefont
{J.}~\bibnamefont {Chollet}}, \bibinfo {author} {\bibfnamefont
{C.}~\bibnamefont {Conta}}, \bibinfo {author} {\bibfnamefont
{G.}~\bibnamefont {Costa}},\ and\ \bibinfo {author} {\bibfnamefont
{E.}~\bibnamefont {Iacopini}},\ }\href@noop {} {\bibfield {journal}
{\bibinfo {journal} {Physics Letters B}\ }\textbf {\bibinfo {volume}
{263}},\ \bibinfo {pages} {563–572} (\bibinfo {year} {1991})}\BibitemShut
{NoStop}%
\end{thebibliography}%

\end{document}